\documentclass[a4paper,fleqn,usenatbib]{mnras}
\pdfoutput=1 

\usepackage[T1]{fontenc}
\usepackage{ae,aecompl}

\usepackage{graphicx}	
\usepackage{amsmath}	
\usepackage{amssymb}	

\newcommand{\myemail}{mpound@umd.edu}

\newcommand{\sio}{\ifmmode{{\rm SiO(J=2-1)}} \else{SiO(J=2-1)\/}\fi}
\newcommand{\sioA}{\ifmmode{{\rm SiO}} \else{SiO\/}\fi}
\newcommand{\hcn}{\ifmmode{{\rm HCN(J=1-0)}} \else{HCN(J=1-0)\/}\fi}
\newcommand{\hcnA}{\ifmmode{{\rm HCN}} \else{HCN}\fi}
\newcommand{\hcop}{\ifmmode{{\rm HCO^{+}(J=1-0)}} \else{HCO$^{+}$(J=1-0)\/}\fi}
\newcommand{\hcopA}{\ifmmode{{\rm HCO^{+}}} \else{HCO$^{+}$\/}\fi}
\newcommand{\ntwohp}{\ifmmode{{\rm N_{2}H^{+}(J=1-0)}} \else{N$_{2}$H$^{+}$(J=1-0)\/}\fi}
\newcommand{\ntwohpA}{\ifmmode{{\rm N_{2}H^{+}}} \else{N$_{2}$H$^{+}$\/}\fi}
\newcommand{\cs}{\ifmmode{{\rm CS(J=2-1)}} \else{CS(J=2-1)\/}\fi}
\newcommand{\vlsr}{\ifmmode{\rm V_{LSR}} \else{V$_{\rm LSR}$\/}\fi}
\newcommand{\feka}{\ifmmode{{\rm Fe~K}\alpha} \else{Fe~K$\alpha$\/}\fi}
\newcommand{\nexpo}[2]{\ifmmode{#1 \times 10^{#2}}\else{$#1 \times 10^{#2}$}\fi}
\newcommand{\expo}[1]{\ifmmode{10^{#1}}\else{$10^{#1}$}\fi}

\newcommand\pasec           {$.\negthinspace^{\prime\prime}$}
\newcommand\pdeg           {$.\kern-.25em ^{^\circ}$}
\newcommand\dV{\ifmmode{\,{\Delta V}}\else{{$\Delta V$}}\fi}
\newcommand\kms{\ifmmode{\,{\rm km~s^{-1}}}\else{{${\rm km~s^{-1}}$}}\fi}
\newcommand\arcdeg{\mbox{$^\circ$}}%

\title[CARMA 3~mm Survey of the CMZ]{The CARMA 3~mm Survey of the Inner $0.7^\circ\times0.4^\circ$ of the Central Molecular Zone}

\author[Pound and Yusef-Zadeh]{
Marc W. Pound,$^{1}$\thanks{E-mail: \myemail}
and Farhad Yusef-Zadeh$^{2}$\thanks{E-mail: zadeh@northwestern.edu}
\\
$^{1}$Department of Astronomy, University of Maryland,College Park, MD 20742\\
$^{2}$Department of Physics and Astronomy, Northwestern University, Evanston, IL 60208
}

\date{Accepted 2017 September 22. Received 2017 September 18; in original
form 2017 July 19}

\pubyear{2017}

\begin{document}
\label{firstpage}
\pagerange{\pageref{firstpage}--\pageref{lastpage}}
\maketitle

\begin{abstract} 
The Central Molecular Zone (CMZ) of the Galactic Center has to date only been fully mapped at mm wavelengths with singledish telescopes, with resolution about 30\arcsec\ (1.2 pc).  Using CARMA, we mapped the innermost 0.25 square degrees of the CMZ over the region between --0\pdeg{2}$\leq l \leq $0\pdeg{5} and --0\pdeg{2}$ \leq b \leq $0\pdeg{2} (90$\times$50 pc) with spatial and spectral resolution of $\sim 10\arcsec$ (0.4 pc) 
and $\sim 2.5~\kms$, respectively.  We provide a catalog of 3~mm continuum sources as well as spectral line images of \sio, \hcop, \hcn, \ntwohp, and \cs, with velocity coverage \vlsr = --200 to 200~\kms.  To recover the large scale structure resolved out by the interferometer, the continuum-subtracted spectral line images were combined with data from the Mopra 22-m telescope survey, thus providing maps containing all spatial frequencies down to the resolution limit. We find that integrated intensity ratio of I(HCN)/I(\hcopA) is anti-correlated with the intensity of the 6.4 keV Fe K$\alpha$, which is excited either by high energy photons or low energy cosmic rays, and the gas velocity dispersion as traced by \hcopA\ is correlated with Fe K$\alpha$ intensity.  The intensity ratio and velocity dispersion patterns are consistent with variation expected from the interaction of low energy cosmic rays with molecular gas.
\end{abstract}


\section{Introduction}

The inner few hundred pc of the Galactic Center differs from the rest of the Galaxy in its ISM properties. The 
central molecular zone (CMZ) is occupied by an impressive collection of massive molecular clouds 
which are characterized by a rich chemistry, broad linewidths, elevated temperatures and high density compared to 
those in the disk of the Galaxy. There are several observations suggesting that the ISM in this region is also 
characterized by strong nonthermal emission compared to the rest of the Galaxy.  First, radio continuum observations 
have uncovered a population of filamentary structures within the two degrees of the Galactic Center 
\citep[e.g.,][]{1984Natur.310..557Y,2004AJ....128.1646N}. Their transverse dimensions are of the order of 1 pc and their length is of the order of tens 
of parsecs. Most of these filaments run perpendicular to the Galactic plane. Their strongly linear polarized
emission and their radio spectral index confirm that the filaments are produced by synchrotron emission from 
relativistic electrons in a highly organized linear magnetic field. In addition, low frequency radio observations at 327 
and 74 MHz emission suggest that the Galactic Center hosts a population of low energy relativistic particles 
\citep{2003ANS...324...17B,2005ApJ...626L..23L,2013JPCA..117.9404Y}. Second, X-ray and $\gamma$-ray observatons detect strong sources of 6.4 keV 
line, and GeV and TeV $\gamma$-ray emission \citep{2010ApJ...714..732P,2016Natur.531..476H,2016ApJ...821..129A}. The neutral Fe 6.4 keV K$\alpha$ line 
emission traces neutral cold gas. One possible interpretion of high energy activity in this region is discussed in 
terms a bremsstrahlung emission due to the interaction of cosmic rays with molecular gas \citep{2013ApJ...762...33Y}. Lastly, 
recent measurements indicate a vast amount of H$_3^+$ \citep{2005ApJ...632..882O,2014ApJ...786...96G,2016A&A...585A.105L} distributed in the Galactic Center 
region, indicative of a high rate of ionization of H$_2$. The inferred 
cosmic ray ionization rate, $\zeta\sim 10^{-15}$\, s$^{-1}$\, H$^{-1}$, is one to two 
orders of magnitude higher in the Galactic Center region than in the Galactic disk. 
A possible cause of the abundance of molecular gas in the Galactic Center has also been 
discussed in the context of cosmic rays \citep{2017MNRAS.467..737C}.  So, a key question is 
whether there are other  chemical signatures that would support 
the interaction of molecular gas  with relativistic particles. 


Numerous molecular line survey observations of the CMZ over the last
30 years have solely been carried out with single dish telescopes with
moderate resolutions.  Comprehensive surveys of the CMZ have been 
out recently by the Mopra telescope at 20-28 GHz \citep{2011MNRAS.416.1764W}, 42-50 GHz \citep{2013MNRAS.433..221J} and 85-93 GHz \citep{2012MNRAS.419.2961J}.
This latter survey fully mapped 18 molecular lines emitting
from the inner 2\pdeg5$\times$0\pdeg5 ($l\times b$) and has both spatial and spectral overlap with our work. 

In this paper, we present the first interferometric 
molecular line survey of  the innermost region of the CMZ. 
One of the key motivations for the survey is to examine the relationship
between the distribution of the molecular line  and nonthermal 
emission in this complex region of the Galaxy.
We don't fully understand 
the complex gas chemistry in this region of the Galaxy but 
assuming that  there is interaction between 
cosmic rays and  molecular gas, 
the  gas chemistry is expected to be  driven by cosmic rays \citep{2017MNRAS.467..737C}.
One recent example shows 
the evidence for thousands of collisionally excited methanol masers  in the central molecular zone \citep{2015ApJ...805...72M,2016ApJS..227...10C}.
It is not clear if these masers are produced by star formation activity or
by the interaction between enhanced cosmic rays and molecular gas in the Galactic Center region.


We present molecular line maps of the survey at one transition
of each of 6 molecular species.  Combining these data with future multi-transition maps at high
frequencies could determine the abundance of different species, and the
gas density and temperature, which are needed to examine  the unusual gas chemistry in this region. 


We have also carried out 3 mm continuum observations of the same region observed in spectral lines. 
Although there is a high concentration of diffuse HII complexes distributed
in the Galactic Center region, there is no published large-area, high frequency, continuum data toward this region. 
Previous continuum surveys of the Galactic Center have not studied  
the class of  compact HII regions at 3 mm, which generally trace high emission measure HII regions. 
Our continuum measurements
identify a population of ultracompact HII regions
and new sites of early star formation.  Again, the images presented here 
need multi-frequency continuum images to separate the contribution of free-free and dust emission. 
These measurements can then
be correlated with other tracers of young massive star formation such
as H$_2$O, methanol masers, green fuzzies, and 24~$\mu$m survey data.

\section{Observations}

The survey was conducted using the Combined Array for Research in Millimeter-wave Astronomy \cite[CARMA,][]{2006SPIE.6267E..13B}.
For this survey, we used two of CARMA's science subarray modes.  The first,
called CARMA-15, is the 15-element array comprised of the six 10.4~m
antennas and nine 6.1~m antennas.  The second, CARMA-8, is the 8-element
array of 3.5~m antennas.  We chose CARMA-15 plus CARMA-8 over CARMA-23
(all 23 antennas cross-correlated) because the continuum bandwidth
available in CARMA-23 mode was half that in CARMA-15 plus CARMA-8.

\subsection{CARMA-15}
During 2012-2013, we observed 527 fields with CARMA-15 in the compact
D-array configuration on a full beam width hexagonal mosaic. The area covered
by these fields is --0\pdeg{2}~$\leq~l~\leq~$0\pdeg{5}
and --0\pdeg{2}$~\leq~b~\leq $0\pdeg{2}.
By sampling on
the full beam width (60\arcsec) of the 10.4~m antennas rather than Nyquist sampling,
we cover much larger area, paying a modest price in image fidelity.
The CARMA-15 signals were transmitted  the eight-band spectral line correlator.
In the spectral correlator, the eight bands are independently positioned
within the 1-9 GHz IF space, providing eight windows in each side band.
Individual bands can be set to differing bandwidths between 2 MHz and
500 MHz, and with selected channel spacings.  These data contain both
spectral line and continuum windows. The spectral lines \sio, \hcn,
\hcop, \ntwohp, and \cs, were observed in four 125 MHz bandwidths with
spectral resolution 781 kHz/channel.  The spectral bands cover roughly
\vlsr = --200 to 200 \kms\ with velocity resolution $\Delta$V $\sim$2.5 \kms.
The remaining four bands were used to measure continuum, each with 500 MHz
bandwidth and 31.25 MHz/channel.  


We added to these data another epoch of CARMA data taken in the C and D configurations of CARMA  in 2009-2010.  
These observations obtained 
continuum,
\ntwohp, and \sio\, data. 
The 2009-2010 data are a 37-point hexagonal mosaic with total diameter $\sim 380\arcsec$, Nyquist-sampled on the
10.4~m beam and centered on Sgr A*.  



\subsection{CARMA-8}
Using the CARMA-8 in the SL configuration, which is optimized for low declination
sources, we obtained shorter {\it uv} spacing continuum data over 273 Nyquist-sampled
fields covering the same area as the CARMA-15 data.  The CARMA-8 signals
were transmitted to the seven-band 7 GHz continuum correlator.  In this correlator,
seven 500 MHz bands with 31.25 MHz/channel each are independently set to
broadly cover the IF space, providing seven windows in each side band.



\subsection{Mapping Strategy}
All 2012-2013 observations were obtained using an on-the-fly (OTF) mosaic
technique in which data are continuously integrated while the antennas
move to each map grid point, and integrations between map grid points are 
blanked and thus do not contribute to the map.
This method significantly reduces observation overhead compared to the
traditional ``point-and-shoot'' mosaic method \citep{2014ApJ...794..165S}. Because
this technique allowed us to cover all mosaic fields at least once in
each sidereal pass, it eliminates weather-related variations that could
occur if different portions of the full map were covered on different days.
By starting each day's observations at a different map field, or rastering
in the opposite direction, we ensured similar {\it uv} coverage in
every field of the final map.

\begin{table*}
\centering
\caption{Observation Summary}
\label{t-obssummary}
\begin{tabular}{cccccccc}
\hline
 & 
Array & 
 & 
Number & 
\multicolumn{2}{c}{Baseline (k$\lambda$)} & 
Mapping & 
Flux \\
Date & 
Config. & 
Correlator & 
of Tracks & 
Min. &  
Max. &  
Mode & 
Calibrator(s) \\
\hline
2009/02/23 - 2010/05/10 & CARMA-15 D & Spectral+Continuum & 6 & 2.96 & 34.60 & Mosaic & 3C273,MWC349\\
2009/05/05 - 2010/03/28 & CARMA-15 C & Spectral+Continuum & 9 & 7.08 & 87.14 & Mosaic & Neptune,3C273,MWC349 \\
2010/05/18 - 2010/05/26 & CARMA-15 D & Continuum          & 3 & 3.14 & 36.52 & Mosaic & Neptune, MWC349\\
2012/04/18 - 2012/05/07 & CARMA-15 D & Spectral+Continuum & 7 & 3.14 & 36.70 & OTF    & Neptune, MWC349,3C273 \\
2012/10/19 - 2013/02/02 & CARMA-8 SL & Continuum          & 18 & 1.58 & 22.54 & OTF    & Neptune, Venus, MWC349, \\
                        &            &                    &    &      &       &        & 3C279, 2015+372\\
2014/12/20 - 2015/01/12 & CARMA-15 D & Continuum          & 3  & 3.14 & 36.70 & OTF    & MWC349 \\
\hline
\end{tabular}
\end{table*}

Details of the observations are provided in Table \ref{t-obssummary}.
Entries in columns 1-8 respectively of Table \ref{t-obssummary} correspond
to date range of observations, array configuration, correlator mode, number
of sidereal tracks observed, minimum and maximum baseline separations,
whether the observations were traditional mosaic or OTF, and sources used
as flux calibrators.  For all observations, the phase calibration source
was the quasar 1733-130. The flux density of 1733-130 varied between 1.9 and 3.5 Jy over the observing period. 
The uncertainty in absolute flux calibration is about 10\%.

The data were reduced using MIRIAD \citep{1995ASPC...77..433S}.  After phase,
amplitude, passband, and flux calibration, and flagging of bad data, visibilities
were inverted onto a 1\arcsec\ grid using a robust weighting value of
zero \citep{1995AAS...18711202B}.  Deconvolution was done 
with the MIRIAD task {\it mossdi}
which uses CLEAN algorithm of \citet{1984A&A...137..159S}.  The deconvolved maps
were then restored with a fitted 2D Gaussian beam.

For spectral line images, continuum subtraction was done in the image
domain by making a dirty map of the first several emission-free channels,
and subtracting that image from each plane of the dirty spectral map before
deconvolution.  The continuum-subtracted, restored spectral line images
were combined with the singledish survey from the Mopra 22-m telescope
(HPBW=39\arcsec) using the MIRIAD task {\it immerge}, except for CS which was
not covered by the Mopra survey \citep{2012MNRAS.419.2961J}.  In this process, the
Mopra antenna temperature maps were regridded to the CARMA pixel size, map center, and spectral
resolution, and multiplied by 22 Jy/K \citep{2012MNRAS.419.2961J} to convert them
to Jy beam$^{-1}$.  We applied flux scale correction factors to put the Mopra
data after conversion to Jy on the CARMA flux scale, derived from the
overlapping Fourier annulus (6.1m - 22m). Specifically the Mopra intensities were 
multiplied by 0.83 (SiO), 0.84 (HCN), and
0.95 (HCO$^+$ and N$_2$H$^+$).  Final map parameters are given in Table
\ref{t-mapsummary}.  Columns 1-5 of Table \ref{t-mapsummary}
list the observed species, synthesized beam size, synthesized beam position
angle, spectral resolution, and average 1$\sigma$ rms noise per channel, respectively.
\begin{table*}
\caption{Summary of parameters of final maps. $^1$Observed rest frequency. 
$^2$Spectral resolution.}
\label{t-mapsummary}
\begin{tabular}{llcccc}
\hline
       &
$\nu_{rest}$\footnotemark[1] &
Beam Size&
Beam PA  &
dV\footnotemark[2]&
RMS     \\
Species & 
(GHz)   &
(\arcsec) & 
(degrees) & 
(\kms)    &
(mJy beam$^{-1}$) \\
\hline
Continuum & 90         &  $7.6\times3.5$ & 67  &  --  & 2-4 \\
\sio      & 86.846998  & $13.1\times5.9$ & 8.5 & 2.70 & 250 \\
\hcn      & 88.631847  & $13.2\times5.9$ & 8.7 & 2.64 & 270 \\
\hcop     & 89.188518  & $13.1\times5.9$ & 8.7 & 2.63 & 280 \\
\ntwohp   & 93.9713505 & $12.3\times5.6$ & 8.0 & 2.51 & 240 \\
\cs       & 97.980968  & $11.8\times5.3$ & 8.2 & 2.39 & 310 \\
\hline

\end{tabular}
\end{table*}

\section{Results}

\subsection{3 mm Continuum Data}\label{s-3mmcont}

\begin{figure*}
\includegraphics[width=2\columnwidth,trim={1cm 0cm 1cm 0cm}]{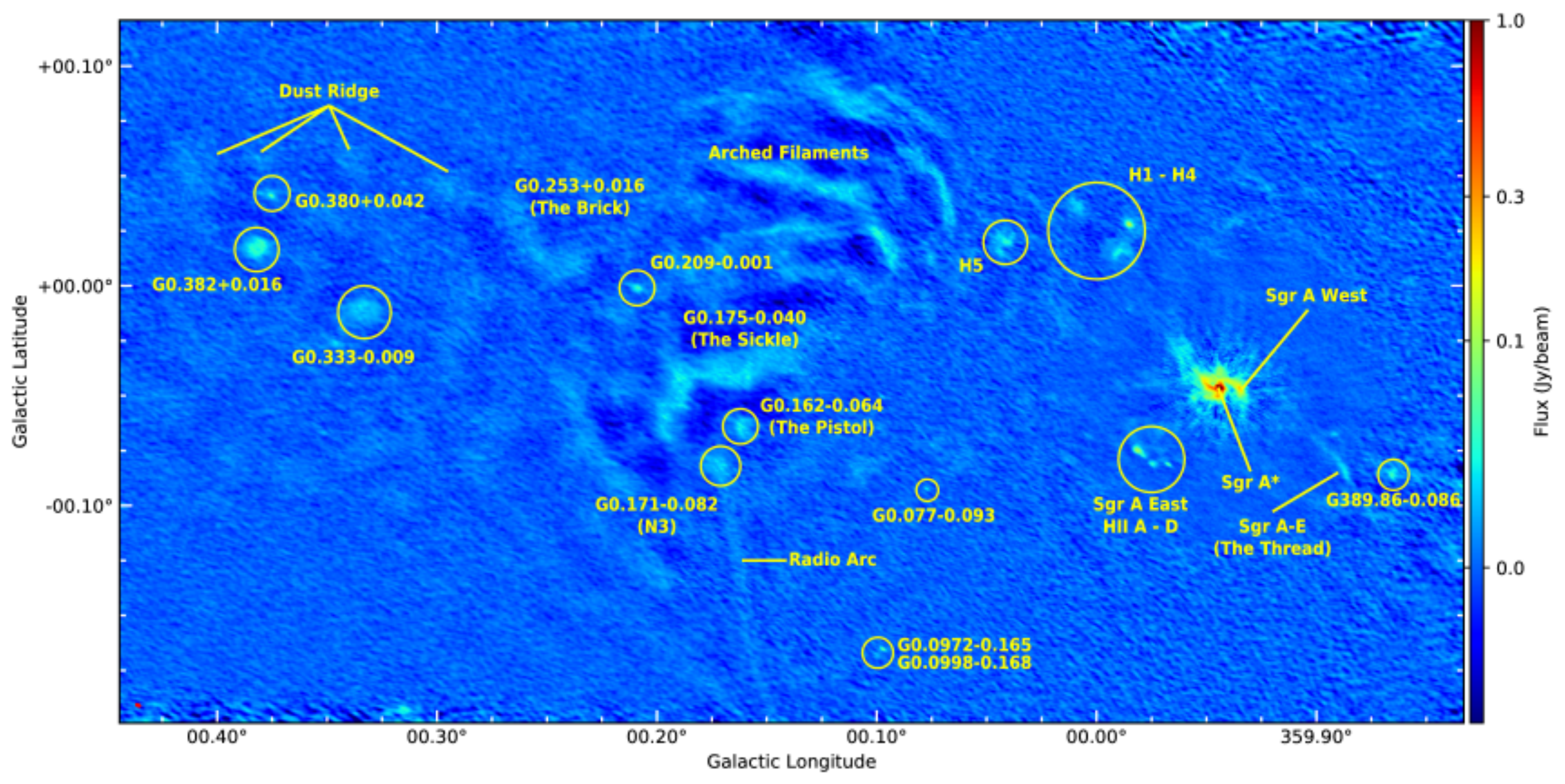} 
\caption{CARMA 15+8 3~mm continuum map with a  spatial resolution is $7.62\arcsec
\times 3.51\arcsec$ (synthesized beam indicated in red in lower left corner). Prominent features are labeled, including the thermal Arched Filaments, Sickle, Pistol,
the Brick, nonthermal radio filaments the Radio Arc and the Thread, Sgr A* and Sgr A West, several HII regions, and the
clouds associated with the Dust Ridge.}
\label{f-continuum-15+8}
\end{figure*}

\begin{figure*}
\includegraphics[width=2\columnwidth,trim={1cm 0cm 1cm 0cm}]{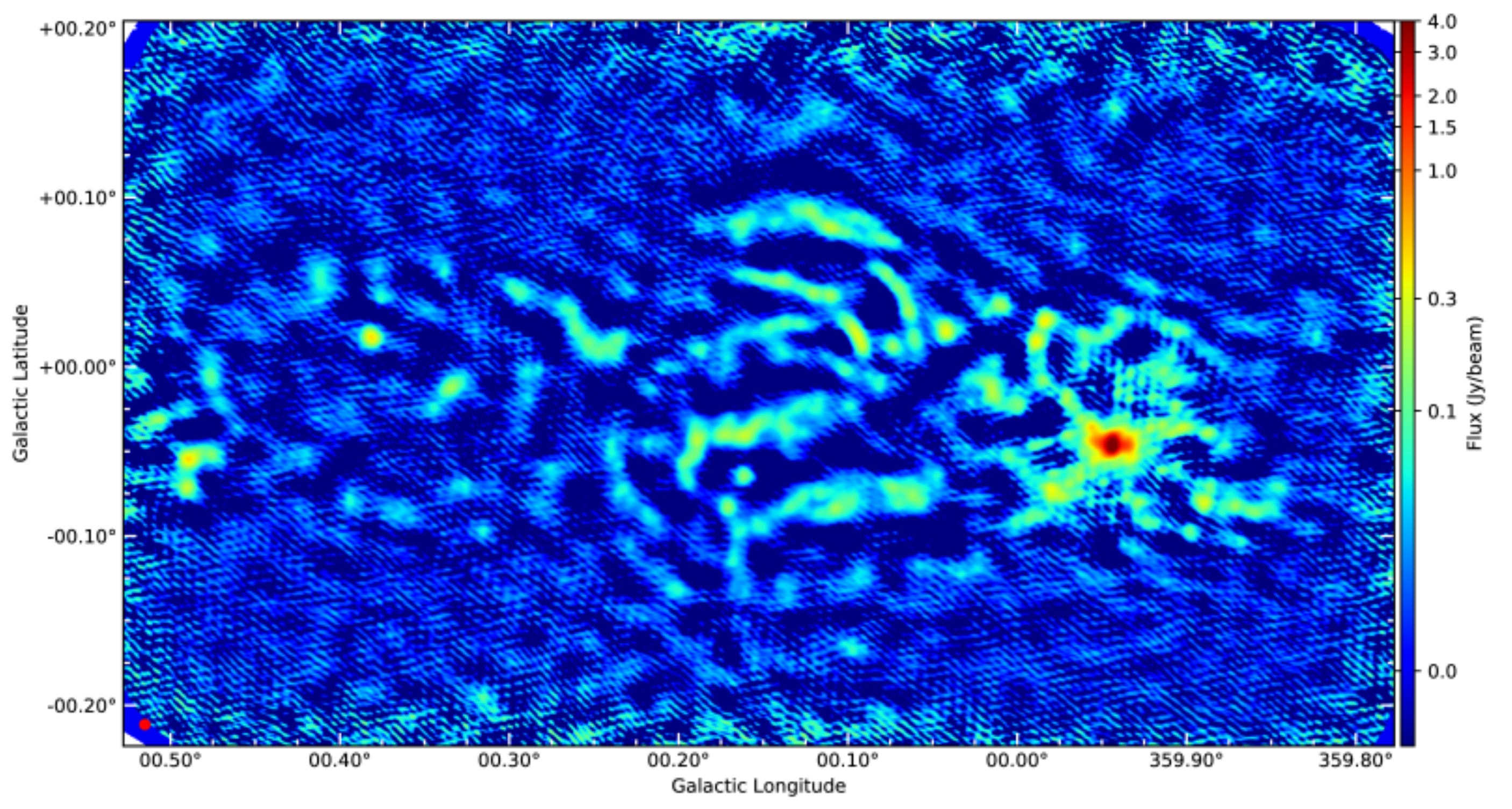}
\caption{CARMA-8 3~mm continuum map, smoothed to a round 20\arcsec\ beam (synthesized beam indicated in red in lower left corner).
Because of the larger primary beam of the 3.5-m antennas, this map slightly
larger than the CARMA 15+8 map displayed in Fig. 1.  Diffuse features with surface brightness, 
such as the Dust Ridge and the Brick, have comparatively higher flux density. There are high frequency striations 
in the vertical direction near Sgr A*. These are artifacts of CLEANing.}
\label{f-continuum-8}
\end{figure*}

The combined CARMA-15 plus CARMA-8 map with spatial resolution $7.62\arcsec
\times 3.51\arcsec$ and as shown in Fig. \ref{f-continuum-15+8} is sensitive
to both small-scale and extended features.  The CARMA-8 image, with
spatial resolution 20\arcsec\, highlights more extended diffuse emission
(Fig. \ref{f-continuum-8}).  

One of the advantages of observing the Galactic Center at 3~mm with CARMA
is that it allows identification of different types of emission from
this confusing region of the Galaxy. In continuum, we detect emission
from dust grains, optically thin  free-free emission from ionized gas,
synchrotron emission from nonthermal filaments, and 
ultra-compact HII regions distributed throughout the region.

Because the electron density is proportional to the observed frequency,
3~mm observations are sensitive to compact and
ultra-compact (UC) HII regions with high densities.  Our 3~mm continuum
survey can study  the class of compact HII regions with emission measures
$\le 3\times$ \expo{10} pc~cm$^{-6}$ at the turnover frequency  $\tau\le1$.
We detect the chain of compact Sgr A East HII regions, which are the
closest massive star forming regions to Sgr A*, and several isolated
compact HII regions.

The Galactic Center also hosts two  well-known young stellar clusters:
the Arches and the Quintuplet clusters.  There is an additional cluster
of stars, the so-called the central cluster, that  orbits Sgr A* 
\citep{2006ApJ...643.1011P}.
Diffuse and bright sources excited by hot members of these young clusters
are Sgr A West, the Arched Filaments, and the Sickle, all of which are
visible at 3 mm.

The physical characteristics of molecular gas in the CMZ are similar to
those of infrared dark clouds (IRDCs). Herschel observations show that IRDCs
have low dust temperatures \citep{2000ApJ...545L.121P,2011ApJ...735L..33M,2013ApJ...767L..13R}.
We identify several of these IRDCs in the chain of clouds forming the
so-called Dust Ridge \citep{1994ApJ...424..189L,2001ApJ...550..761L} of molecular gas distributed
between G0.253+0.016/  and Sgr B2. G0.253+0.016 has
been the focus of numerous recent papers, both suggesting it is the
precursor cloud to an Arches-like stellar cluster \citep{2012ApJ...746..117L, 2015ApJ...802..125R}
and that it is not \citep{2013ApJ...765L..35K,2014A&A...568A..56J}.  In Figs. \ref{f-continuum-15+8} and \ref{f-continuum-8},
the cloud is seen with uniform intensity $\sim 0.01$ Jy beam$^{-1}$.

Low frequency radio continuum surveys of the Galactic Center show a
population of magnetized filamentary structures throughout the CMZ,
many of which have a steep spectrum, and thus cannot be detected at 3 mm
\citep{1999ApJ...526..727L,2000AJ....119..207L,1984Natur.310..557Y,2004ApJS..155..421Y}. However, some filaments have a flat spectrum, such as
the ones in the Radio Arc at $l \sim $0\pdeg2.  In Figs. \ref{f-continuum-15+8} and \ref{f-continuum-8} show clearly
the brightest filament running perpendicular to the Galactic plane,
passing through nonthermal source N3 \citep{1987AJ.....94.1178Y,2016ApJ...826..218L}, through the Sickle, and curving to pass across the Arched Filaments.
The mean 3~mm flux density of this filament is 3.5 mJy beam$^{-1}$. 
The origin of these unique flat-spectrum filaments is unknown.

\begin{figure}
\includegraphics[width=\columnwidth,trim={0cm 4cm 3cm 2cm}]{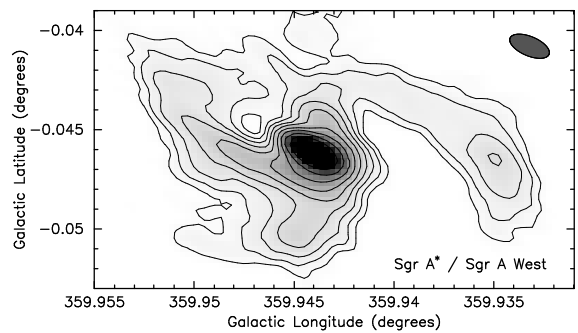}
\caption{3 mm continuum map of Sgr A* and Sgr A West aka the mini-spiral.
The peak flux intensity of Sgr A* is 2.8 Jy beam$^{-1}$.  Contours are 
at 0.080, 0.114, 0.163, 0.232, 0.332, 0.473, 0.675, 0.964,  1.38,  1.96,  2.80 Jy beam$^{-1}$
(1$\sigma$ rms noise is 0.002 Jy beam$^{-1}$). The grayscale ranges from 0.05 to 1.5 Jy beam$^{-1}$.
In this and the following figures, the synthesized beam is shown in the upper right corner.
}
\label{f-sgrastar}
\end{figure}

We have measured the parameters at 3~mm for both compact and extended sources.
For sources which appeared unresolved or marginally resolved, we used
MIRIAD's {\it imfit} task to fit a 2-dimensional Gaussian, excluding from
the fit pixels with intensities below 
the 1$\sigma$ rms noise level measured locally. Such exclusion produced quantitatively better fits to the measured peak intensities than clipping at zero flux density.
The deconvolved source size and peak intensity are derived from the Gaussian fit.
Brightness temperatures were computed using equation 2.18 of \cite{1987soap.conf...35B}.  The low derived brightness temperatures indicate that all of these sources are optically thin at 3 mm.
For the compact sources, entries in columns 1-6 respectively of Table \ref{t-compactfluxes} list source name,
Galactic coordinates, fitted peak intensity, brightness temperature, and fitted source size and position angle. 
For extended sources, we measured the
total flux density in a box around each source, clipping the background at twice the 1$\sigma$
rms noise level which was measured locally (Table \ref{t-extendedfluxes}).  
Entries in columns 1-5  of Table \ref{t-extendedfluxes} are
source name, Galactic coordinates, total flux density, the size of
the box within which the flux was measured, and area of pixels greater than
2$\sigma$ that contribute to the total flux density, respectively. Individual sources
are shown in Figs. \ref{f-sgrastar} to \ref{f-sickle} and are briefly described below.
\begin{table*}
\caption{Parameters of Compact 3~mm Sources. $^1$Peak Intensity and deconvolved quantities are derived from Gaussian fit. See \S\ref{s-3mmcont}.}
\label{t-compactfluxes}
\begin{tabular}{lccccccc}
\hline
    &
$l$ &
$b$ &
Peak Intensity\footnotemark[1]  &
Brightness Temp. &
Deconvolved Size\footnotemark[1]  &
Deconvolved PA\footnotemark[1]  &
 \\
Source & 
(\arcdeg) & 
(\arcdeg) & 
(Jy beam$^{-1}$) &
(K) &
(\arcsec) &
(\arcdeg) &
Fig. \\
\hline
H7           & 0.0331   & +0.0295 & 0.013 $\pm$ 0.001 & 0.07 $\pm$ 0.01 & 7.4 $\times$ 1.1 & 56.2 &\ref{f-h1-8} \\  
G0.077-0.093 &   0.0773 & -0.0925 & 0.028 $\pm$ 0.005 & 0.16 $\pm$ 0.03 & 3.6 $\times$ 1.4 & 26.3 &\ref{f-c1g07g011} \\
G0.075-0.073 &   0.0749 & -0.0726 & 0.016 $\pm$ 0.002 & 0.09 $\pm$ 0.01 & 8.2 $\times$ 4.9 & 44.8 &\ref{f-c1g07g011} \\
G0.098-0.050 &   0.0975 & -0.0504 & 0.022 $\pm$ 0.005 & 0.12 $\pm$ 0.03 & 7.5 $\times$ 1.4 & 51.7 &\ref{f-c1g07g011} \\
G0.0972-0.165 &   0.0972 & -0.1654 & 0.068 $\pm$ 0.007 & 0.38 $\pm$ 0.04 & 4.1 $\times$ 1.8 & 38.3 &\ref{f-a1a2} \\
G0.0998-0.168 &   0.0998 & -0.1680 & 0.014 $\pm$ 0.002 & 0.08 $\pm$ 0.01 &13.7 $\times$ 8.5 & 37.6 &\ref{f-a1a2} \\
G0.209-0.001 &   0.2083 & -0.0012 & 0.038 $\pm$ 0.003 & 0.21 $\pm$ 0.02 &17.5 $\times$ 9.2 & 70.8 &\ref{f-g021} \\
G0.323+0.108 &   0.3228 & +0.0183 & 0.010 $\pm$ 0.002 & 0.06 $\pm$ 0.01 &15.7 $\times$ 6.5 & 75.3 &\ref{f-g035}\\
G0.352-0.067 &   0.3515 & -0.0669 & 0.015 $\pm$ 0.001 & 0.08 $\pm$ 0.01 &10.8 $\times$ 4.2 & 78.0 &\ref{f-g035}\\
G0.380+0.042 &   0.3802 & +0.0418 & 0.016 $\pm$ 0.002 & 0.09 $\pm$ 0.01 &18.6 $\times$ 7.6 & 73.6 &\ref{f-g038} \\

\hline
\end{tabular}

\end{table*}

\begin{table*}
\caption{Flux Densities of Extended 3~mm Sources.
$^1$Denotes the area of pixels greater than the cutoff that contribute to the total flux density.
$^2$While this box includes The Pistol, we masked out emission from the Pistol box when measuring the flux of The Sickle.}
\label{t-extendedfluxes}
\begin{tabular}{lccccc}
\hline
Source & 
$l$     &
$b$     &
Flux Density & 
Box size &
Area\footnotemark[1]     \\
     &
(\arcdeg) & 
(\arcdeg) & 
(Jy) &
$\Delta~l(\arcsec)\times\Delta~b(\arcsec)$ &
(\sq\arcsec) \\
\hline
 G359.86-0.086    & 359.8656 & -0.0862 & 0.375 $\pm$ 0.041 & 61$\times$ 37 & 709\\   
 G359.89-0.068    & 359.8917 & -0.0676 & 0.052 $\pm$ 0.011 & 31$\times$ 21 & 153\\   
 SgrA-E           & 359.8870 & -0.0847 & 0.392 $\pm$ 0.030 & 79$\times$ 81 & 1070\\   
 Parachute        & 359.9217 &  0.0435 & 0.405 $\pm$ 0.042 & 101$\times$ 101 & 1328\\   
Sgr A East HII A/B& 359.9815 & -0.0751 & 0.486 $\pm$ 0.020 & 32$\times$ 32 & 570\\   
 Sgr A East HII C & 359.9745 & -0.0812 & 0.107 $\pm$ 0.012 & 21$\times$ 18 & 205\\   
 Sgr A East HII D & 359.9674 & -0.0814 & 0.088 $\pm$ 0.010 & 23$\times$ 16 & 164\\   
               H1 & 359.9903 & +0.0152 & 0.510 $\pm$ 0.047 & 64$\times$ 61 & 1114\\   
               H2 & 359.9852 & +0.0281 & 0.538 $\pm$ 0.038 & 56$\times$ 48 & 721\\   
               H3 &   0.0084 & +0.0356 & 0.272 $\pm$ 0.034 & 49$\times$ 48 & 602\\   
               H4 &   0.0223 & +0.0297 & 0.080 $\pm$ 0.022 & 51$\times$ 61 & 248\\   
               H5 &   0.0415 & +0.0198 & 0.636 $\pm$ 0.049 & 51$\times$ 61 & 1205\\   
               H8 & 359.9853 & -0.0110 & 0.027 $\pm$ 0.009 & 31$\times$ 26 & 122\\
 G0.109-0.084     &  0.1089 & -0.0841  & 0.226 $\pm$ 0.034 & 81$\times$ 69 & 756\\   
 Arched Filaments &  0.125 & +0.045    & 12.073 $\pm$ 0.213 & 461$\times$ 391 & 29852 \\
 Pistol           &  0.1618 & -0.0644  & 0.537 $\pm$ 0.034 & 71$\times$ 61 & 1251\\  
 N3               &  0.1714 & -0.0816  & 0.686 $\pm$ 0.046 & 81$\times$ 71 & 1775\\  
 Sickle           &  0.175 & -0.040    & 6.651 $\pm$ 0.133 &271$\times$ 191\footnotemark[2] & 14953\\ 
 G0.333-0.009     &  0.3331 & -0.0086  & 0.968 $\pm$ 0.052 & 91$\times$ 66 & 2579\\  
 G0.346-0.026     &  0.3460 & -0.0260  & 0.182 $\pm$ 0.024 & 55$\times$ 36 & 537\\   
 G0.375+0.041     &  0.3754 & +0.0408  & 0.243 $\pm$ 0.023 & 52$\times$ 28 & 519\\   
 G0.382+0.016     &  0.3820 & +0.0164  & 1.193 $\pm$ 0.043 & 83$\times$ 66 & 1829\\   
\hline

\end{tabular}
\end{table*}

\subsubsection{Notes On Individual Continuum Sources}

\paragraph{Sgr A* and Sgr A West}
The mini-spiral HII region in Figure 3 shows  three arms of ionized gas orbiting Sgr A*. Low-frequency  emission from the southern arm of the mini-spiral seen to the west is optically thick \citep{2004ApJS..155..421Y}, whereas the emission at 3mm is most likely  a mixture of dust and ionized gas. Recent ALMA observations  of the mini-spiral show clearly that cool dust dominates the emission  at submm wavelengths \citep{2016PASJ...68L...7T}.

\begin{figure}
\includegraphics[width=\columnwidth,trim={0cm 5cm 3cm 5cm}]{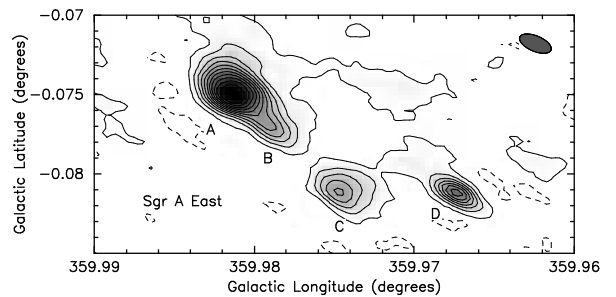}
\caption{3 mm continuum map of the Sgr A East A-D complex of compact HII regions.
Contours are in steps of $4\sigma$ from
-2$\sigma$ to 46$\sigma$, where the 1$\sigma$ rms noise is 0.002 Jy beam$^{-1}$. The
grayscale ranges from 0.005 to 0.1 Jy beam$^{-1}$.
}
\label{f-sgraeast}
\end{figure}

\paragraph{Sgr A East A-D (Fig. \ref{f-sgraeast})}

The Sgr A East A-D complex of compact HII regions was first
noted by \citet{1983A&A...122..143E}.  These HII regions  photoionized by young 
O8-B0 stars 
are associated with the 50
\kms\ molecular cloud \citep[][and references therein]{2010ApJ...725.1429Y,2011ApJ...735...84M}.  At 3 mm, sources A, B, and D
appear centrally peaked, while source C has a peak offset to the east. The peak is coincident with 6 cm source C1 and the
broadening towards the west coincident with 6 cm source C2 \citep{1987ApJ...320..545Y}.

\begin{figure}
\includegraphics[width=\columnwidth,trim={2cm 2cm 2cm 2cm}]{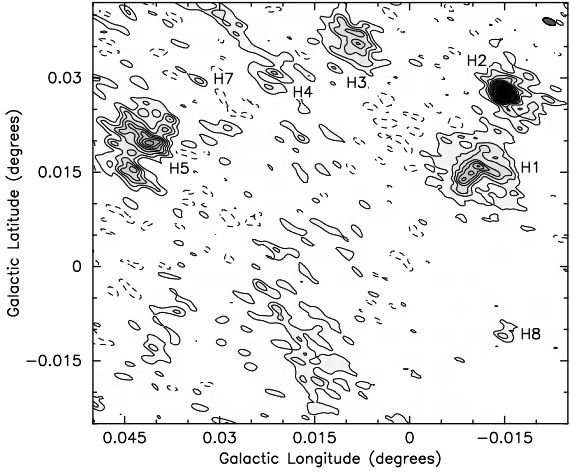}
\caption{3 mm continuum image of  sources H1 through H8 (H6 was not detected). 
Contours are at -0.0075, 0.005, 0.01, 0.015, 0.02, 0.025, 0.03, 0.035,
0.04, 0.045, 0.055, 0.065, 0.075, 0.085, 0.095, 0.105, 0.115, 0.125
Jy beam$^{-1}$. The $1\sigma$ rms noise is 0.002 Jy beam$^{-1}$ and the grayscale ranges
from 0.005 to 0.12 Jy beam$^{-1}$.
}
\label{f-h1-8}
\end{figure}

\paragraph{Sources H1 through H8 (Fig. \ref{f-h1-8})}

H1 through H5 are known to be HII regions and at 3~mm their morphologies
correspond well to 6 cm continuum maps \citep{1987ApJ...320..545Y, 1993ApJ...418..235Z}.  H1 has a shell-like
morphology, while H2 is very bright and compact with a narrow, $\sim 25\arcsec$-long ridge to
the south.  H4 is a comparatively weak shell with a $\sim 40\arcsec$-long low intensity
extension to the north.  H5 breaks up into bright east and west components
surrounded by lower intensity emission. A hint of the 6 cm tongue of
emission extending southward is seen at 3 mm.  H7 appears as a
compact sources H7 (intensity per unit frequency $I_\nu = 13 \pm 1$ mJy beam$^{-1}$). H8 ($S_\nu = 27 \pm 9$ mJy) is resolved with a slight extension westwards.
We also see in this region broad-scale, low intensity emission at positive Galactic longitude that is similar to features seen at 6 cm. 

\begin{figure}
\includegraphics[width=\columnwidth,trim={1cm 2cm 2cm 2cm}]{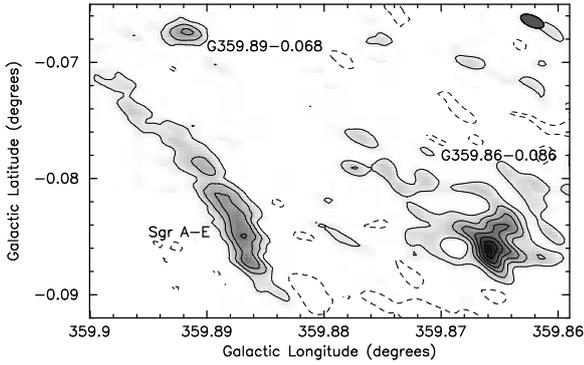}
\caption{Continuum 3~mm close-up of sources SgrA-E (also known as ``The Thread''), G359.89-0.068, and
G359.86-0.086.  The  $1\sigma$ rms noise in this map is
about 0.003 Jy beam$^{-1}$ (somewhat lower near SgrA-E, somewhat higher near
G359.89-0.068).  Contours are -0.007, 0.007, 0.014, 0.021, 0.028, 0.035,
0.042 Jy beam$^{-1}$ and the grayscale ranges from 0.0035 to 0.05 Jy beam$^{-1}$.
}
\label{f-g359}
\end{figure}

\paragraph{Sources SgrA-E, G359.89-0.068, and G359.86-0.086 (Fig. \ref{f-g359})} 

SgrA-E is a nonthermal filament associated with X-ray source XMM
J174540-2904.5 and believed to be interacting with a molecular cloud
\citep{2005AdSpR..35.1074Y}.  In our map, it is nearly 100\arcsec\, long with a peak
intensity of 0.028 Jy beam$^{-1}$.  The brightest region in 3 mm 
($-0.09^\circ \la b \la -0.08^\circ$) corresponds to the regions of most intense 2 cm and X-ray emission.

G359.89-0.068 is a compact 3~mm source, coincident with submillimeter
source JCMTSF J174537.6-290350 from the SCUBA survey catalog of
\citet{2008ApJS..175..277D} that has an 850~\micron\ flux density of 9.62 Jy.
We find no counterpart to G359.89-0.068 at 2 cm or 6 cm. G359.86-0.086,
also known as SgrA-G, is an HII region \citep{1985ApJ...288..575H,1987ApJ...320..545Y} near the
edge of a bright submillimeter cloud in the SCUBA map.  At 3 mm, it is
centrally peaked with a maximum intensity of 0.047 Jy beam$^{-1}$ and at 2 cm,
the source  appears clumpy, possibly shell-like \citep{1985ApJ...288..575H}.

\begin{figure}
\includegraphics[width=\columnwidth,trim={2cm 2cm 2cm 2cm}]{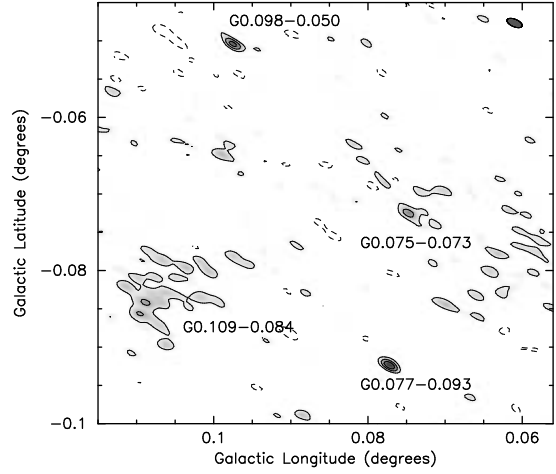}
\caption{
The 3~mm continuum map of  G0.075-0.073, G0.077-0.093, G0.098-0.050, and G0.109-0.084.
Contours are -0.007, 0.007, 0.014, 0.021 Jy beam$^{-1}$.  The $1\sigma$ rms is
0.0035 Jy beam$^{-1}$ and the grayscale ranges from 0.005 to 0.035 Jy beam$^{-1}$.
}
\label{f-c1g07g011}
\end{figure}

\paragraph{Sources G0.075-0.073, G0.077-0.093, G0.098-0.050, and G0.109-0.084 (Fig. \ref{f-c1g07g011})} 

In our 3~mm map, G0.075-0.073, G0.077-0.093 and G0.098-0.050 are isolated,
unresolved sources, with peak intensities between 16 and 28 mJy beam$^{-1}$.
G0.077-0.093 and G0.098-0.050 are identified as compact sources B1 and C1,
respectively, in \citet{2014MsT..........2T} with varying spectral indices between 2.8
GHz and 5.3 GHz.  G0.077-0.093 appears in the 2LC catalog \citep{2008ApJS..174..481L}
with a 1.4 GHz flux density of 0.00292 Jy.  It is also 1.6\arcsec\ from
Chandra source CXOGCS J174609.7-285505 \citep{2006ApJS..165..173M}.  We find no clear
match for G0.098-0.050 in the SIMBAD database.  

G0.109-0.084 is a diffuse,
low intensity source with a peak intensity of 0.015 Jy beam$^{-1}$, while  G0.075-0.073 is a
compact peak in a more diffuse region.
These two diffuse regions coincide with
BGPS G000.106-00.085 and BGPS G000.066-00.079, respectively, which in the BOLOCAM 1.1 mm 
survey image are teardropped shaped peaks within more extended nebulosity.
The 1.1 mm flux density of BGPS G000.106-00.085 is $13.584\pm0.88$ Jy and of 
BGPS G000.066-00.07 is $11.237\pm0.723$ Jy  \citep{2010ApJS..188..123R}.  
G0.075-0.073 is coincident with the narrow end of the BGPS G000.066-00.079
``teardrop'' and may represent an embedded source or simply a local
peak. These diffuse clouds are clearly seen in the CARMA-8 map (Fig.
\ref{f-continuum-8}).

\begin{figure}
\includegraphics[scale=0.4,trim={0cm 0cm 0cm 0cm}]{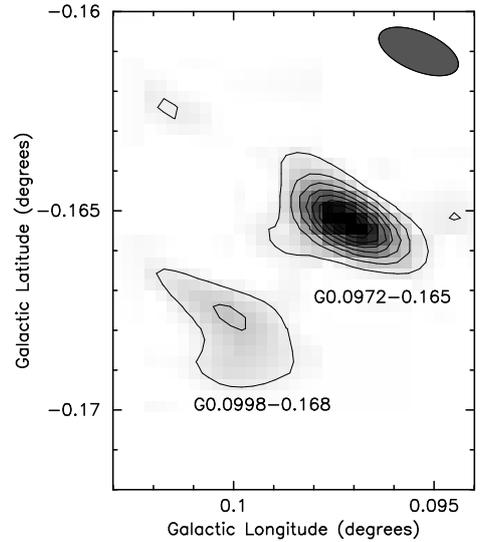}
\caption{The 3~mm CARMA15+8 continuum map of G0.0972-0.165 and G0.0998-0.168
corresponding to sources A1 and A2, respectively, in \citet{2014MsT..........2T}.  
Contours are 0.008, 0.016, 0.024, 0.032, 0.040, 0.048, 0.056, 0.064 Jy beam$^{-1}$.
The $1\sigma$ rms is 0.004 Jy beam$^{-1}$ and the grayscale ranges from 0.004 to 0.06 Jy beam$^{-1}$.}
\label{f-a1a2}
\end{figure}

\paragraph{Sources G0.0972-0.165 and G0.0998-0.168 (Fig. \ref{f-a1a2})}  

G0.0972-0.165 and G0.0998-0.168 correspond to sources A1 and A2, respectively,
in \citet{2014MsT..........2T}.  They are both unresolved at 3 mm. 
They coincide with the unresolved BOLOCAM source BGPS G000.098-00.163,
which has a 1.1 mm flux density of $1.272\pm0.133$ Jy.
At 4.8GHz, G0.0998-0.168 appears
as a clumpy shell about 12\arcsec\ in diameter \citep{2014MsT..........2T}.

\begin{figure}
\includegraphics[width=\columnwidth,trim={6cm 2cm 7cm 2cm}]{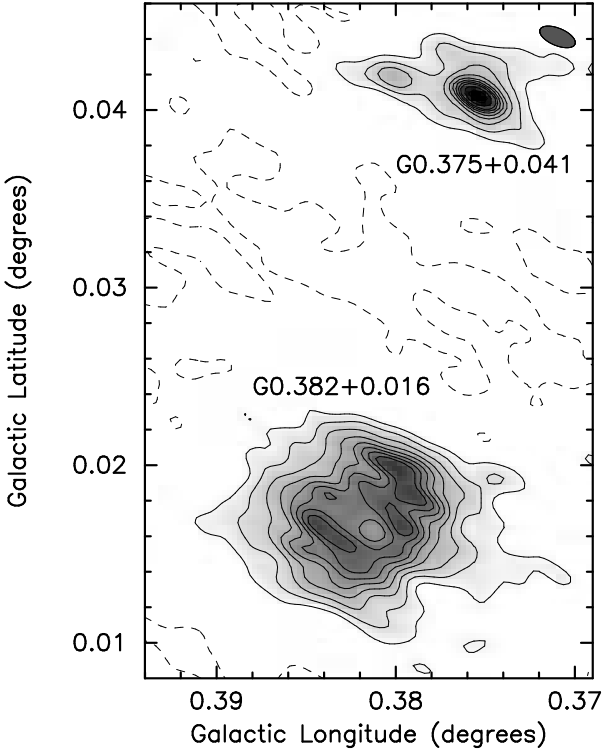}
\caption{Sources G0.382+0.016 and G0.375+0.041. 
Contours are -0.0056, 0.0056, 0.0112, 0.0168, 0.0224, 0.028,
0.0336, 0.0392, 0.0448, 0.056, 0.0672 Jy beam$^{-1}$.  The $1\sigma$ rms noise
is 0.0028 Jy beam$^{-1}$ and the grayscale ranges from 0.003 to 0.06 Jy beam$^{-1}$.
}
\label{f-g038}
\end{figure}

\paragraph{Sources G0.382+0.016 and G0.375+0.041 (Fig. \ref{f-g038})}

G0.382+0.016 has been variously classified as an HII region 
based on radio recombination and HI spectral lines
\citep{2011ApJS..194...32A}, as an UC HII region 
based on IRAS colors \citep{1994ApJS...91..347B}, and as a supernova remnant 
\citep{2000AJ....119..207L}.  In the 870~\micron\ and 8.4 GHz continuum surveys of
\citet{2012A&A...537A.121I}, it is identified as Source E. With an integrated flux
density of 1.193 Jy, is one of the brightest continuum sources in our 3~mm survey, with
a clear shell structure. 
Assuming $S_\nu \propto \nu^\alpha$, the spectral index $\alpha$ is
defined as $\alpha = \log(S_1/S_2)/\log(\nu_1/\nu_2)$. Using  
the flux density at 9 GHz of 1.513$\pm$0.053 Jy and
our 90 GHz measurement (Table \ref{t-extendedfluxes}), we derive a spectral
index $\alpha = -0.10\pm0.05$ (where $\alpha = \log(S_1/S_2)/\log(\nu_1/\nu_2)$,
given $S_\nu \propto \nu^\alpha$), consistent with optically thin free-free emission and
thus supporting the identification as an HII region.  G0.375+0.041 is
coincident with maser features \citep[e.g.,][]{1995MNRAS.272...96C, 2000ApJS..129..159A} and
Spitzer source SSTGC 639320, and is classifed as a YSO by \citet{2009ApJ...702..178Y}.
At 3 mm, it is resolved into two compact sources with a common envelope
of emission.  The brighter of these has a peak intensity of 0.07 Jy beam$^{-1}$, the weaker 0.02 Jy beam$^{-1}$.

\begin{figure}
\begin{center}
\includegraphics[width=\columnwidth,trim={1cm 1.5cm 1cm 1cm}]{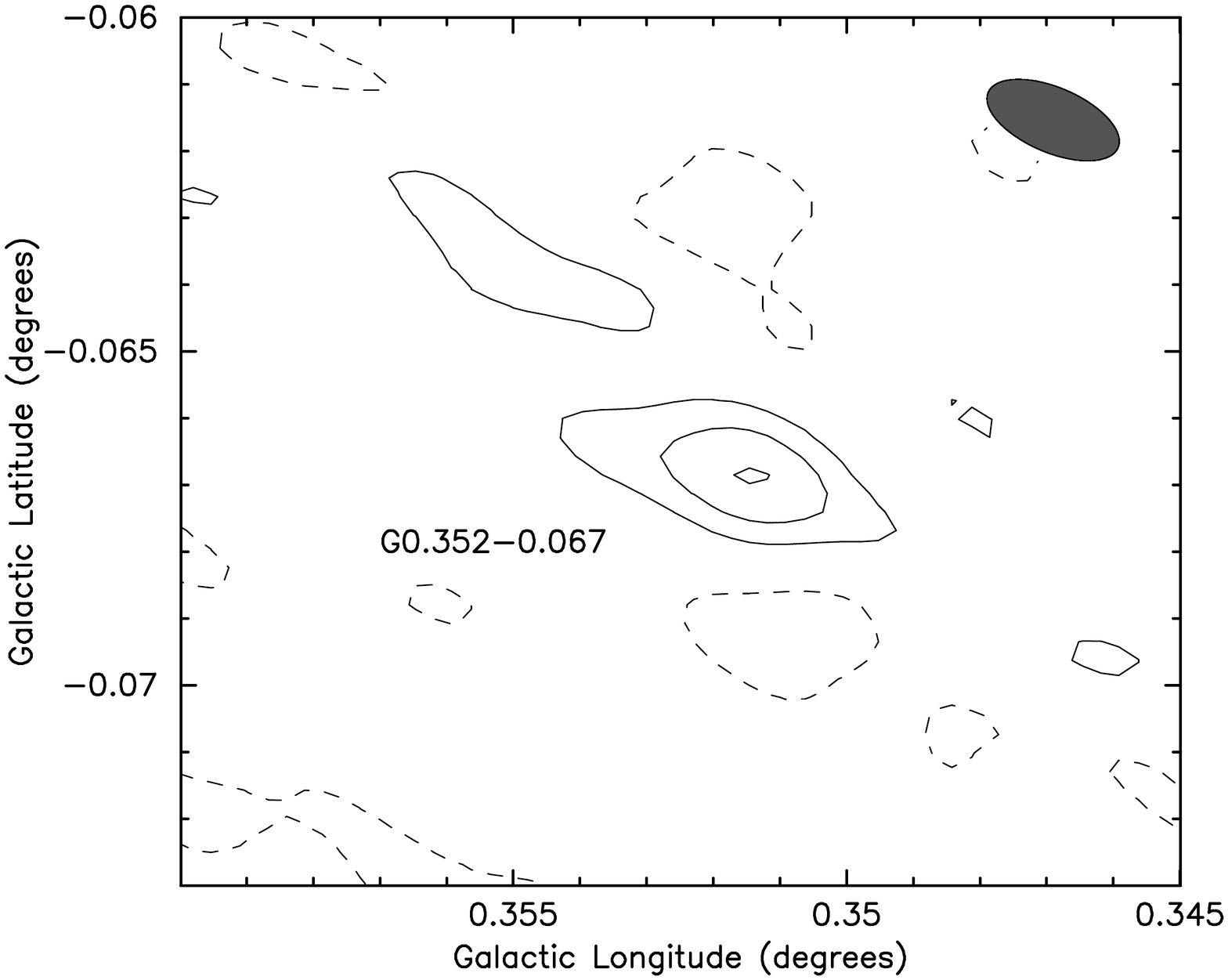}
\includegraphics[width=\columnwidth,trim={2cm 1cm 2cm 1cm}]{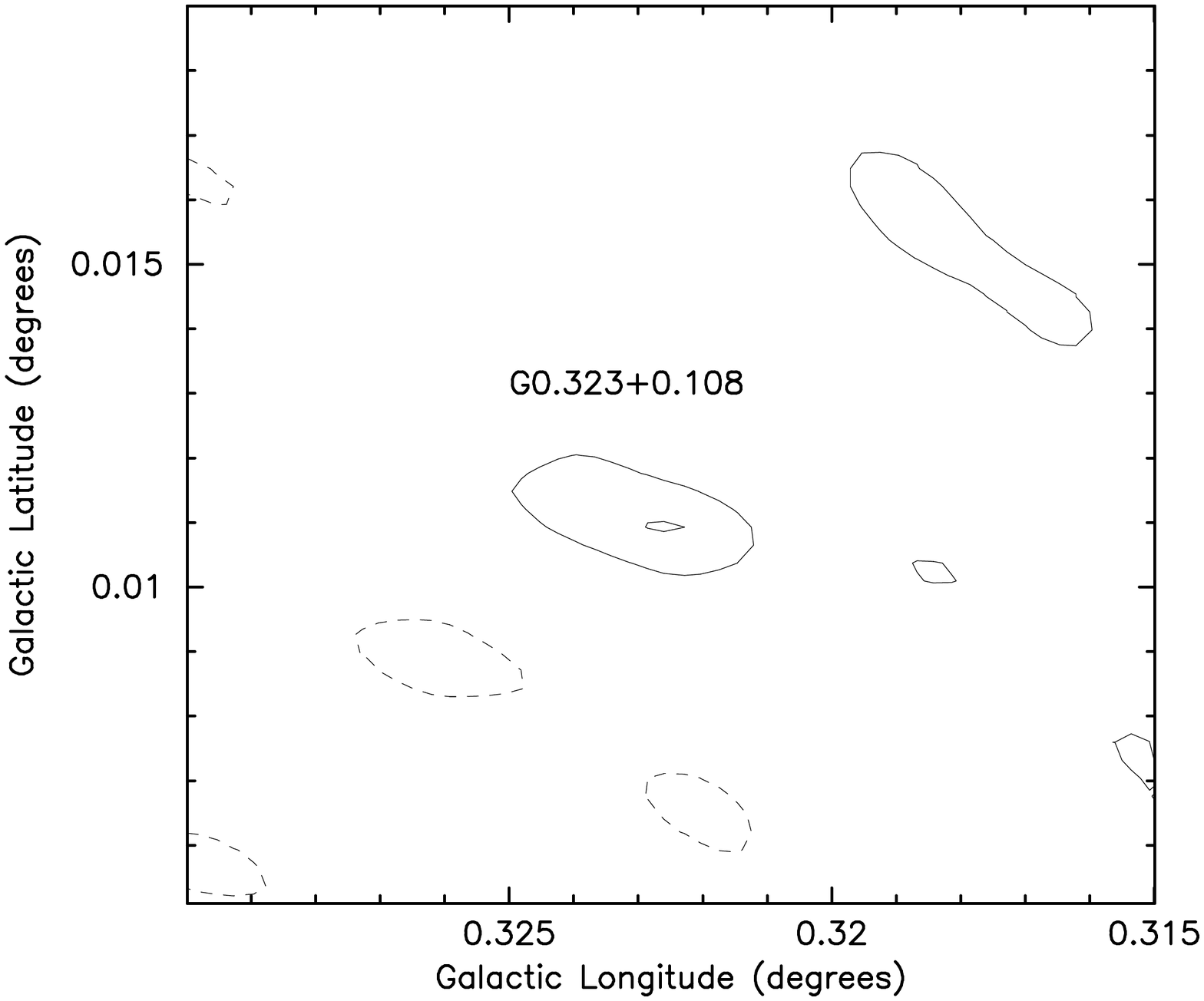}
\end{center}
\caption{Sources G0.352-0.067 (top) and G0.323+0.108 (bottom).
For G0.352-0.067, contours are -0.005, 0.005, 0.01, 0.015 Jy beam$^{-1}$,
the $1\sigma$ rms noise is 0.0025 Jy beam$^{-1}$, and the grayscale ranges from 0.003 to 0.02 Jy beam$^{-1}$.
For G0.323+0.108 contours are -0.006, 0.006, 0.012 Jy beam$^{-1}$,
the $1\sigma$ rms noise is 0.003 Jy beam$^{-1}$, and the grayscale ranges from 0.004 to 0.02 Jy beam$^{-1}$.
}
\label{f-g035}
\end{figure}

\paragraph{Sources G0.352-0.067 and G0.323+0.0108 (Fig. \ref{f-g035})}

G0.352-0.067 is an isolated, compact 3~mm source with peak intensity of 0.015 Jy beam$^{-1}$ .
It appears in the 2LC catalog \citep{2008ApJS..174..481L} with angular diameter of
2\arcsec\ and 1.4~GHz flux density of 0.0277 Jy.  At 5~GHz, it has flux
density of 0.0224~Jy \citep{1994ApJS...91..347B}.   
G0.323+0.0108 is also compact and isolated at 3~mm with a peak intensity of 0.010 Jy beam$^{-1}$.
We find no lower frequency radio counterpart within 30\arcsec\ of
G0.323+0.0108 in the SIMBAD database. An extended submillimeter source JCMTSE
J174620.3-283931 is centered 12.2\arcsec\ away with a total 850~\micron\
flux density of 0.53 Jy \citep{2008ApJS..175..277D}.  The position of G0.323+0.0108 is coincident
with a small tongue of emission extending
from JCMTSE J174620.3-283931.  However, there is no 3~mm emission associated with
JCMTSE J174620.3-283931 in the CARMA-8 map, which is more sensitive to
extended structure.  

\begin{figure}
\includegraphics[width=\columnwidth,trim={3cm 2cm 5cm 2cm}]{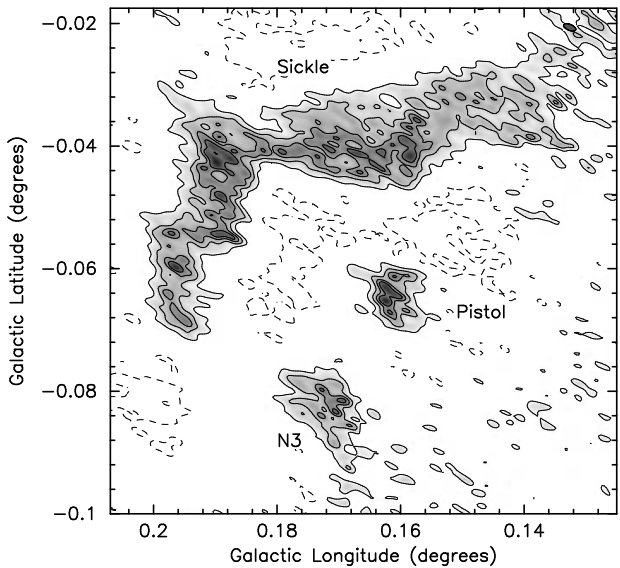}
\caption{The 3~mm CARMA15+8 continuum map of region of the Sickle, the Pistol, and N3.
Contours are -0.009, 0.006, 0.012, 0.018, 0.024 Jy beam$^{-1}$. 
The $1\sigma$ rms is 0.003 Jy beam$^{-1}$ and the grayscale ranges from 0.005 to 0.04 Jy beam$^{-1}$.
}
\label{f-sickle}
\end{figure}

\paragraph{The Sickle, The Pistol, and N3 (Fig. \ref{f-sickle})}
The nonthermal radio filaments of the Arc
appear to cross the sickle  and the Galactic plane.
There are  two prominent HII regions G0.18-0.04 (The Sickle nebula) and G0.16-0.06 (the Pistol nebula).  These  nebulae lie in the region of the Quintuplet Cluster, a young cluster of massive stars in the Galactic Center.  The radio source N3 (G0.171-0.082) appears in Fig. \ref{f-continuum-15+8} to be crossed by the Radio Arc as is also seen in maps at lower frequencies \citep{1987AJ.....94.1178Y,2016ApJ...826..218L}. Between 4.5 and 49 GHz, N3 is a point source \citep{2016ApJ...826..218L}; however at 3mm we see extended emission surrounding the point source location that coincides with one of the 3 mm contour peaks.  The extended continuum emission is coincident with the interacting molecular cloud identified by \citep{2016ApJ...826..218L} which we also clearly detect in CS, SiO, and HCN, and weakly detect in \hcopA\ and \ntwohpA\ (Figs. \ref{f-siomoment0} to \ref{f-csmoment0}).

\begin{figure}
\includegraphics[width=\columnwidth,trim={1cm 2cm 2cm 2cm}]{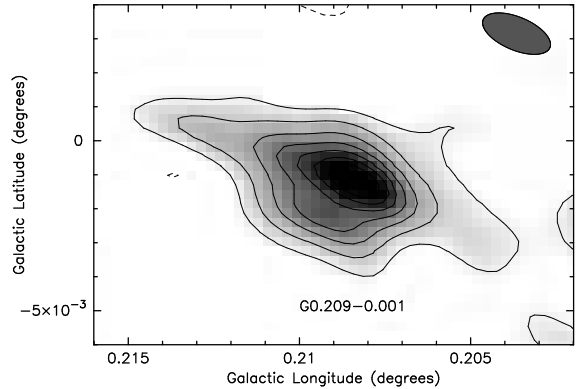}
\caption{CARMA 3~mm continuum map of G0.209-0.001.
Contour levels  are set at -0.006, 0.006, 0.012, 0.018, 0.024, 0.030, 0.036, 0.042, 0.048 Jy beam$^{-1}$.
The $1\sigma$ rms is 0.003 Jy beam$^{-1}$ and the grayscale ranges from 0.004 to 0.04 Jy beam$^{-1}$.
}
\label{f-g021}
\end{figure}

\paragraph{Source G0.209-0.001 (Fig. \ref{f-g021})}

G0.209-0.001 is a compact source north of the Sickle with a peak 3~mm flux
density of 0.05 Jy beam$^{-1}$.  It has a slightly 
curved, lower emissivity ($\sim$0.01 Jy beam$^{-1}$) north-south extension.   \citet{2010ApJS..191..275L} measure
a 1.4~GHz flux density of 0.447$\pm$0.032 Jy. It is identified
as source D1 in \citet{2014MsT..........2T} with a varying spectral index between 2.8 and 5.3 GHz,
as SCUBA source JCMTF J174606.9-284536 with a flux density of 4.14 Jy \citep{2008ApJS..175..277D},
and BOLOCAM source BGPS G000.208-00.003 with a 1.1mm (268 GHz) flux density of 1.506$\pm$0.14 Jy \citep{2010ApJS..188..123R}.
\citet{2005MNRAS.363..405H} note the source is coincident with both radio continuum emission and with
a methanol maser with radial velocity $V = 42.1 - 49.4~\kms$ \citep{1998MNRAS.301..640W},
and is identified as an UC HII region at a distance
of 7.36 kpc \citep{2003A&A...410..597W}.  Its total 3~mm flux density is
0.227 $\pm$ 0.024 Jy, giving a spectral index values 
$\alpha =  -0.16 \pm 0.07$ and  $1.7\pm0.2$ measured between 
 1.4 -- 90 GHz  and 90 -- 268 GHz, respectively.
 
\begin{figure}
\includegraphics[width=\columnwidth,trim={2cm 2cm 2.5cm 2cm}]{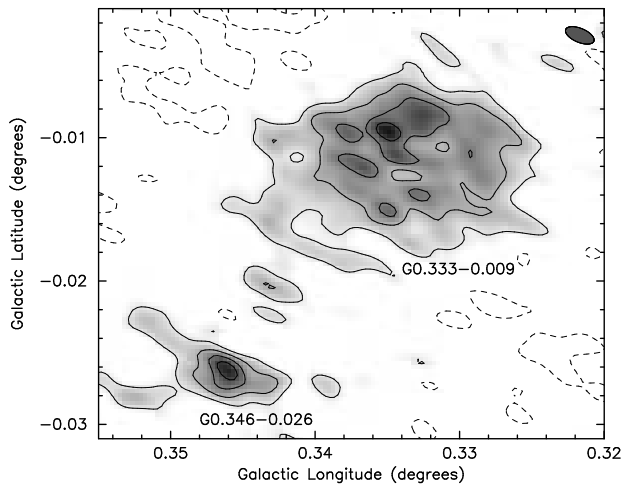}
\caption{Sources G0.333-0.009 and G0.346-0.026.
Contours are -0.0056, 0.0056, 0.0112, 0.0168, 0.0224 Jy beam$^{-1}$.
The $1\sigma$ rms
noise is 0.0028 Jy beam$^{-1}$ and the grayscale ranges from 0.003 to 0.03 Jy beam$^{-1}$.
}
\label{f-g033}
\end{figure}

\paragraph{Sources G0.333-0.009 and G0.346-0.026 (Fig. \ref{f-g033})}

G0.333-0.009 is roughly circular and has a clumpy, shell morphology with
integrated flux density of 0.968$\pm$0.052 Jy at 3 mm.  At 1.4~GHz it has
a total flux density of 1.618$\pm$0.101 Jy \citep{2010ApJS..191..275L}, giving a
spectral index to 90~GHz of $\alpha = -0.12\pm0.05$. This source is likely
an HII region.  G0.346-0.026 is listed with a 1.4 GHz flux density of
0.178$\pm$0.028 Jy in \citet{2010ApJS..191..275L}, a 5 GHz flux density of 0.080$\pm$0.024 Jy 
in \citet{1994ApJS...91..347B}, and categorized as a YSO candidate by \citet{2002A&A...392..971F}.
In our 3~mm map, the source has a central peak intensity of 0.026 Jy beam$^{-1}$ and is elongated north-south.

\subsection{3 mm Spectral Line Data}

\begin{figure*}
\includegraphics[width=2\columnwidth,angle=0,trim={2cm 4cm 2cm 4cm}]{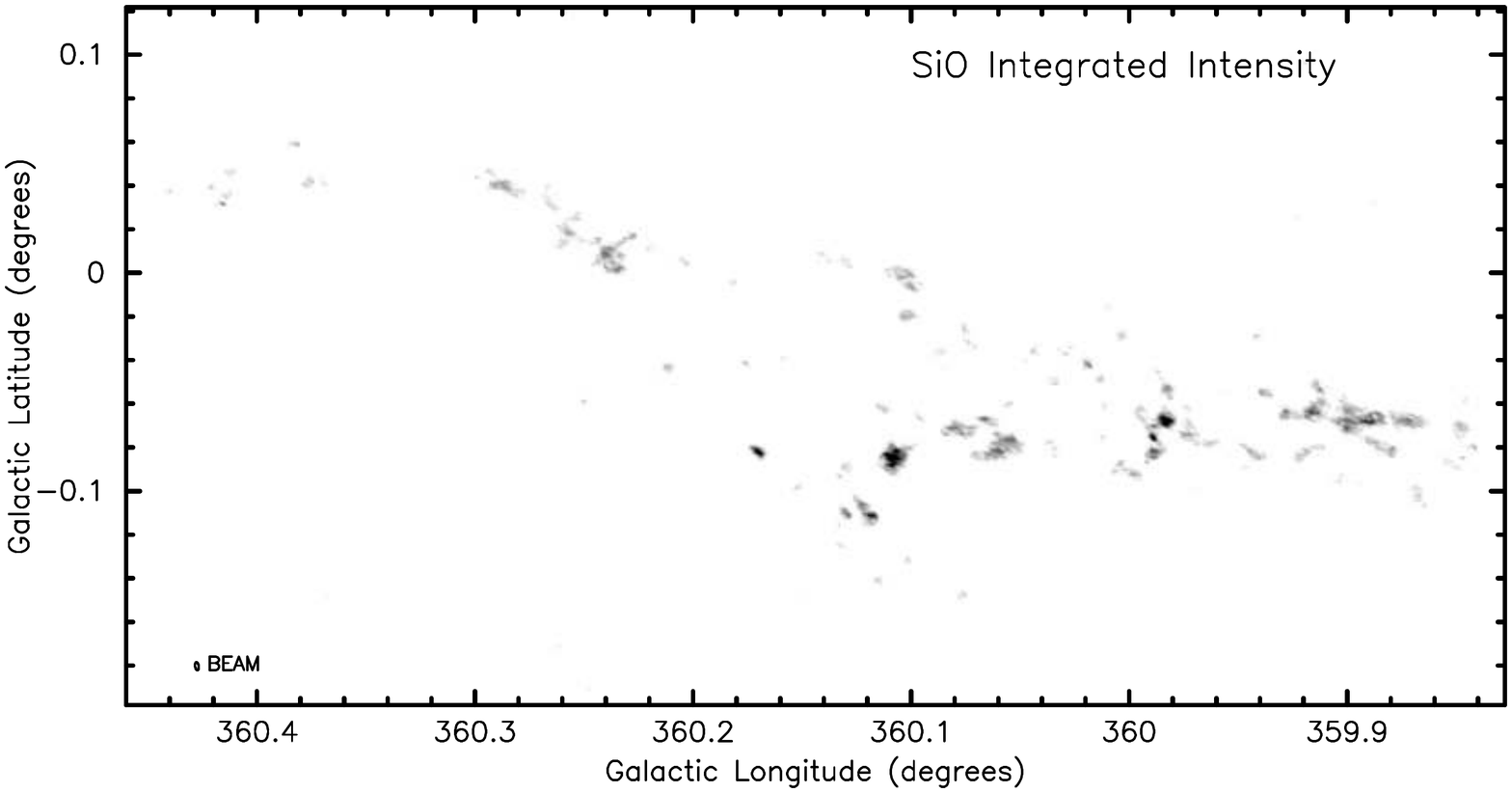}  
\caption{Intensity of \sio\ integrated between \vlsr = 0-60 \kms. This map contains only CARMA-15 interferometric data.  The grayscale range is 0 to 17 Jy beam$^{-1}$ \kms. The synthesized beam, measuring 13\pasec 1\arcsec $\times$ 5\pasec 9 is indicated in the lower left corner.}
\label{f-siomoment0}
\end{figure*}

\begin{figure*}
\includegraphics[width=2\columnwidth,angle=0,trim={2cm 4cm 2cm 4cm}]{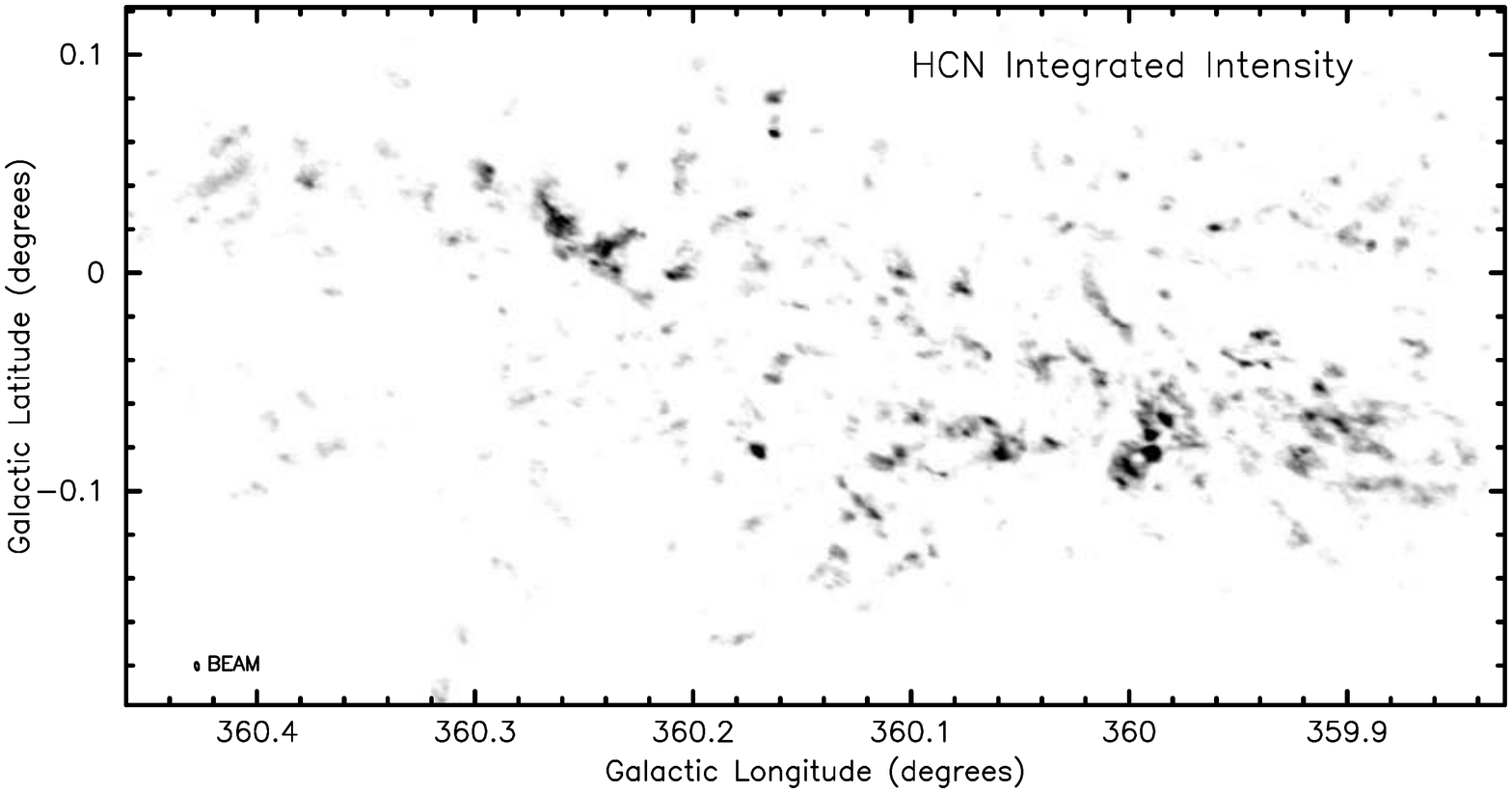}  
\caption{Intensity of \hcn\ integrated between \vlsr = 0-60 \kms. This map contains only CARMA-15 interferometric data.  The grayscale range is 1 to 40 Jy beam$^{-1}$ \kms. The synthesized beam, measuring 13\pasec 2\arcsec $\times$ 5\pasec 9 is indicated in the lower left corner.}
\label{f-hcnmoment0}
\end{figure*}

\begin{figure*}
\includegraphics[width=2\columnwidth,angle=0,trim={2cm 4cm 2cm 4cm}]{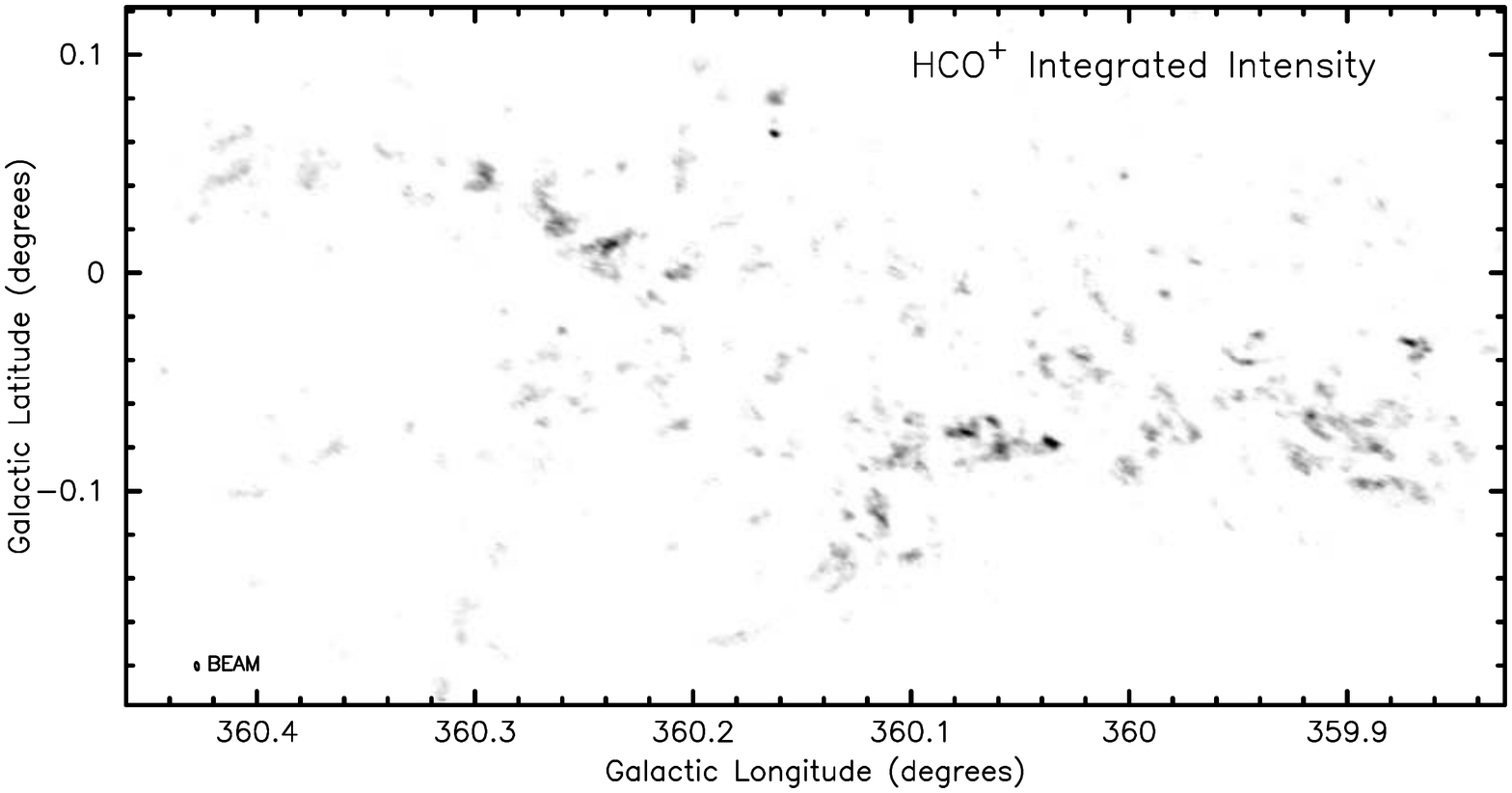}  
\caption{Intensity of \hcop\ integrated between \vlsr = 0-60 \kms. This map contains only CARMA-15 interferometric data. The grayscale range is 1 to 40 Jy beam$^{-1}$ \kms. The synthesized beam, measuring 13\pasec 1\arcsec $\times$ 5\pasec 9 is indicated in the lower left corner.}
\label{f-hcopmoment0}
\end{figure*}

\begin{figure*}
\includegraphics[width=2\columnwidth,angle=0,trim={2cm 4cm 2cm 4cm}]{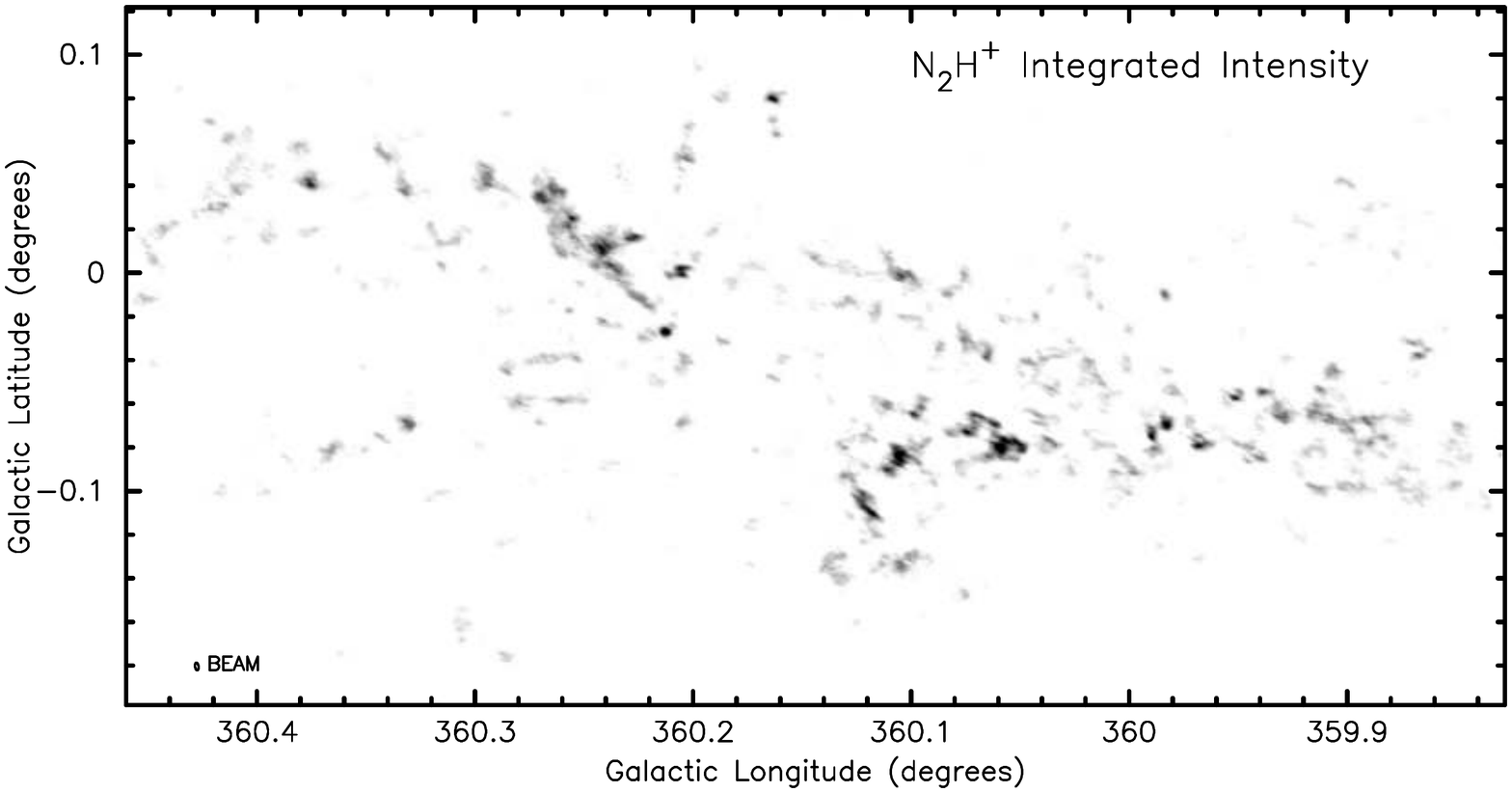}  
\caption{Intensity of \ntwohp\ integrated between \vlsr = 0-60 \kms. This map contains only CARMA-15 interferometric data. The grayscale range is 1 to 40 Jy beam$^{-1}$ \kms. The synthesized beam, measuring 12\pasec 3\arcsec $\times$ 5\pasec 6 is indicated in the lower left corner.}
\label{f-n2hpmoment0}
\end{figure*}

\begin{figure*}
\includegraphics[width=2\columnwidth,angle=0,trim={2cm 4cm 2cm 4cm}]{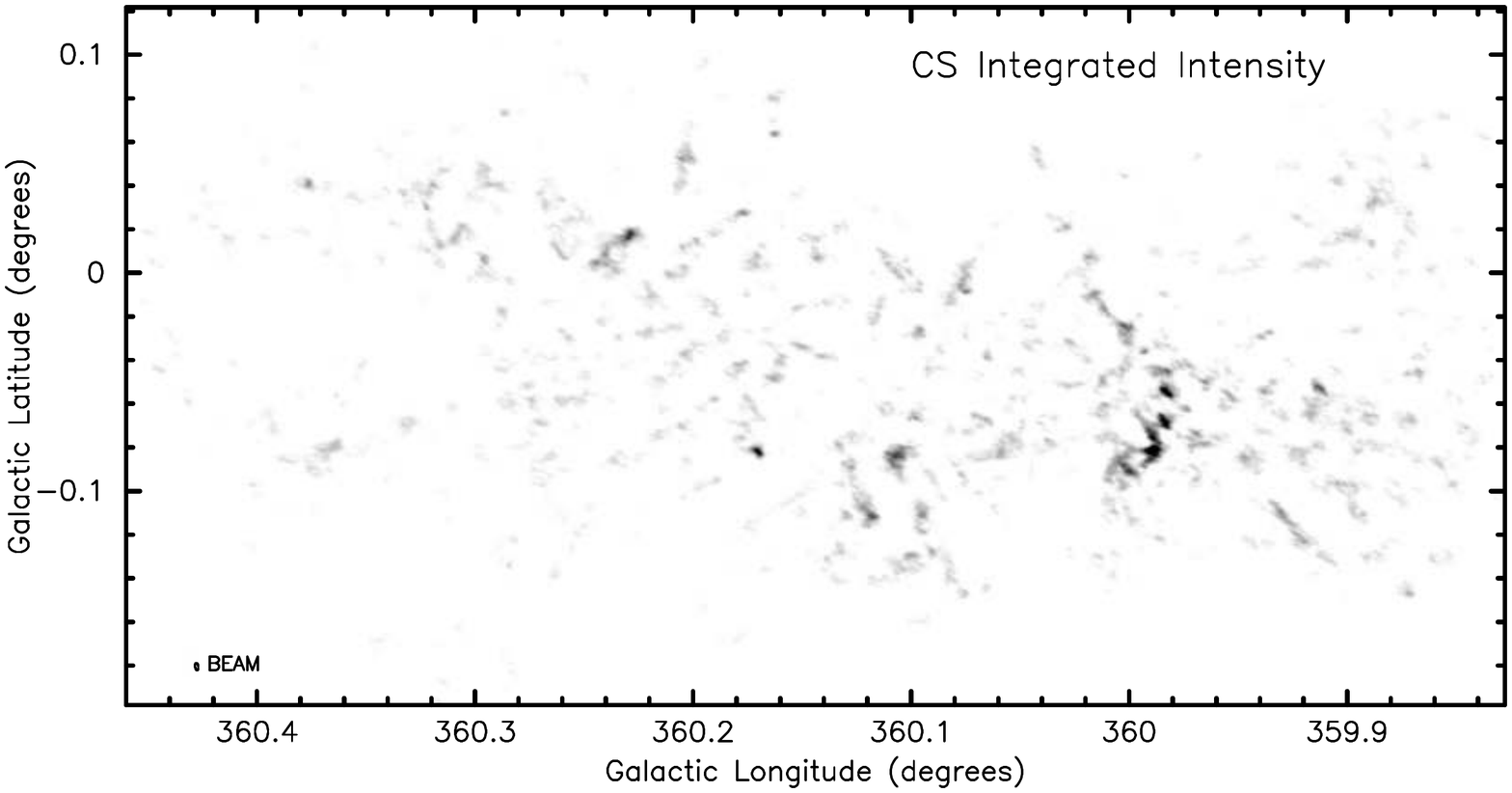}  
\caption{Intensity of \cs\ integrated between \vlsr = 0-60 \kms. This map contains only CARMA-15 interferometric data. The grayscale range is 2 to 50 Jy beam$^{-1}$ \kms. The synthesized beam, measuring 11\pasec 8\arcsec $\times$ 5\pasec 3 is indicated in the lower left corner.}
\label{f-csmoment0}
\end{figure*}

\begin{figure*}
 \includegraphics[scale=0.8,angle=-90,trim={0cm 2cm 0cm 2cm}]{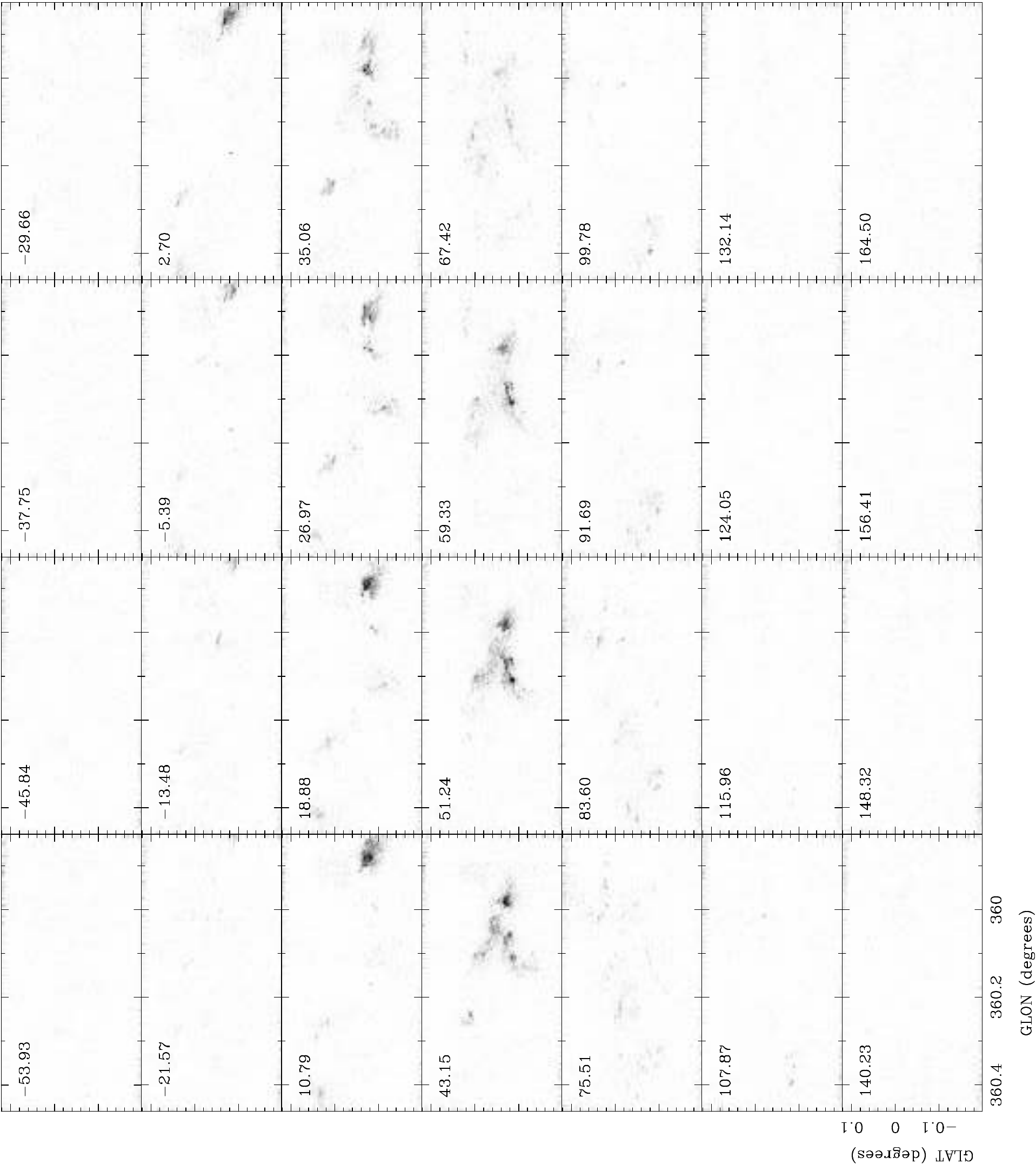}  
\caption{Channel maps of \sio\ of the combined CARMA-15 plus Mopra data. Panels are three channel averages with center LSR velocity indicated in upper left of each panel. The grayscale range is 0.2 to 1.3 Jy beam$^{-1}$. The spatial resolution is 13\pasec 1\arcsec $\times$ 5\pasec 9.}
\label{f-SiOc+m}
\end{figure*}

\begin{figure*}
\includegraphics[scale=0.8,angle=-90,trim={0cm 2cm 0cm 2cm}]{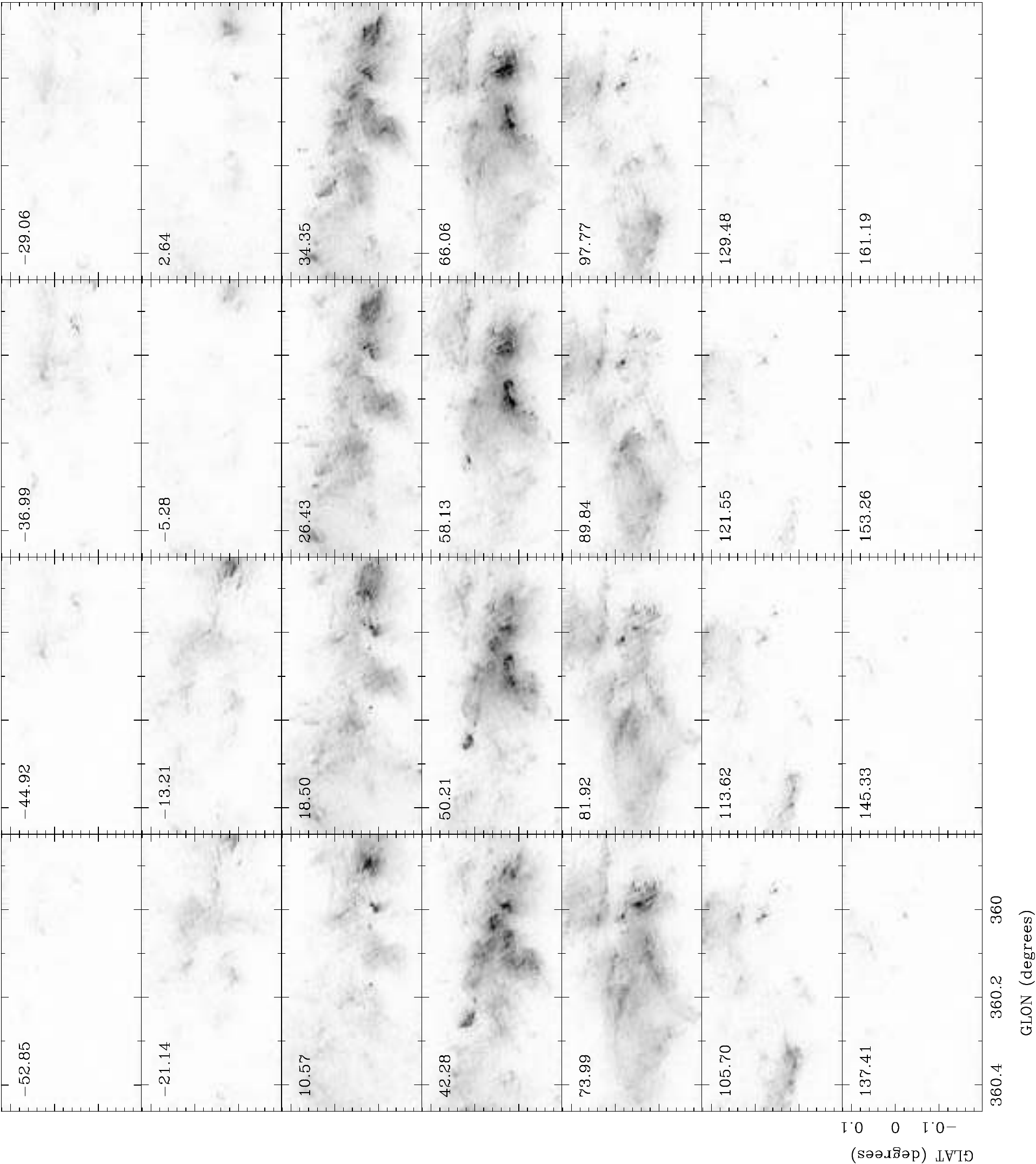}  
\caption{Channel maps of \hcn\ of the combined CARMA-15 plus Mopra data. Panels are three channel averages with center LSR velocity indicated in upper left of each panel. The grayscale range is 0.1 to 4 Jy beam$^{-1}$. 
The spatial resolution is 13\pasec 2\arcsec $\times$ 5\pasec 9\arcsec.
}
\label{f-HCNc+m}
\end{figure*}

\begin{figure*}
\includegraphics[scale=0.8,angle=-90,trim={0cm 2cm 0cm 2cm}]{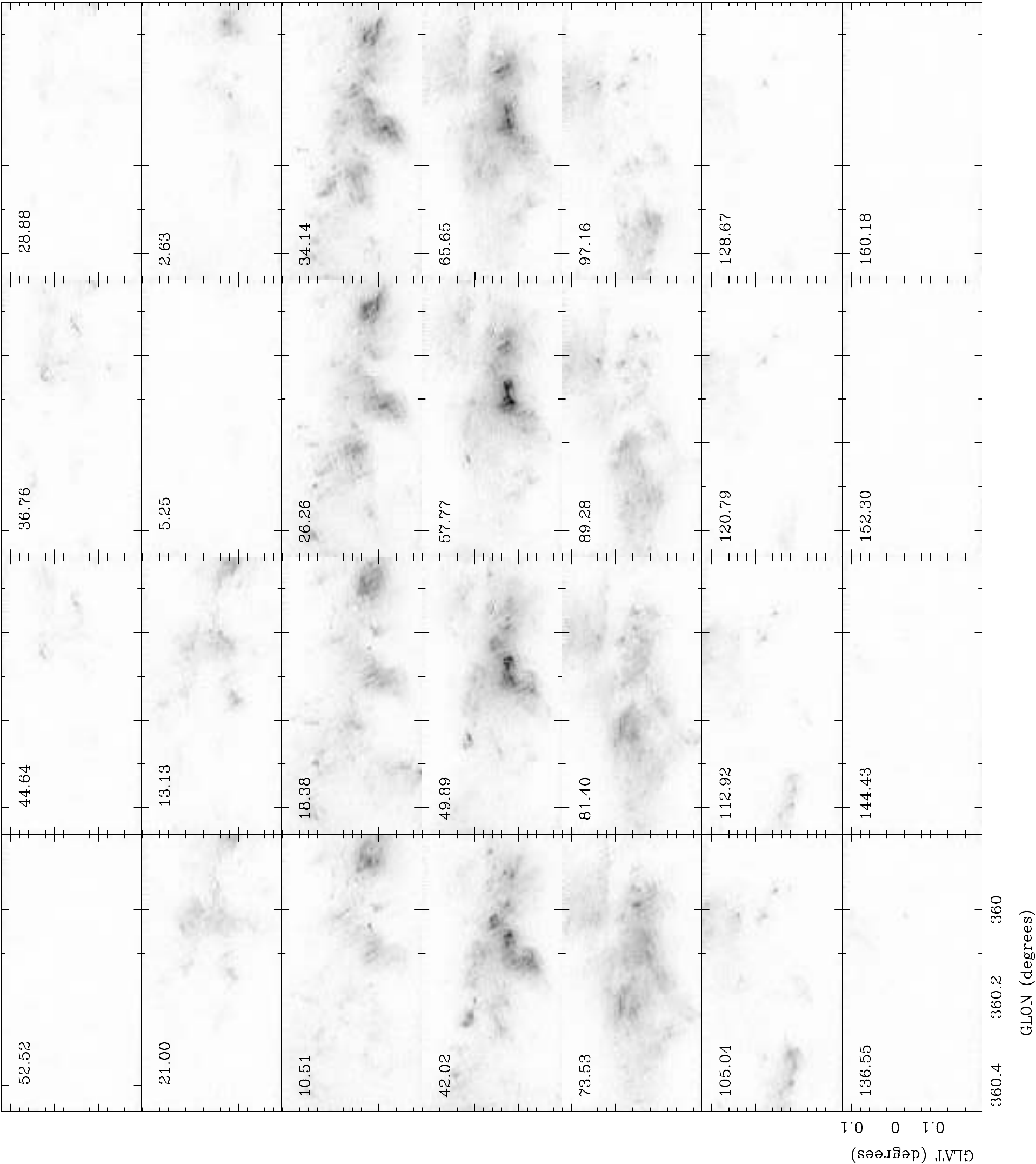}  
\caption{Channel maps of \hcop\ of the combined CARMA-15 plus Mopra data. Panels are three channel averages with center LSR velocity indicated in upper left of each panel. The grayscale range is 0.1 to 4 Jy beam$^{-1}$.
The spatial resolution is 13\pasec 1\arcsec $\times$ 5\pasec 9\arcsec.
}
\label{f-HCOpc+m}
\end{figure*}

\begin{figure*}
\includegraphics[scale=0.8,angle=-90,trim={0cm 2cm 0cm 2cm}]{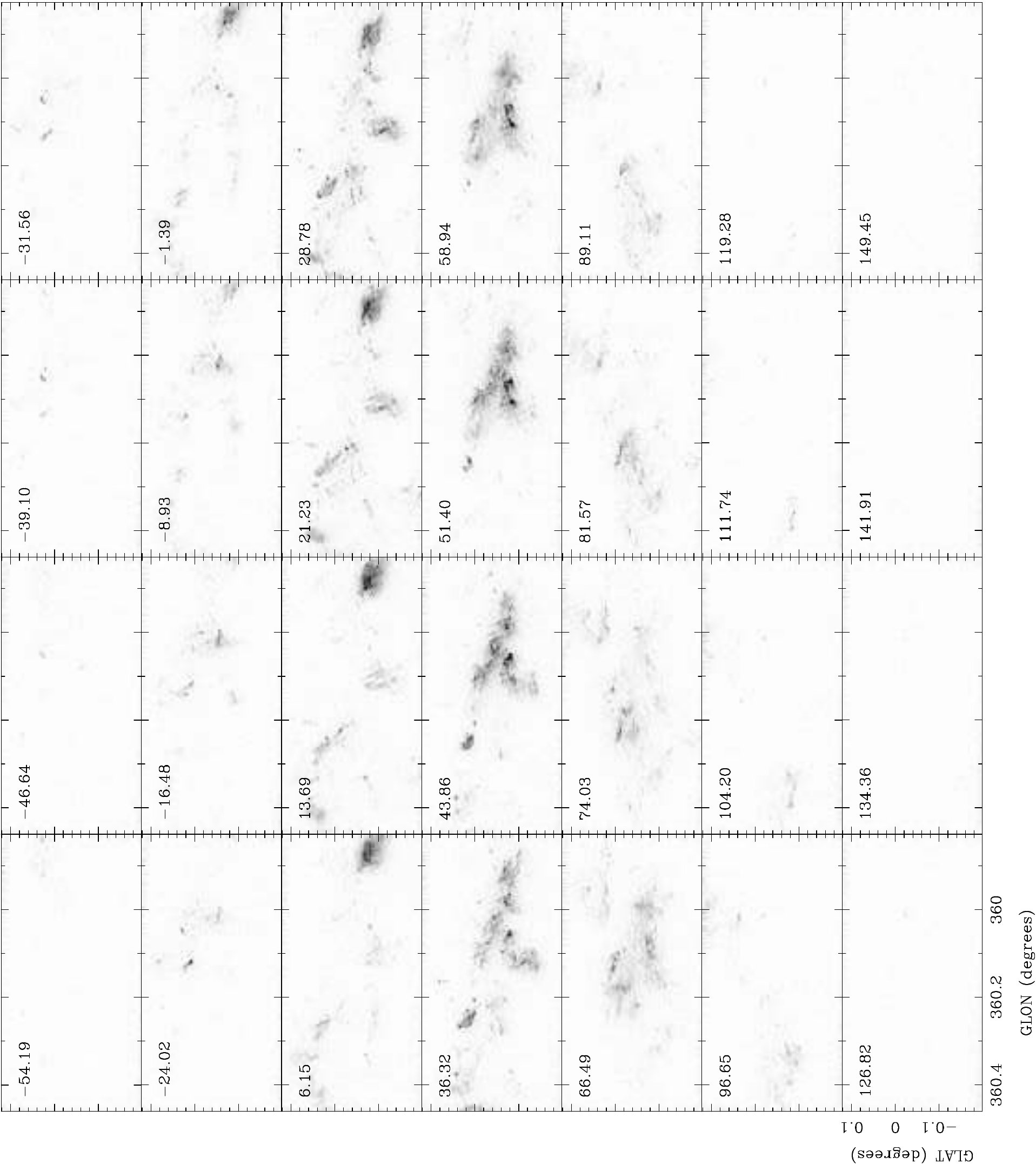}  
\caption{Channel maps of \ntwohp\ of the combined CARMA-15 plus Mopra data. Panels are three channel averages with center LSR velocity indicated in upper left of each panel. The grayscale range is 0.1 to 2.75 Jy beam$^{-1}$.
The spatial resolution is 12\pasec 3\arcsec $\times$ 5\pasec 6\arcsec.
}
\label{f-N2Hpc+m}
\end{figure*}

\begin{figure*}
\begin{tabular}{c}
\includegraphics[scale=0.56,angle=-90]{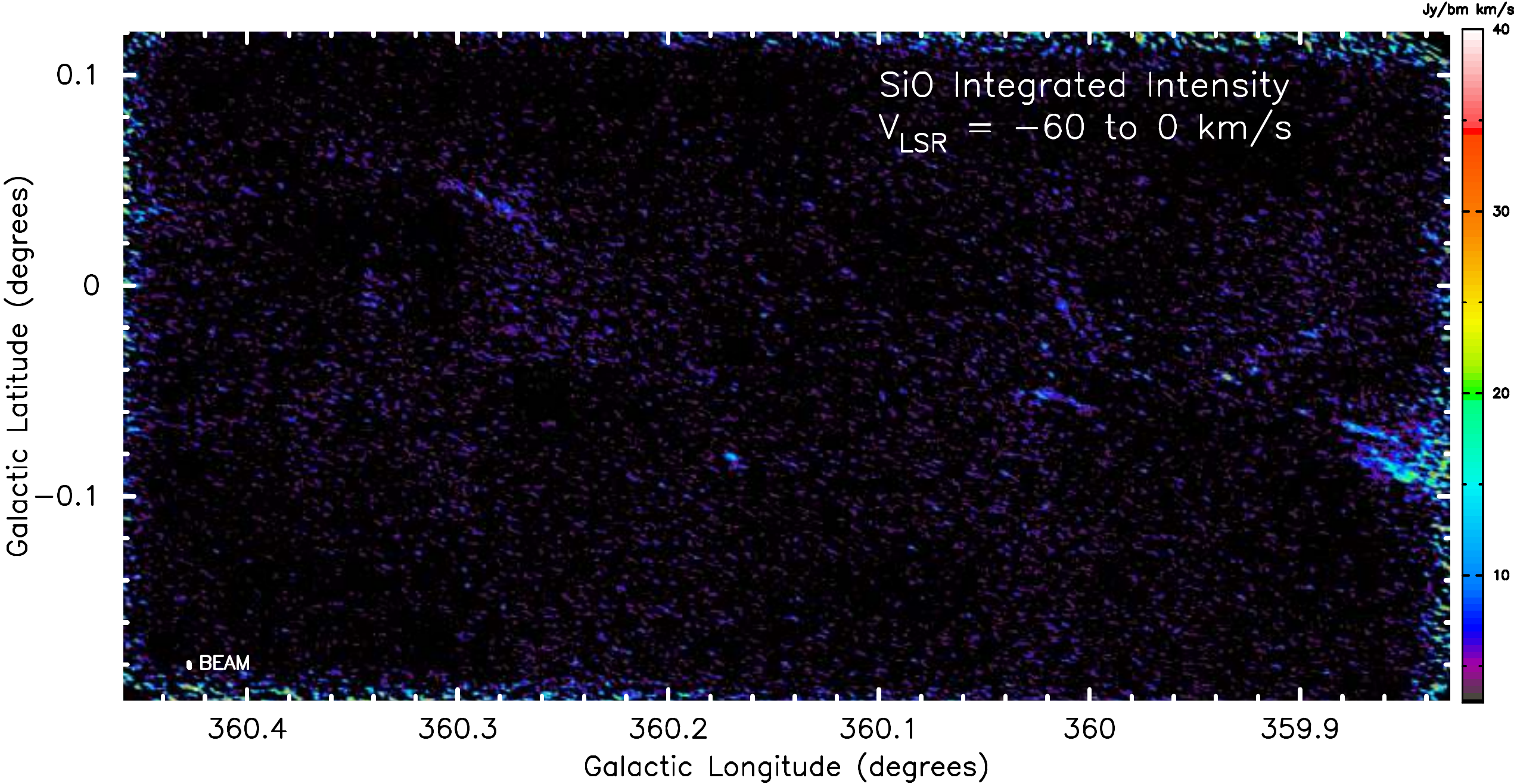} \\
\includegraphics[scale=0.56,angle=-90]{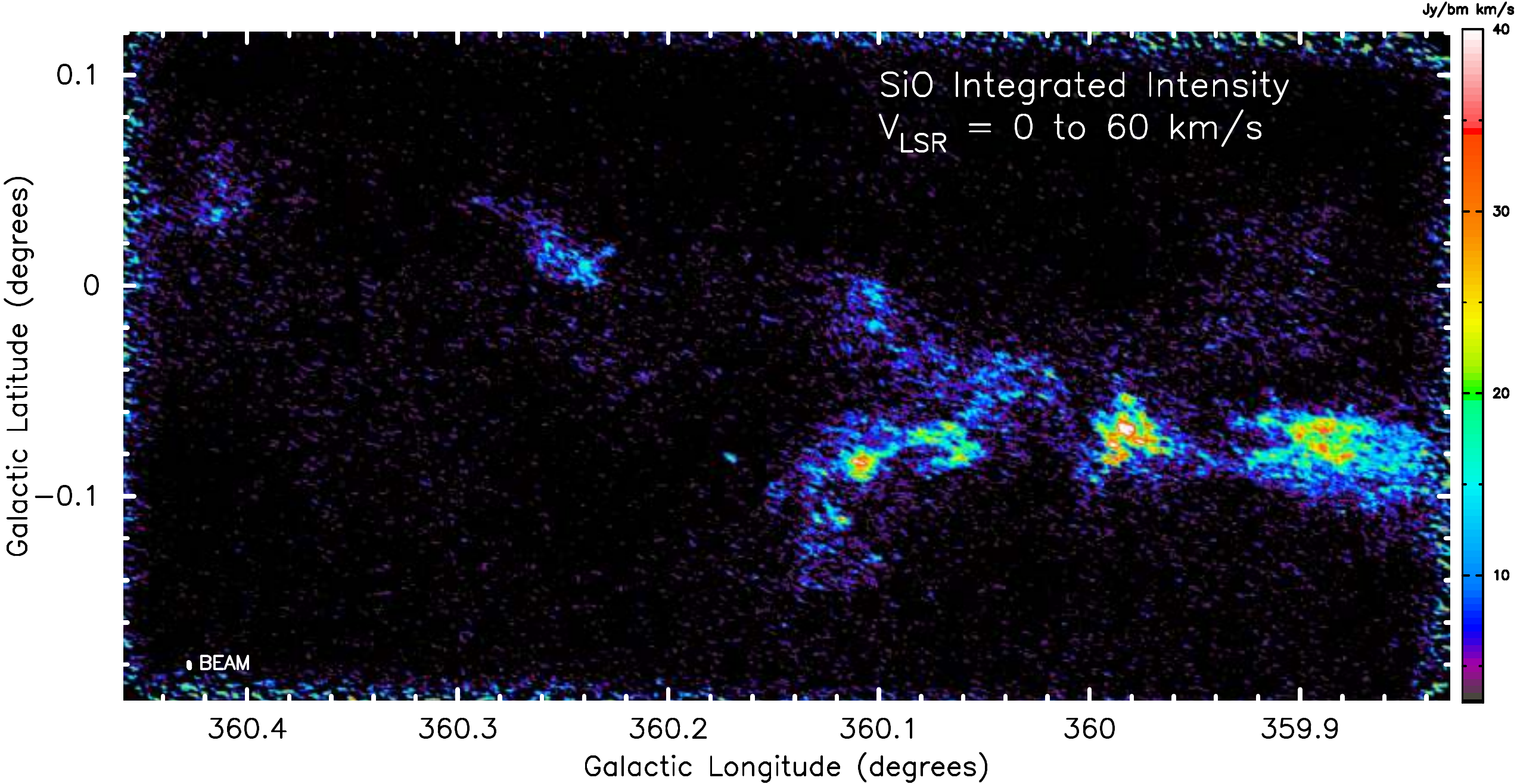} \\
\includegraphics[scale=0.56,angle=-90]{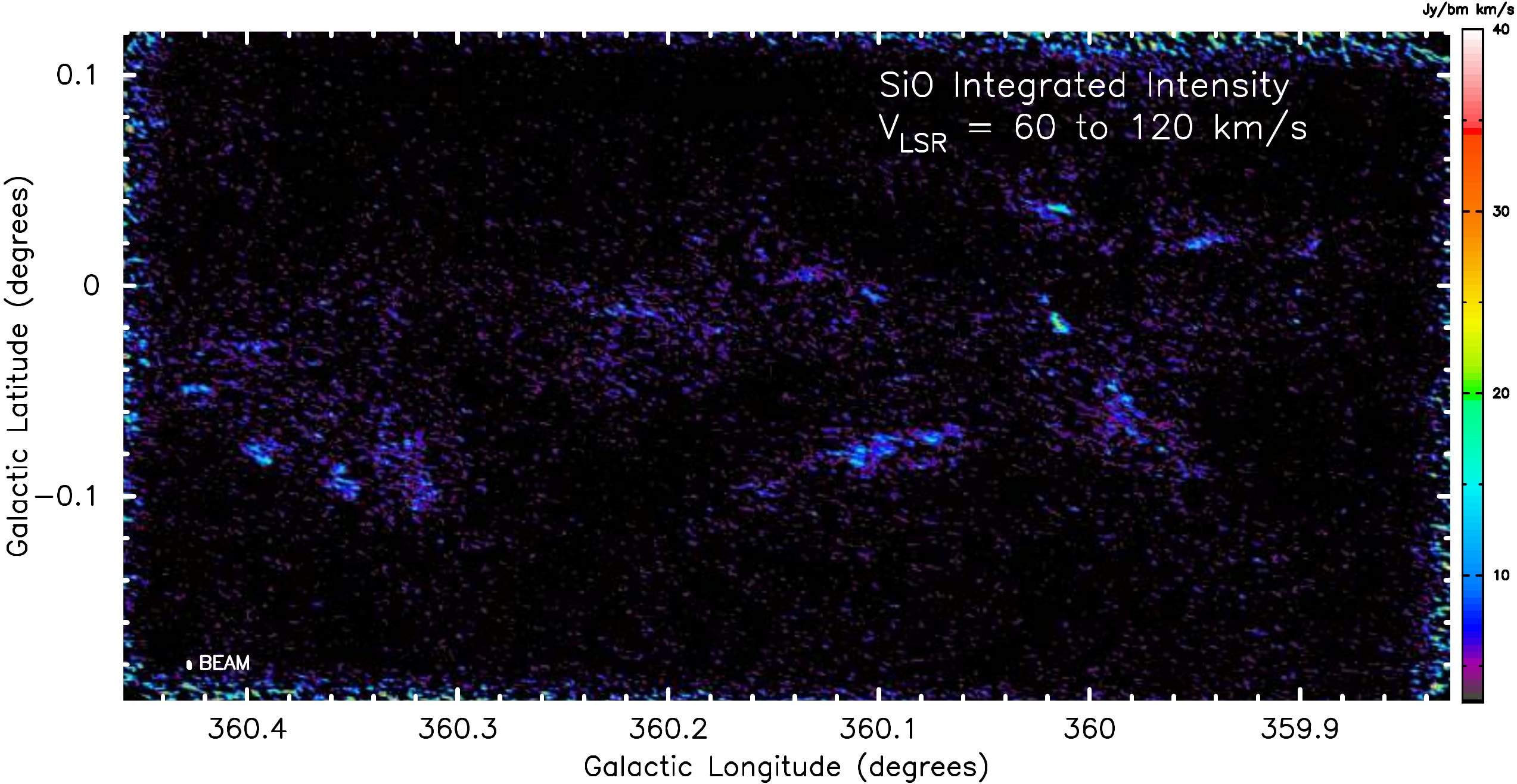} \\
\end{tabular}
\caption{\sio\ integrated intensity of the CARMA-15 plus Mopra data.
Panels show integrals over selected velocity intervals:
\textit{top)} \vlsr = --60 to 0 \kms; 
\textit{middle)} \vlsr = 0 to 60 \kms; 
\textit{bottom)} \vlsr = 60 to 120 \kms. 
The color wedge indicates map intensity values.
}
\label{f-SiOmoments}
\end{figure*}

\begin{figure*}
\begin{tabular}{cc}
\includegraphics[scale=0.56,angle=-90]{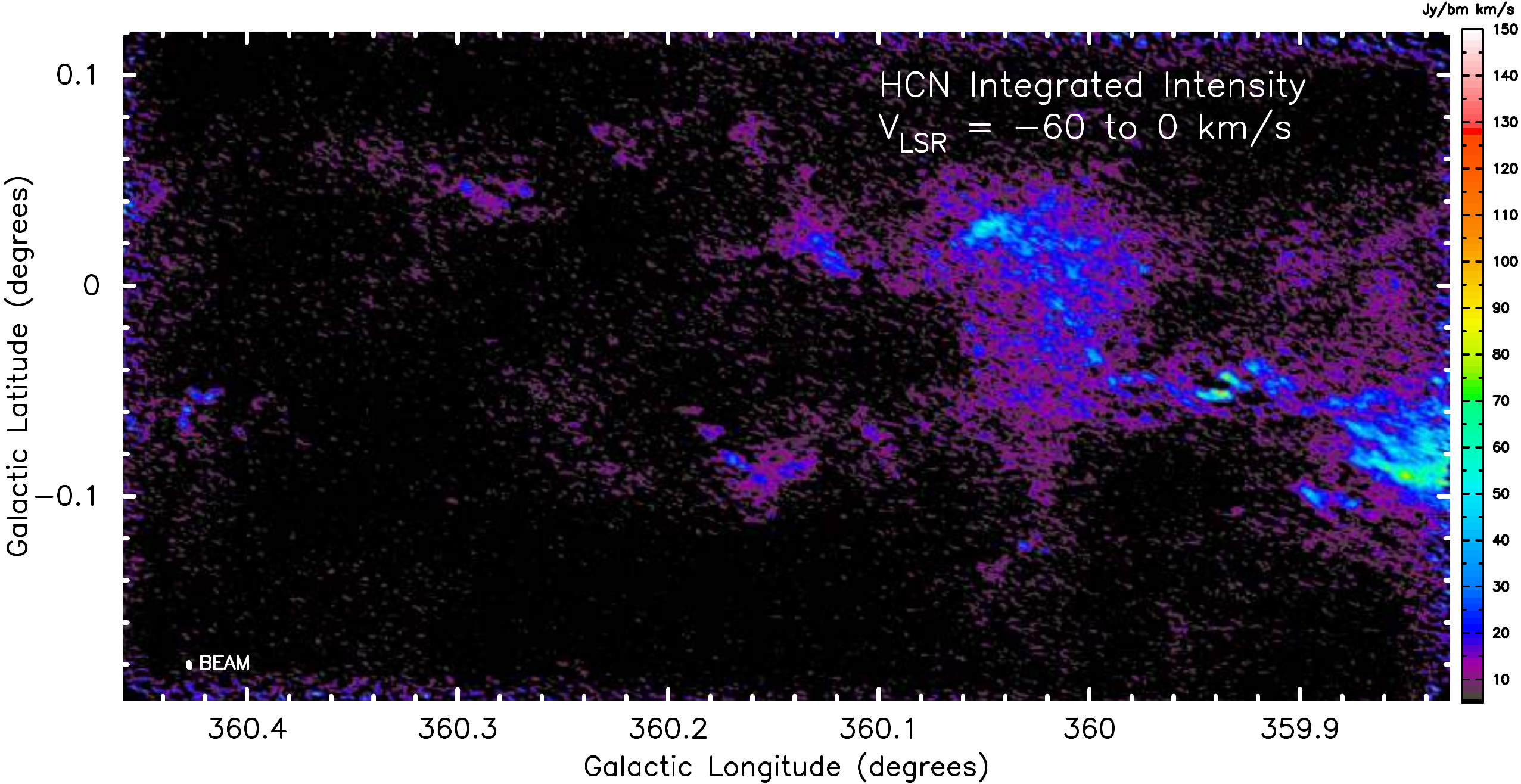} \\
\includegraphics[scale=0.56,angle=-90]{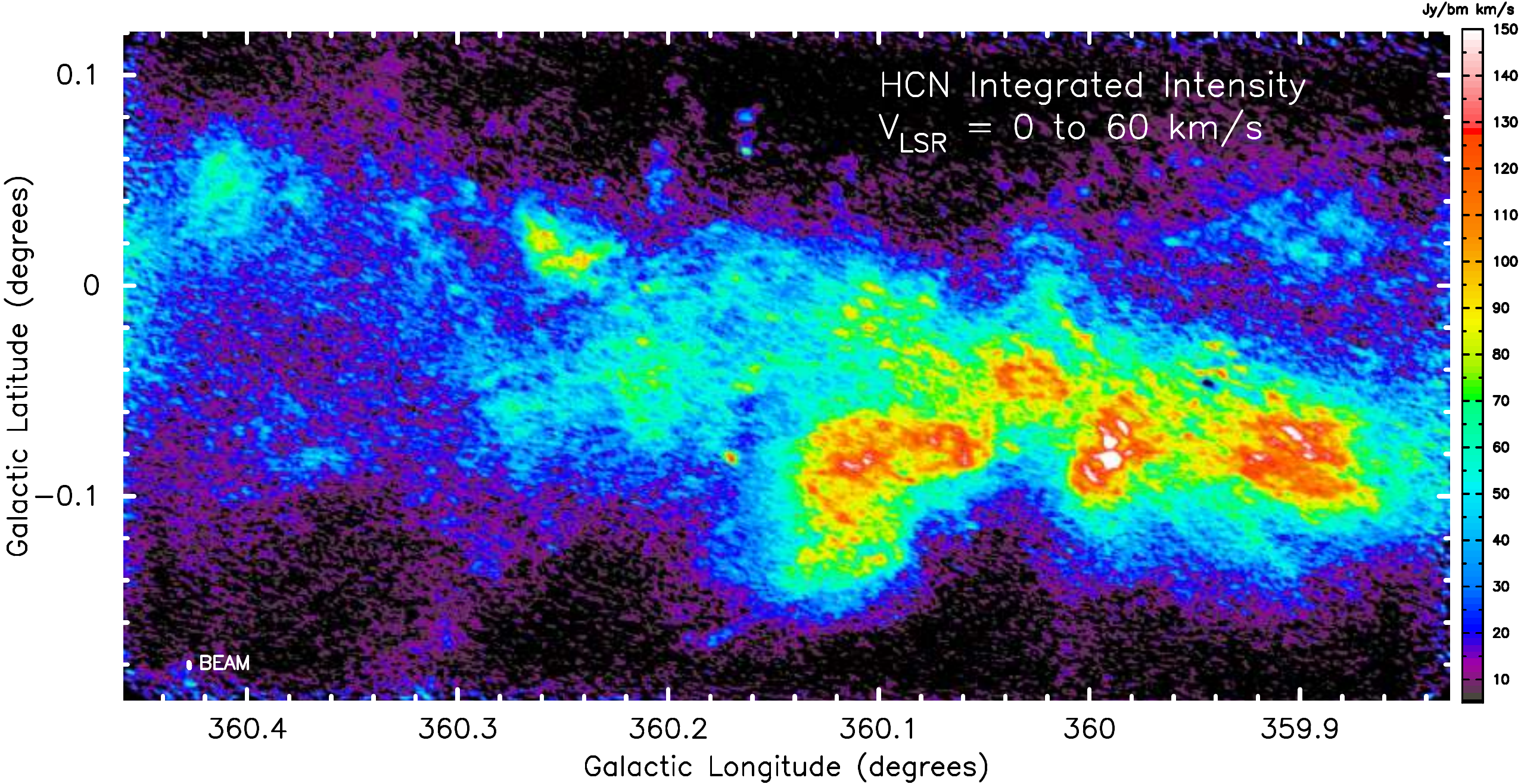} \\
\includegraphics[scale=0.56,angle=-90]{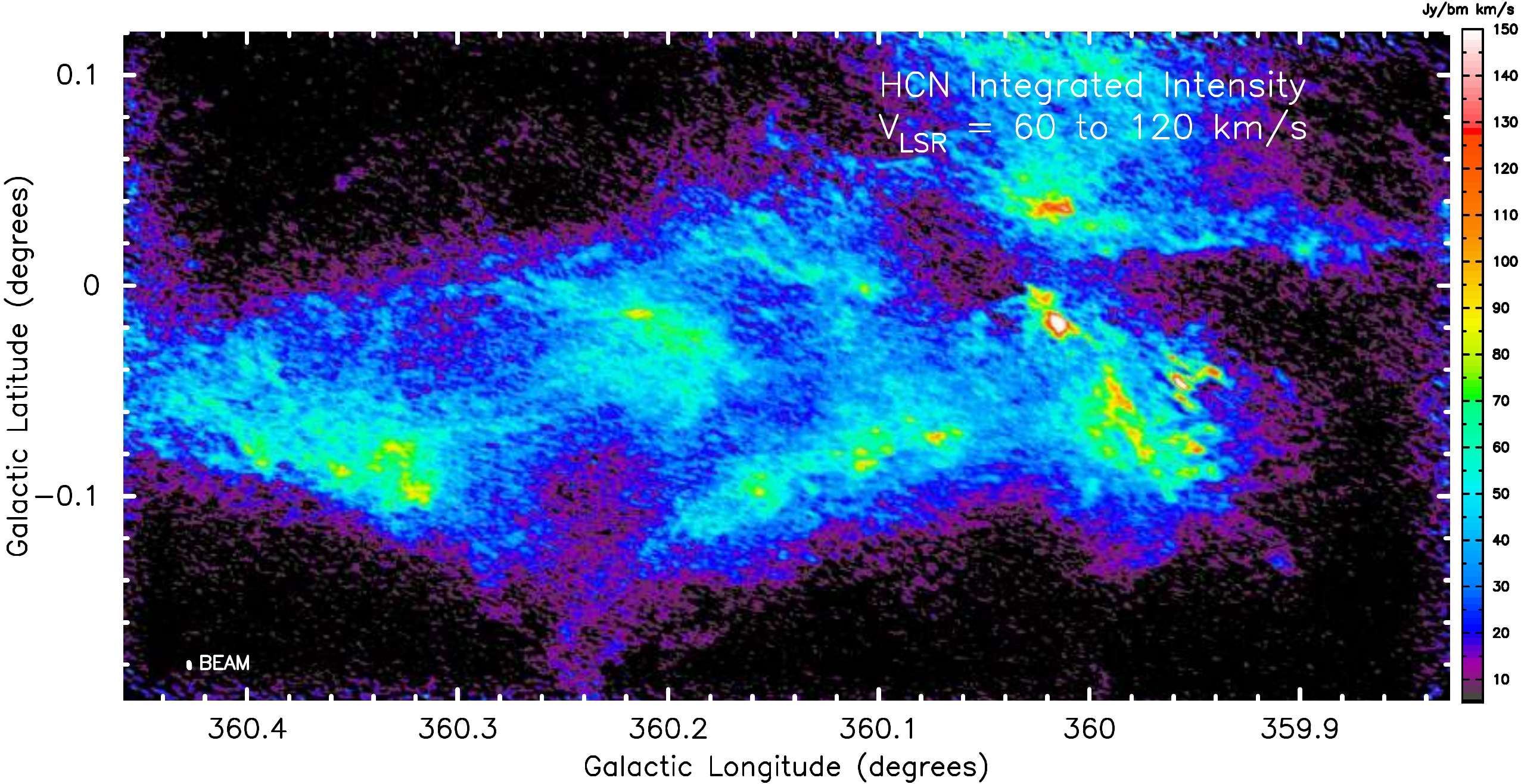} \\
\end{tabular}
\caption{\hcn\ integrated intensity of the CARMA-15 plus Mopra data.
Panels show integrals over selected velocity intervals:
\textit{top)} \vlsr = --60 to 0 \kms; 
\textit{middle)} \vlsr = 0 to 60 \kms; 
\textit{bottom)} \vlsr = 60 to 120 \kms. 
The color wedge indicates map intensity values.
}
\label{f-HCNmoments}
\end{figure*}

\begin{figure*}
\begin{tabular}{c}
\includegraphics[scale=0.56,angle=-90]{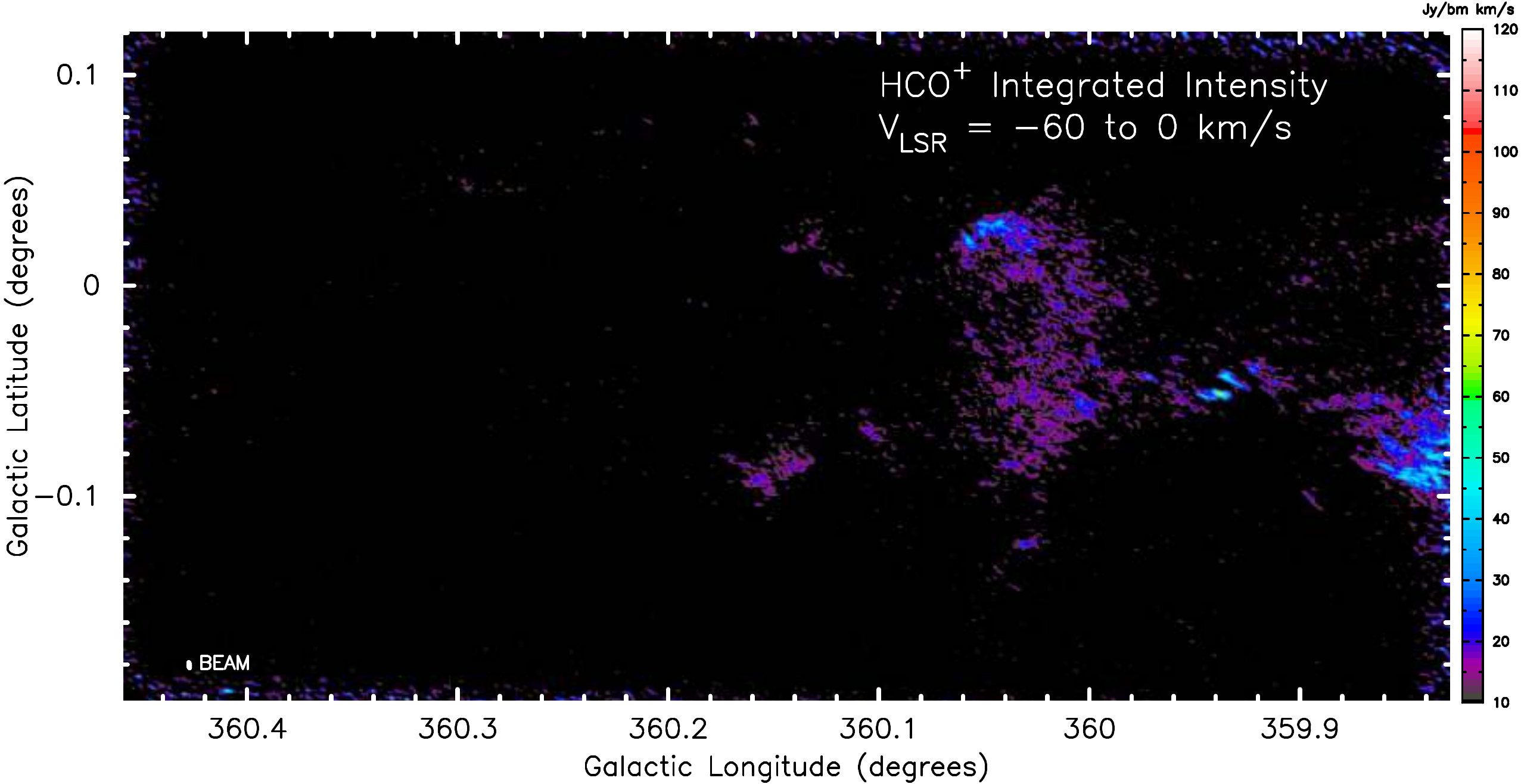} \\
\includegraphics[scale=0.56,angle=-90]{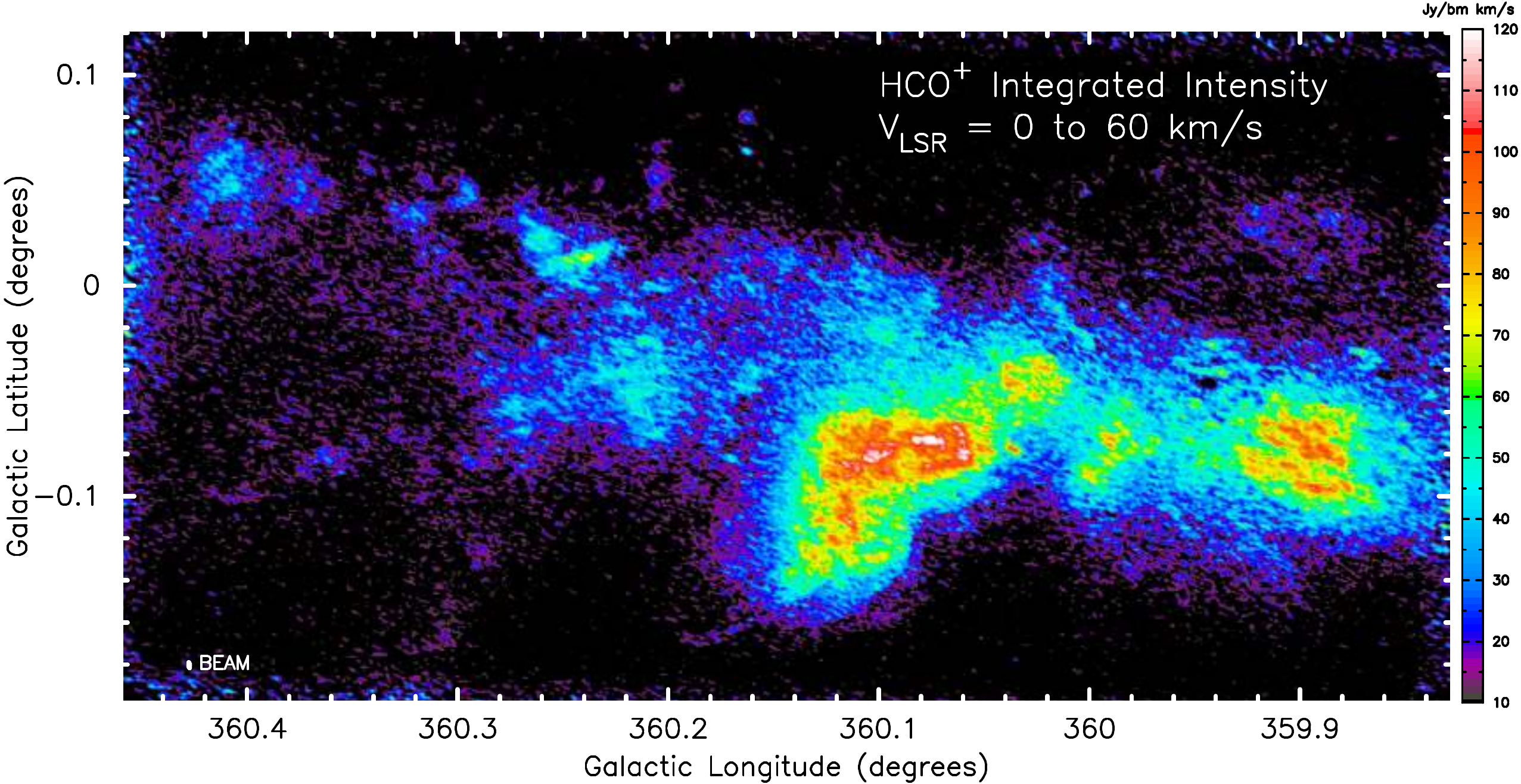} \\
\includegraphics[scale=0.56,angle=-90]{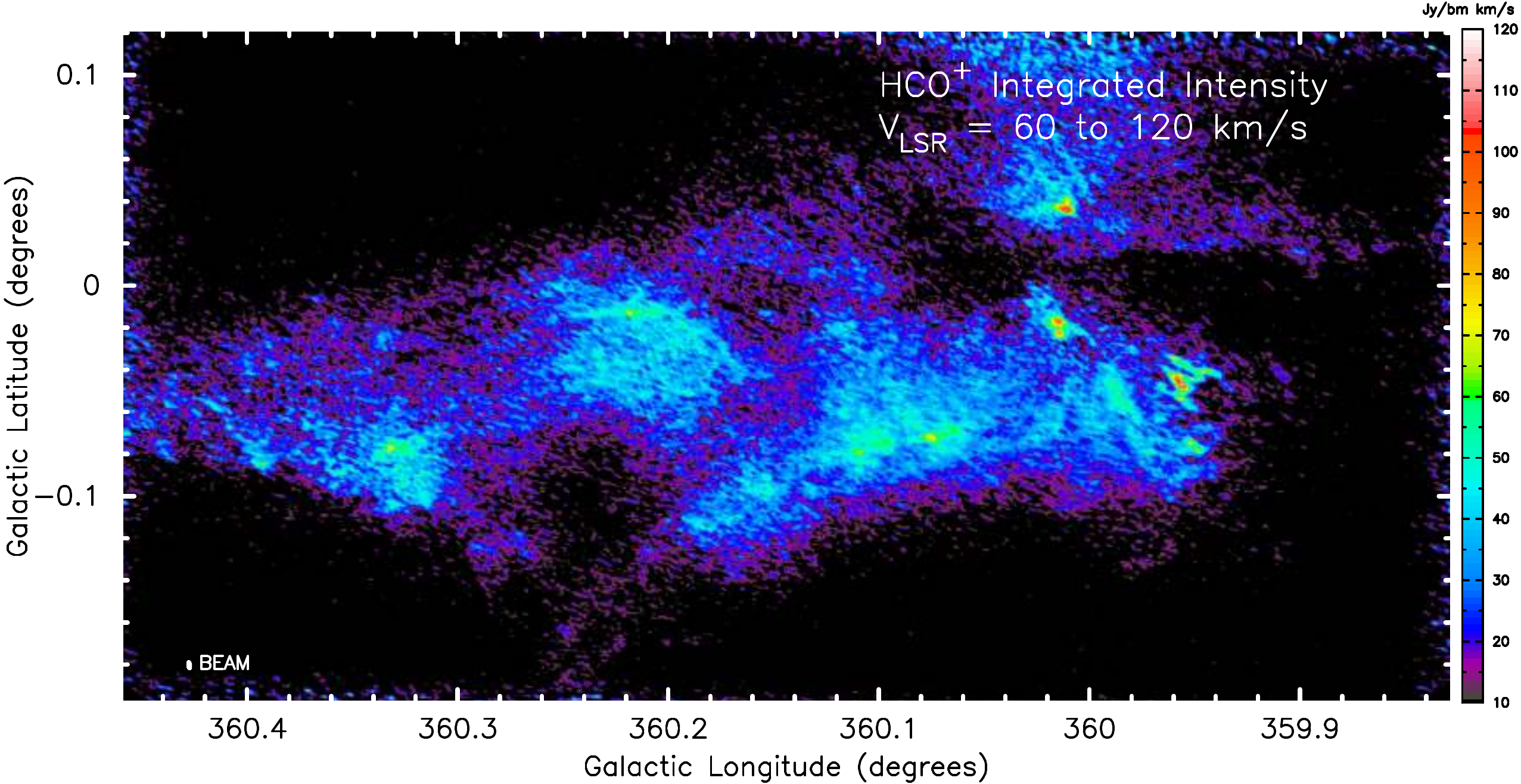} \\
\end{tabular}
\caption{\hcop\ integrated intensity  of the CARMA-15 plus Mopra data.
Panels show integrals over selected velocity intervals:
\textit{top)} \vlsr = --60 to 0 \kms; 
\textit{middle)} \vlsr = 0 to 60 \kms; 
\textit{bottom)} \vlsr = 60 to 120 \kms. 
The color wedge indicates map intensity values.
}
\label{f-HCOpmoments}
\end{figure*}

\begin{figure*}
\begin{tabular}{c}
\includegraphics[scale=0.56,angle=-90]{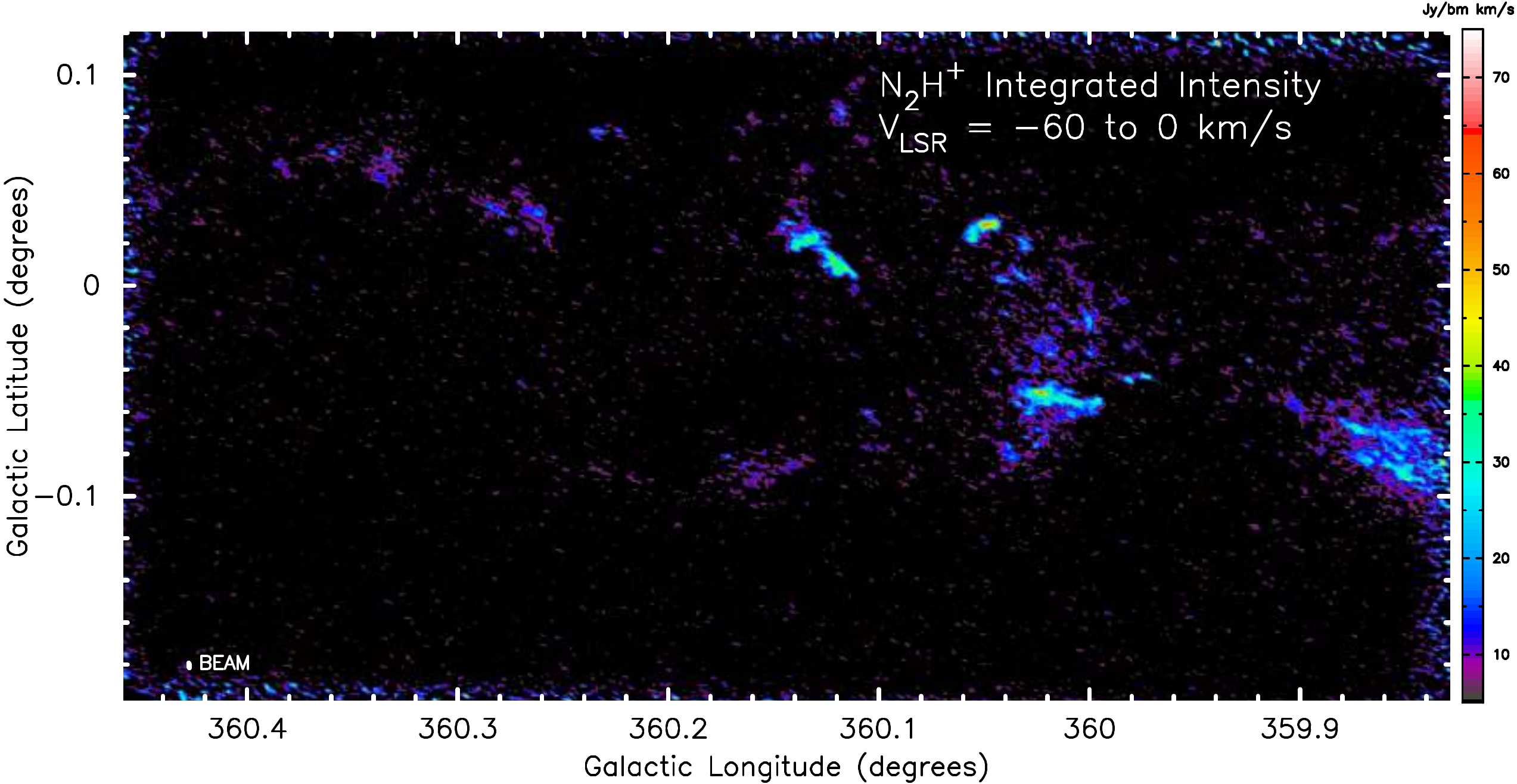} \\
\includegraphics[scale=0.56,angle=-90]{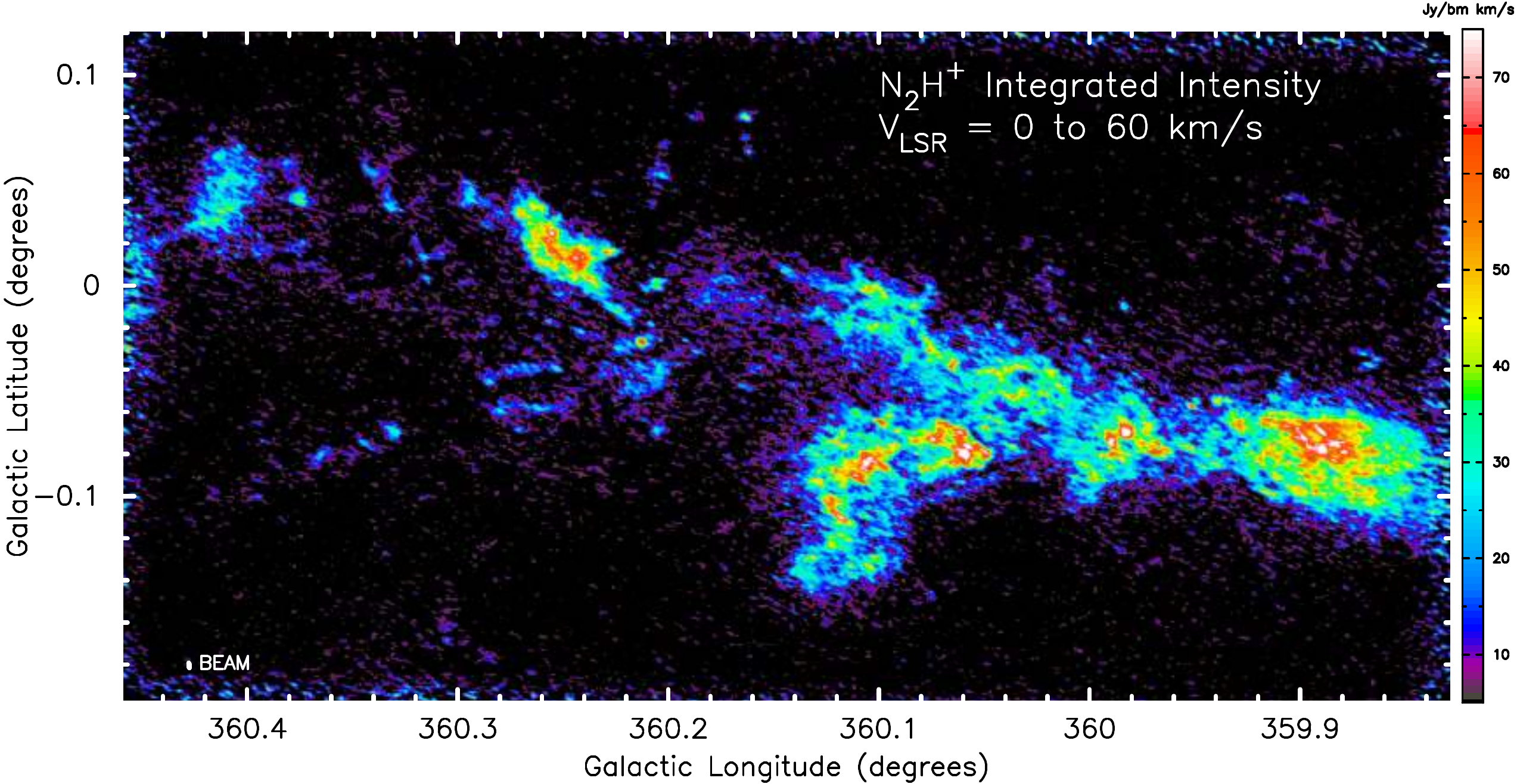} \\
\includegraphics[scale=0.56,angle=-90]{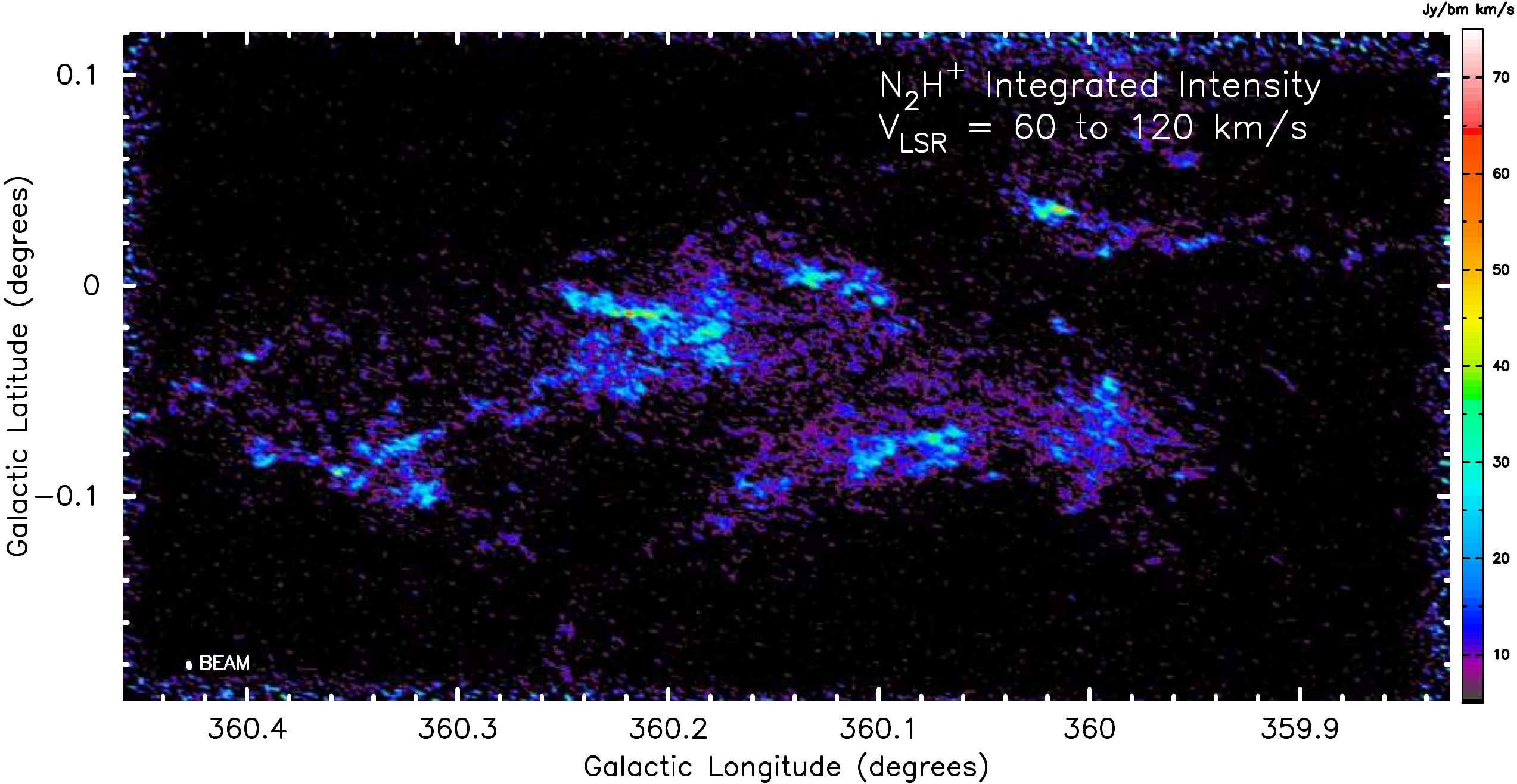} \\
\end{tabular}
\caption{\ntwohp\ integrated intensity  of the CARMA-15 plus Mopra data.
Panels show integrals over selected velocity intervals:
\textit{top)} \vlsr = --60 to 0 \kms; 
\textit{middle)} \vlsr = 0 to 60 \kms; 
\textit{bottom)} \vlsr = 60 to 120 \kms. 
The color wedge indicates map intensity values.
}
\label{f-N2Hpmoments}
\end{figure*}

We present integrated intensity maps from both the CARMA-15 (Figs.
\ref{f-siomoment0} to \ref{f-csmoment0})
and  CARMA-15 plus Mopra data cubes
(Figs. \ref{f-SiOc+m} to \ref{f-N2Hpmoments}).  
In maps containing only CARMA data, compact clouds and patches
of diffuse features are detected whereas, the combined maps, additional large scale
diffuse emission is revealed.  In the channel maps (Figs. \ref{f-SiOc+m} to
\ref{f-N2Hpc+m}), molecular emission is detected from a number of Galactic Cener clouds such as G0.253+0.016
(``The Brick''), G0.13-0.13, the 20 \kms\ and 50 \kms\ clouds, and 
the 2--7 pc circumnuclear ring orbiting Sgr A*. 
In addition, we find a 
wealth of fine scale filamentary structure
throughout the region.  
Also seen are \hcopA\ and \hcnA\ absorption features between $\vlsr \sim -140-100 \kms$ 
at the position of Sgr A*, as previously reported by \citet{2001ApJ...551..254W} (see also upper left panel of Fig. \ref{f-examplespec}).
  
Figs. \ref{f-SiOmoments} to \ref{f-N2Hpmoments} show intensity maps
integrated over 60 \kms\ wide intervals covering the bulk of the emission
in the data cubes. The data were clipped at twice the relevant 1$\sigma$ rms noise indicated in Table \ref{t-obssummary}
before integration.  The \sio\ emission is the weakest, principally seen only in
the range $\vlsr = 0-60$~\kms\ and in only the region where HCN emission is strongest.   The \hcnA\
and \hcopA\ emission are brightest with large regions of diffuse emission.
The \ntwohpA\ emission, in which individual
clouds are most easily distinguished,
has brightness intermediate between \sio\ and \hcopA. 

\begin{figure*}
\begin{tabular}{c}
\includegraphics[scale=0.56,angle=-90]{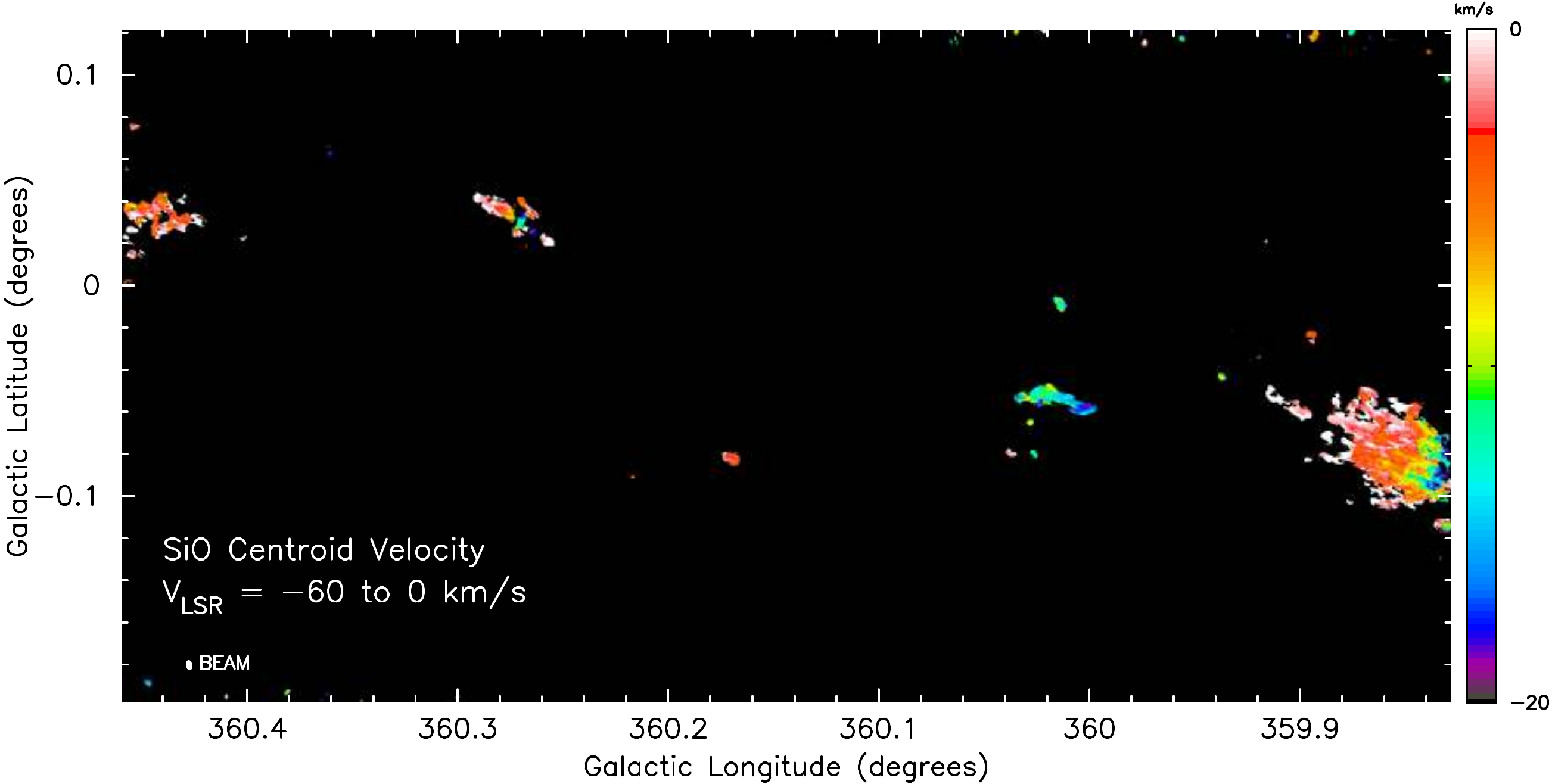} \\
\includegraphics[scale=0.56,angle=-90]{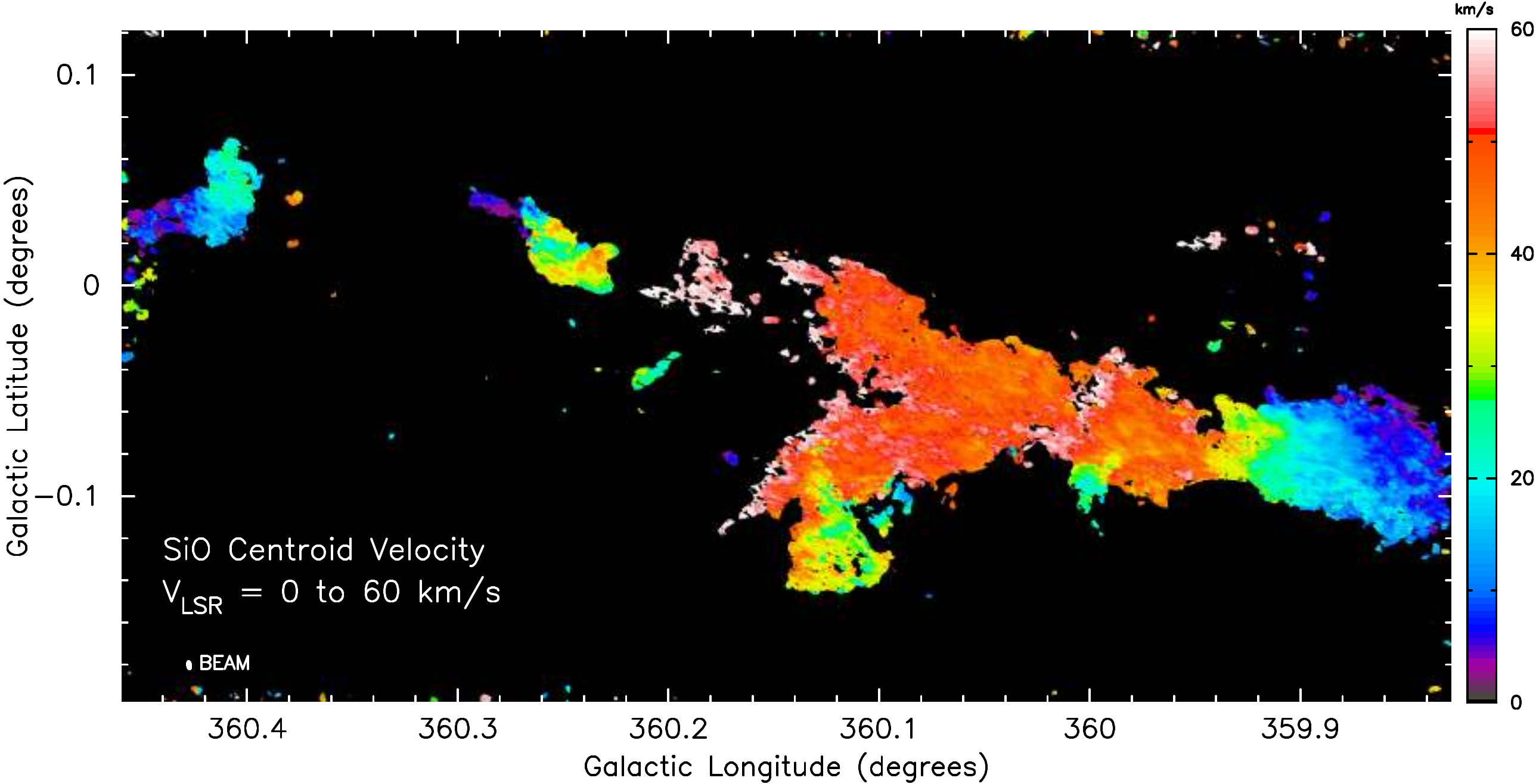} \\
\includegraphics[scale=0.56,angle=-90]{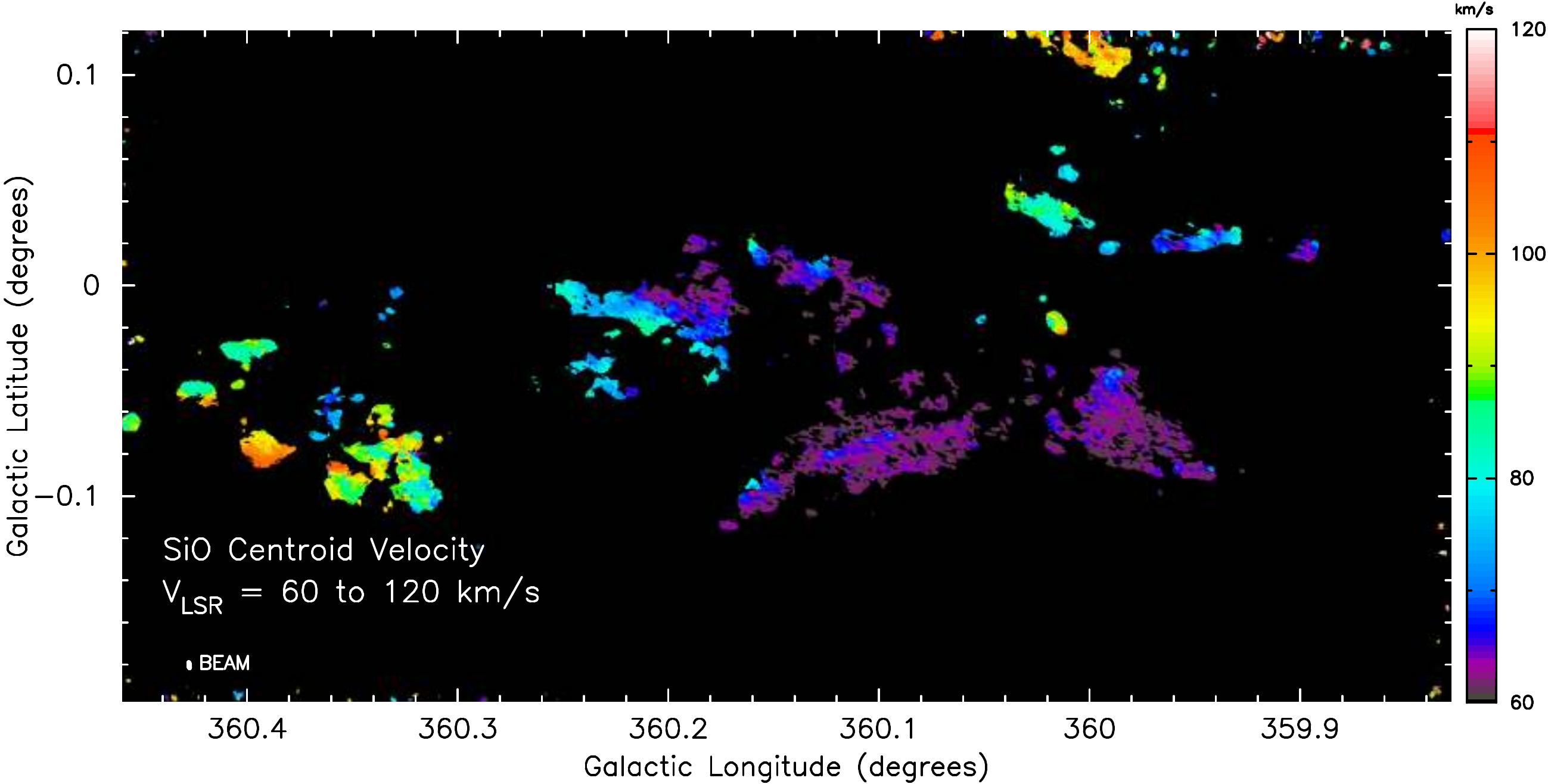} \\
\end{tabular}
\caption{\sio\ centroid velocity  of the CARMA-15 plus Mopra data.
Panels show integrals over selected velocity intervals:
\textit{top)} \vlsr = -60 to 0 \kms; 
\textit{middle)} \vlsr = 0 to 60 \kms; 
\textit{bottom)} \vlsr = 60 to 120 \kms; 
and the color wedge indicates map centroid velocity values.
}
\label{f-SiOcentroid}
\end{figure*}

\begin{figure*}
\begin{tabular}{c}
\includegraphics[scale=0.56,angle=-90]{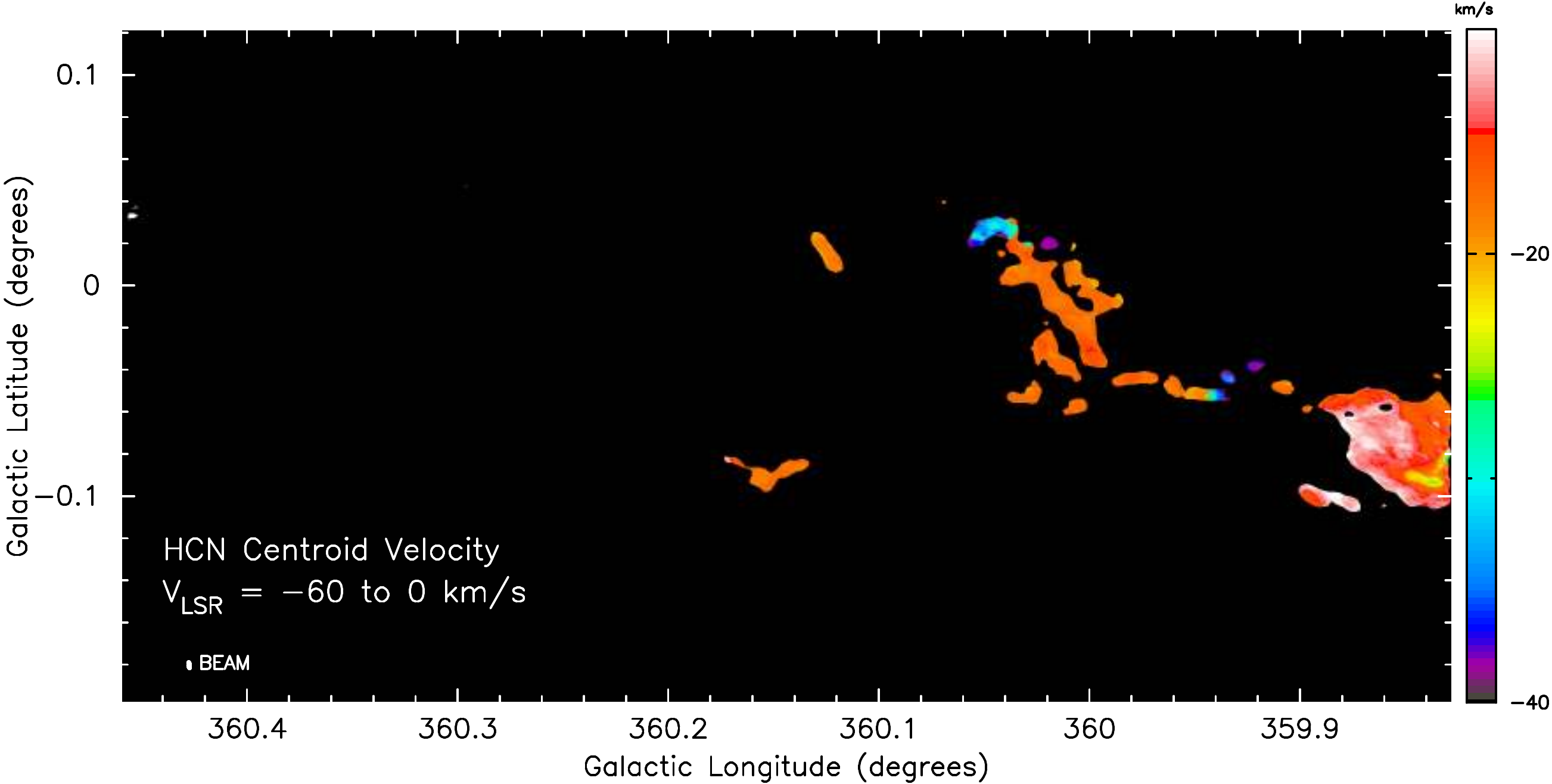} \\
\includegraphics[scale=0.56,angle=-90]{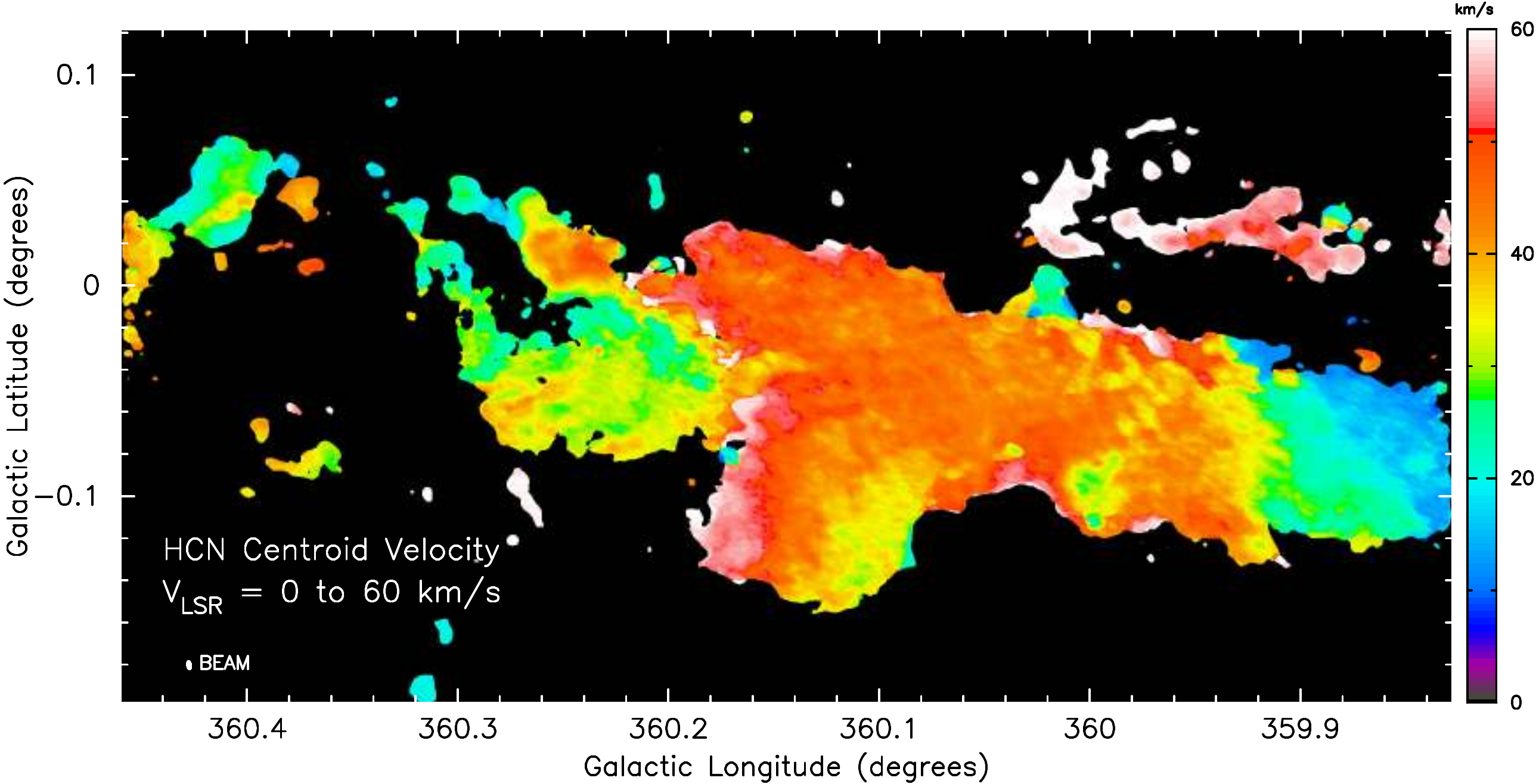} \\
\includegraphics[scale=0.56,angle=-90]{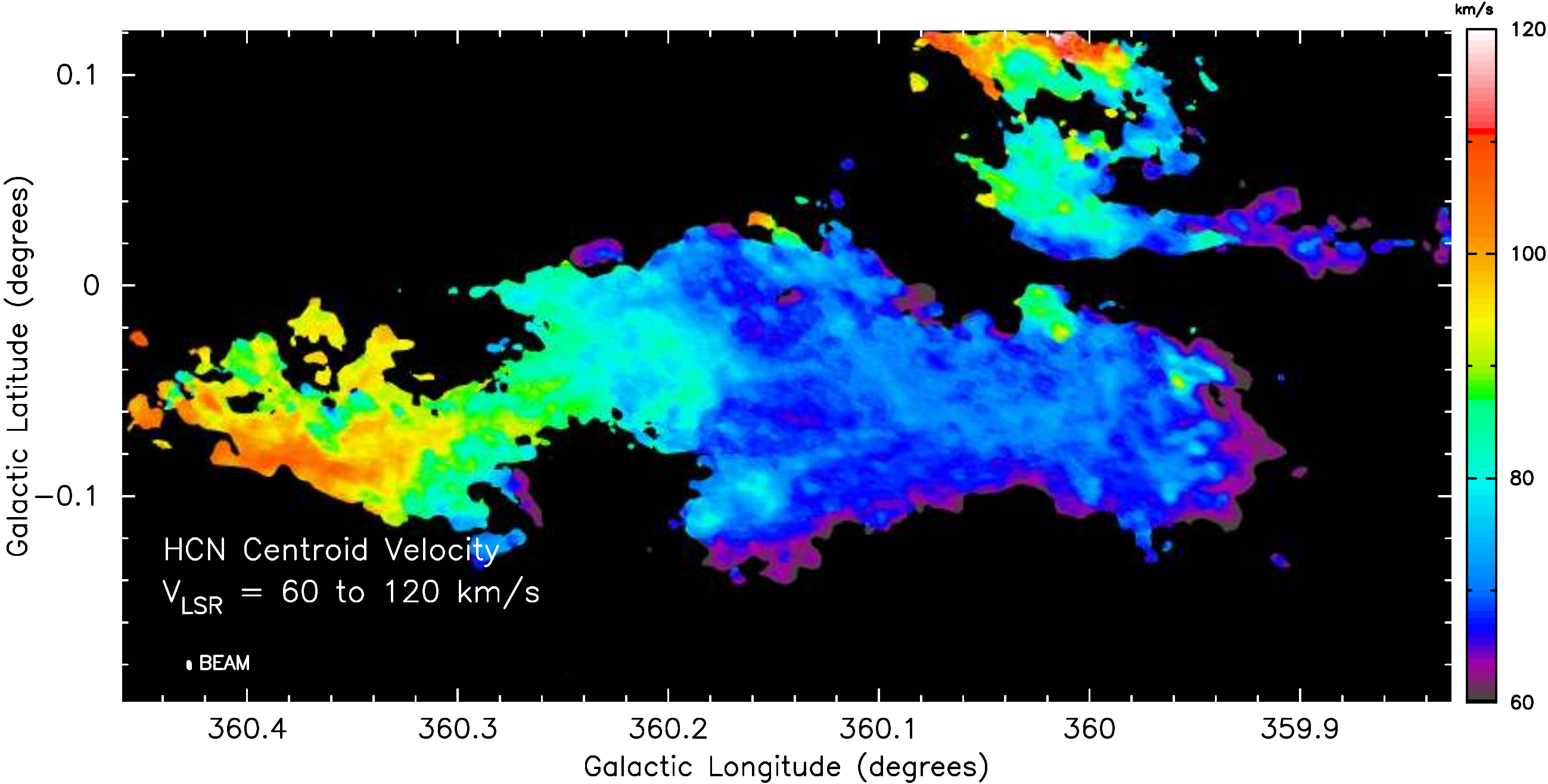} \\
\end{tabular}
\caption{\hcn\ centroid velocity of the CARMA-15 plus Mopra data.
Panels show integrals over selected velocity intervals:
\textit{top)} \vlsr = -60 to 0 \kms; 
\textit{middle)} \vlsr = 0 to 60 \kms; 
\textit{bottom)} \vlsr = 60 to 120 \kms; 
and the color wedge indicates map centroid velocity values.
}
\label{f-HCNcentroid}
\end{figure*}

\begin{figure*}
\begin{tabular}{c}
\includegraphics[scale=0.56,angle=-90]{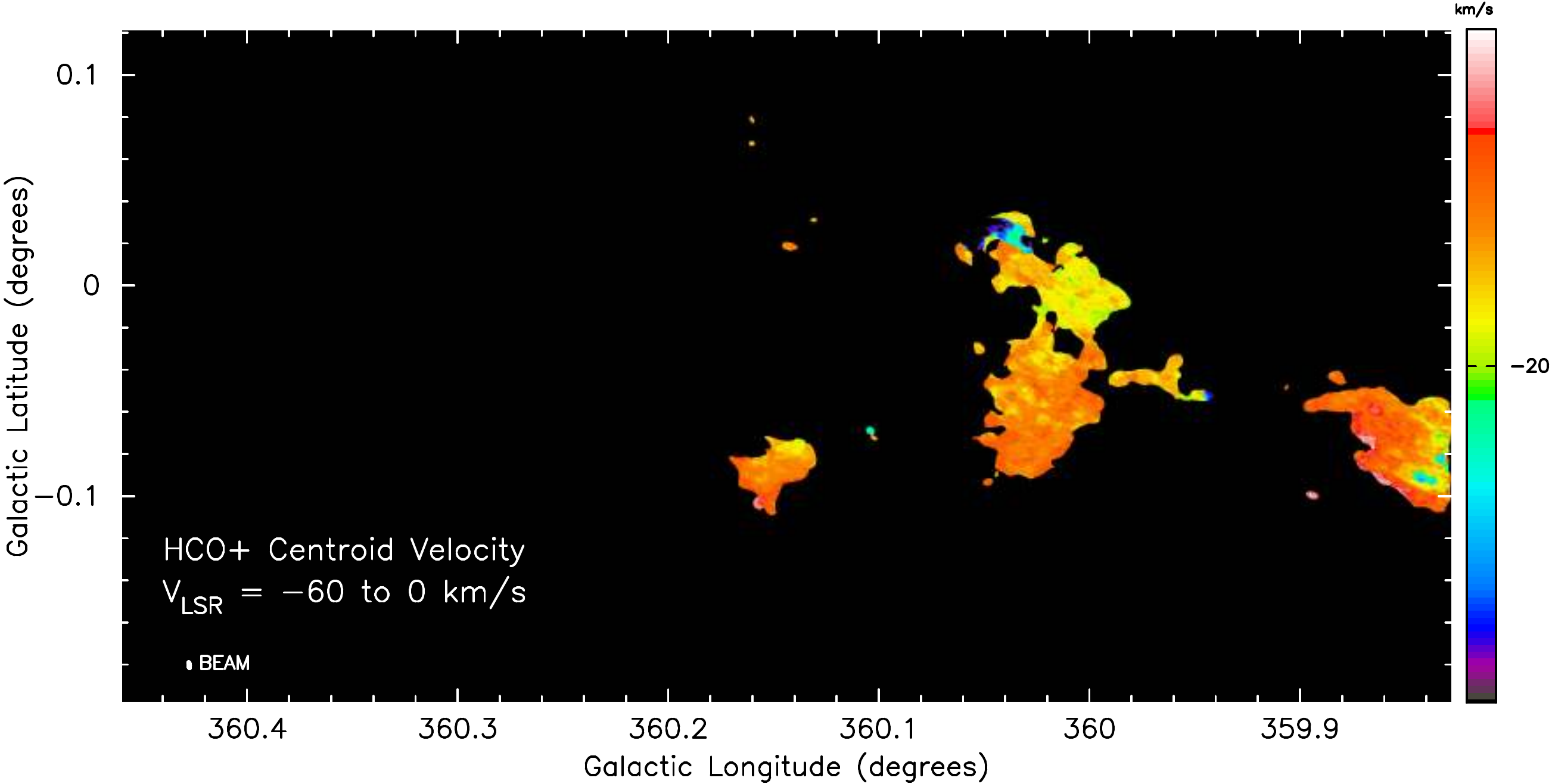} \\
\includegraphics[scale=0.56,angle=-90]{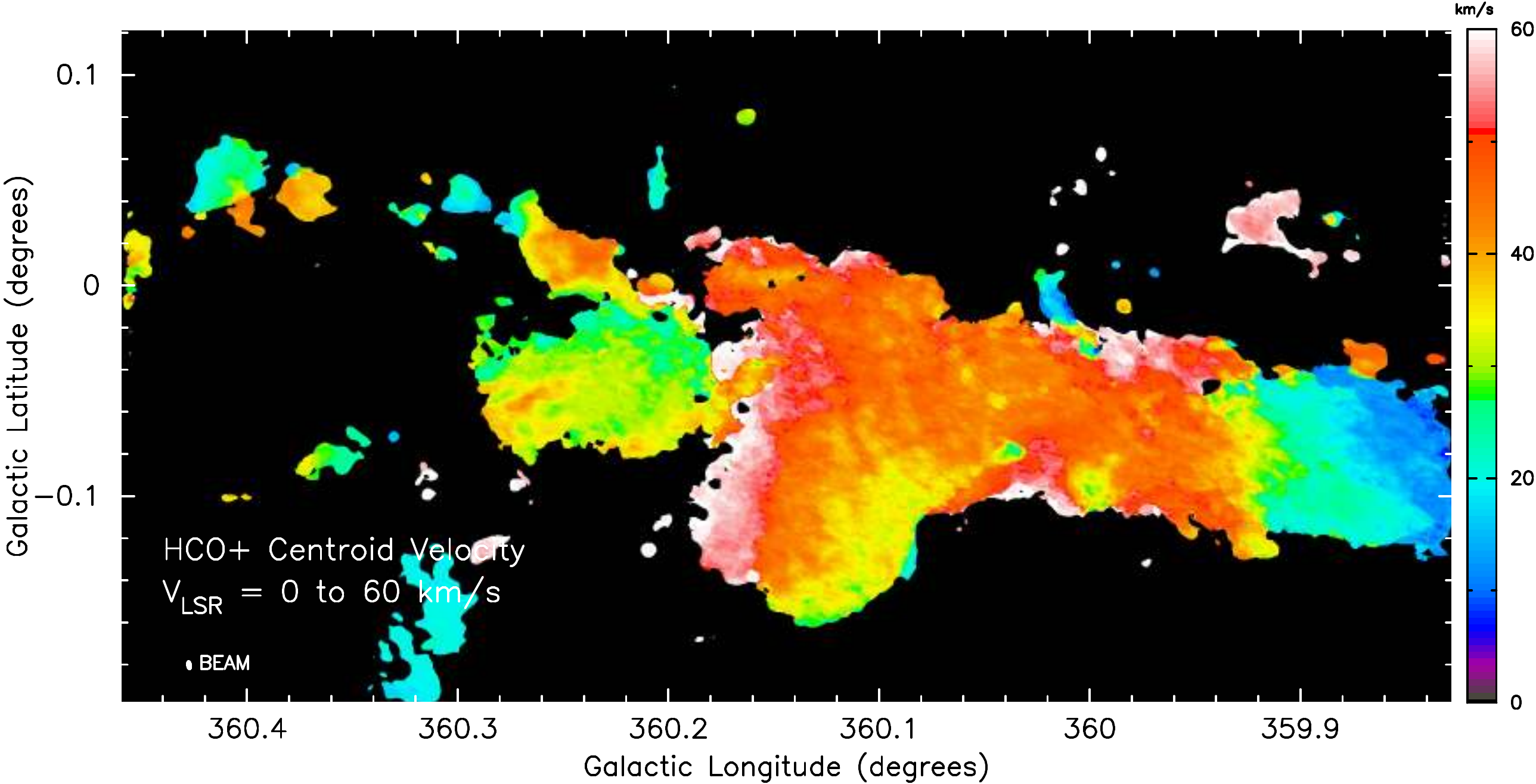} \\
\includegraphics[scale=0.56,angle=-90]{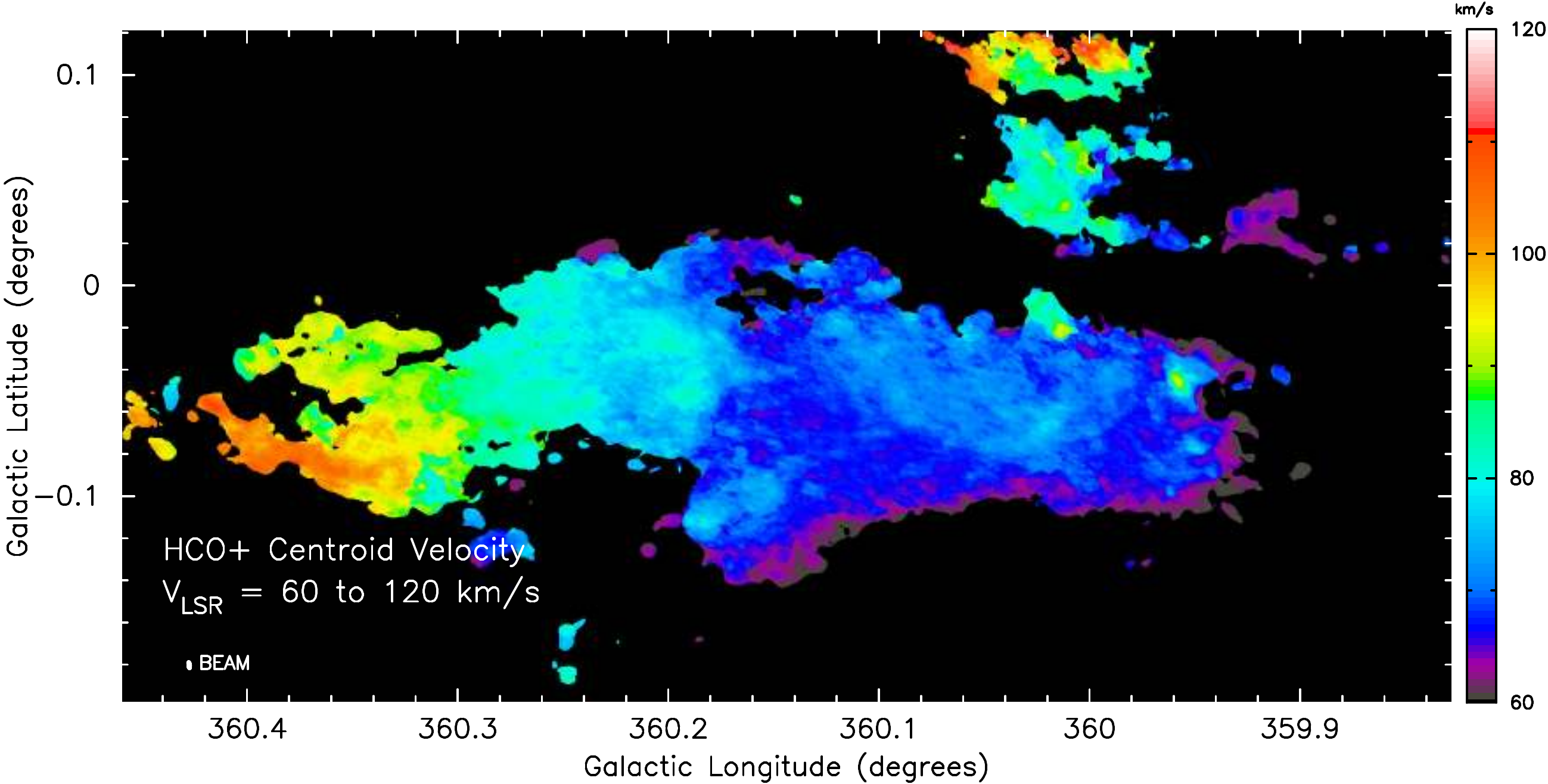} \\
\end{tabular}
\caption{\hcop\ centroid velocity of the CARMA-15 plus Mopra data.
Panels show integrals over selected velocity intervals:
\textit{top)} \vlsr = -60 to 0 \kms; 
\textit{middle)} \vlsr = 0 to 60 \kms; 
\textit{bottom)} \vlsr = 60 to 120 \kms; 
and the color wedge indicates map centroid velocity values;
}
\label{f-HCOpcentroid}
\end{figure*}

\begin{figure*}
\begin{tabular}{c}
\includegraphics[scale=0.56,angle=-90]{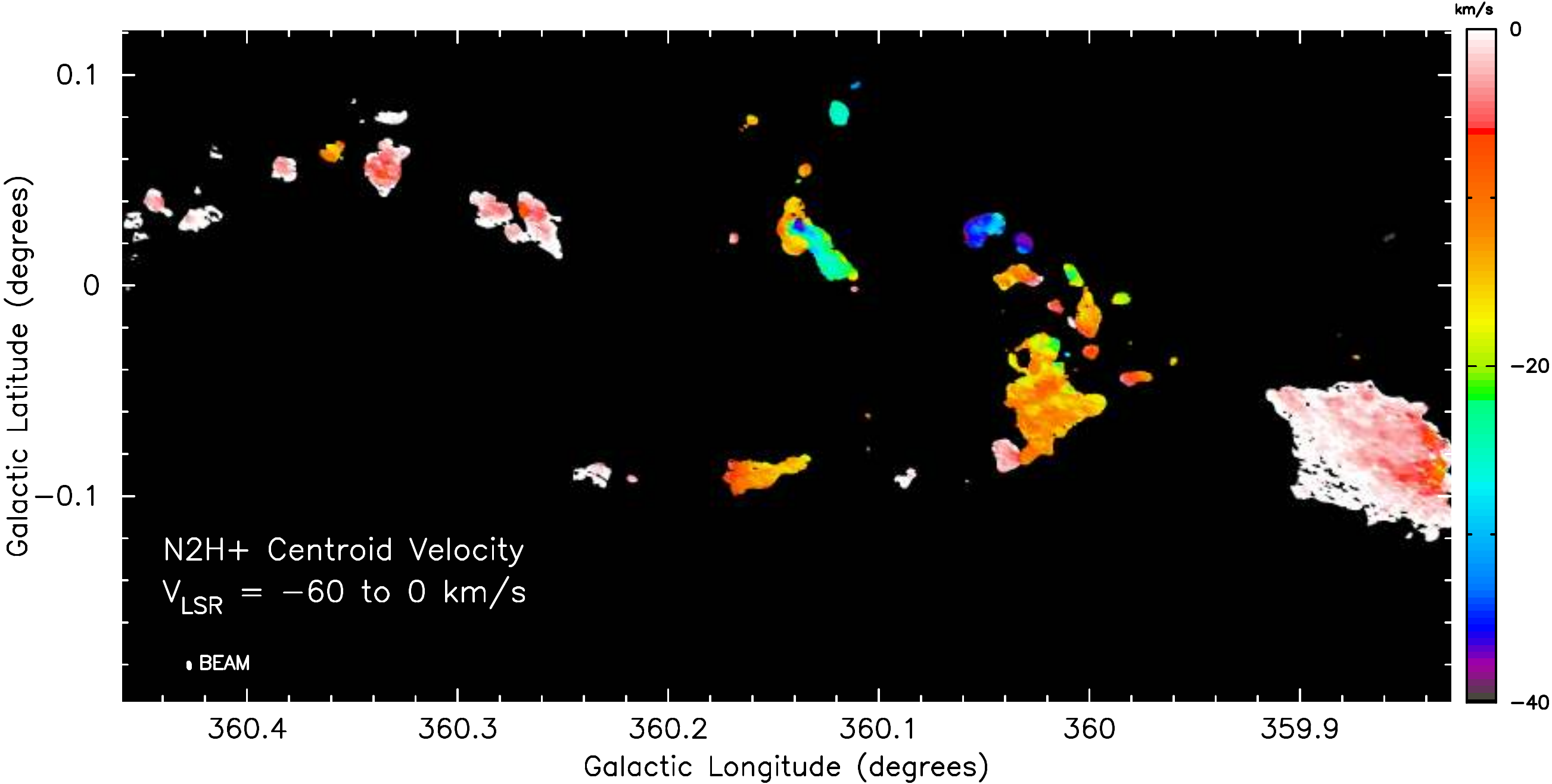} \\
\includegraphics[scale=0.56,angle=-90]{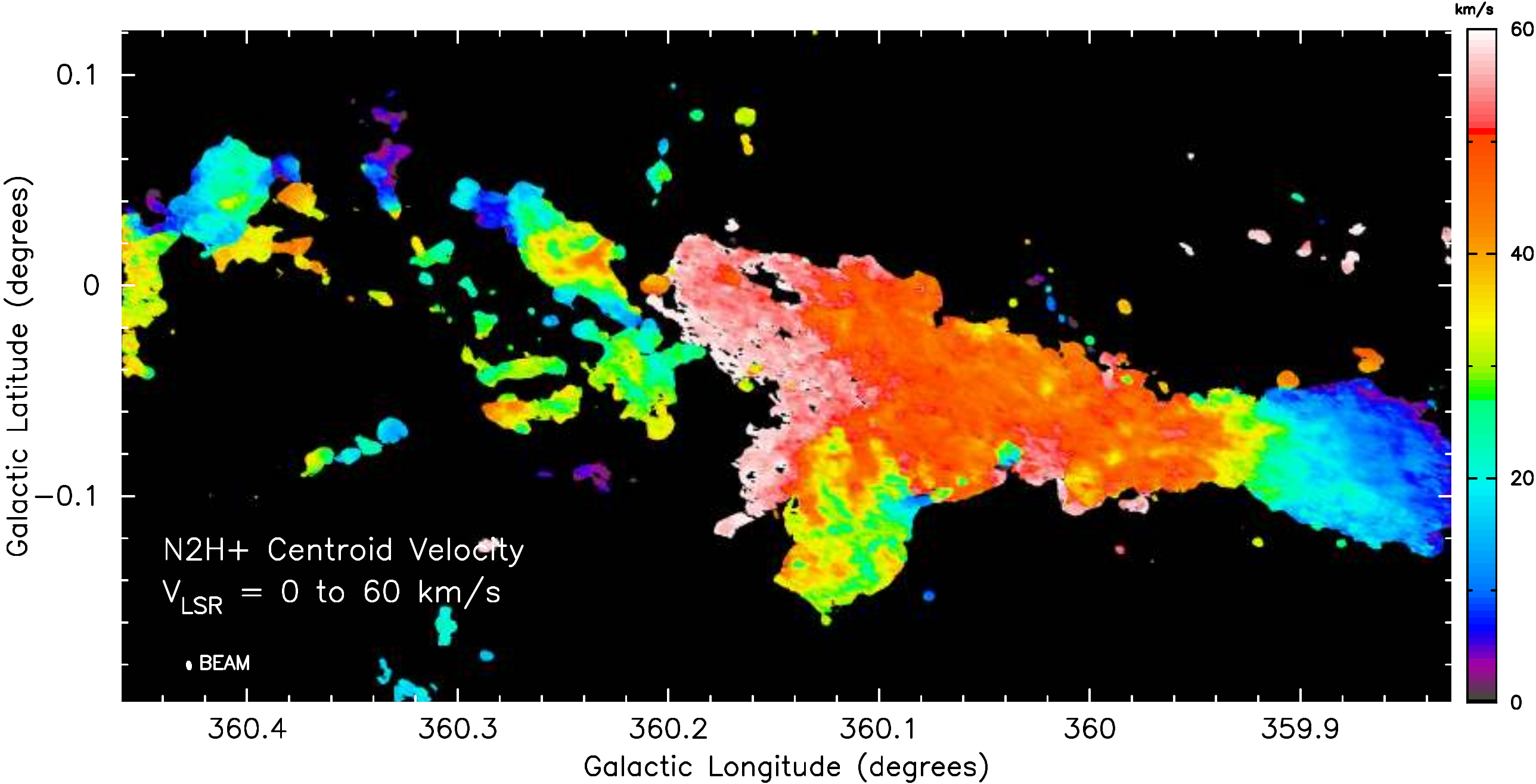} \\
\includegraphics[scale=0.56,angle=-90]{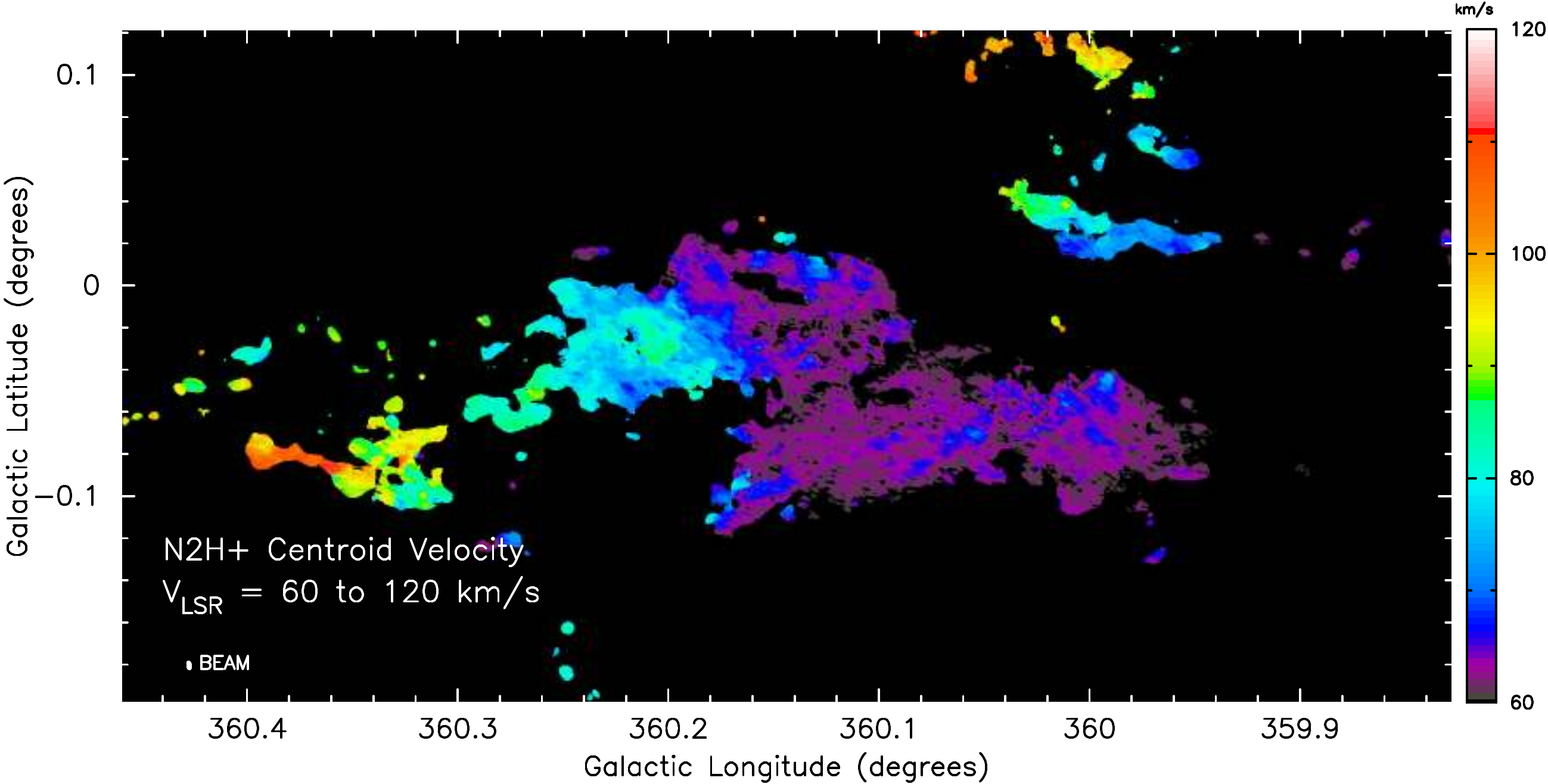} \\
\end{tabular}
\caption{\ntwohp\ centroid velocity of the CARMA-15 plus Mopra data.
Panels show integrals over selected velocity intervals:
\textit{top)} \vlsr = -60 to 0 \kms; 
\textit{middle)} \vlsr = 0 to 60 \kms; 
\textit{bottom)} \vlsr = 60 to 120 \kms;
and the color wedge indicates map centroid velocity values.
}
\label{f-N2Hpcentroid}
\end{figure*}

\begin{figure*}
\begin{tabular}{c}
\includegraphics[scale=0.56,angle=-90]{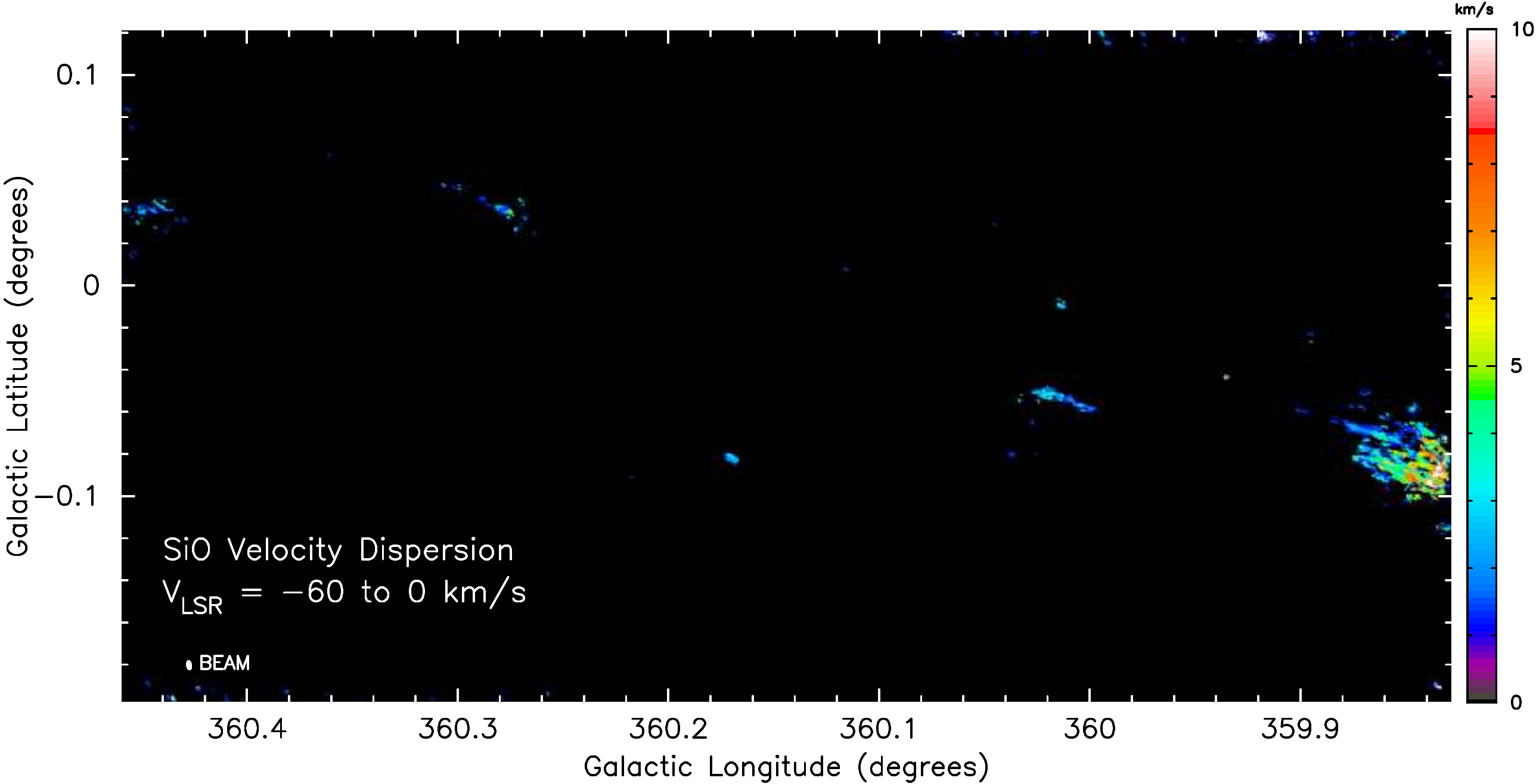}\\
\includegraphics[scale=0.56,angle=-90]{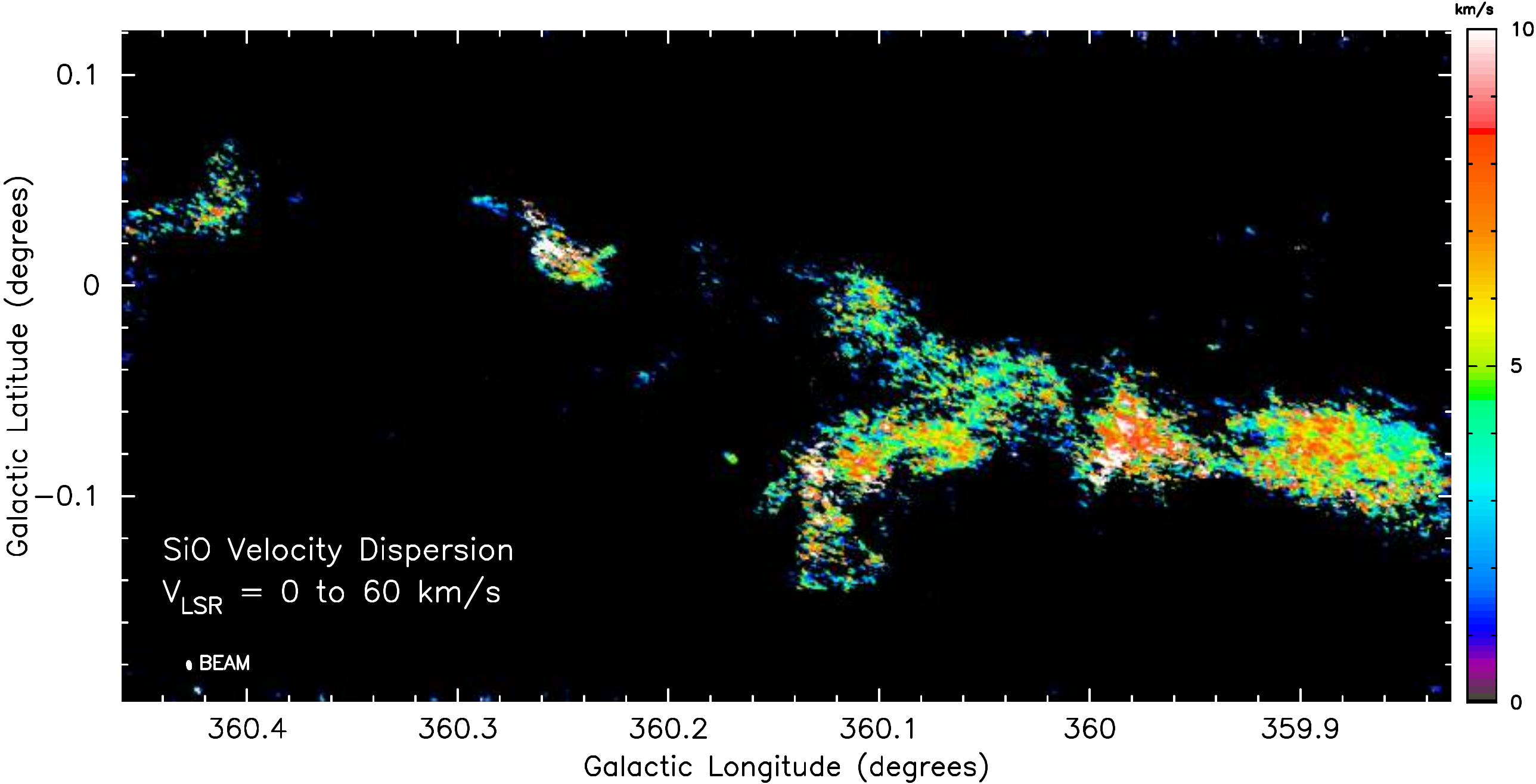} \\
\includegraphics[scale=0.56,angle=-90]{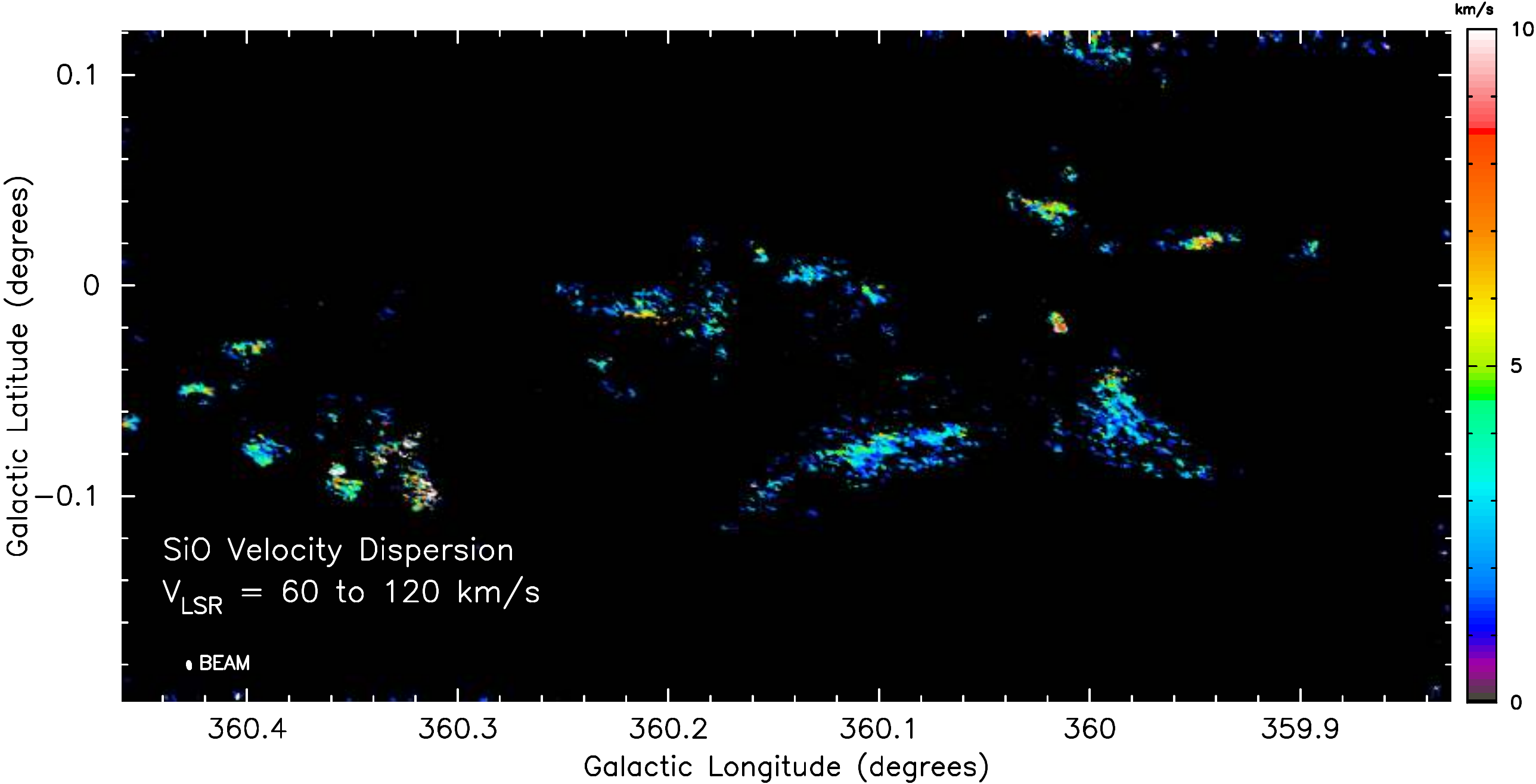}  \\
\end{tabular}
\caption{\sio\ one-dimensional velocity dispersion of the CARMA-15 plus Mopra data.
Panels show integrals over selected velocity intervals:
\textit{top)} \vlsr = -60 to 0 \kms; 
\textit{middle)} \vlsr = 0 to 60 \kms; 
\textit{bottom)} \vlsr = 60 to 120 \kms;
and the color wedge indicates map velocity dispersion values.
}
\label{f-SiOdispersion}
\end{figure*}

\begin{figure*}
\begin{tabular}{c}
\includegraphics[scale=0.56,angle=-90]{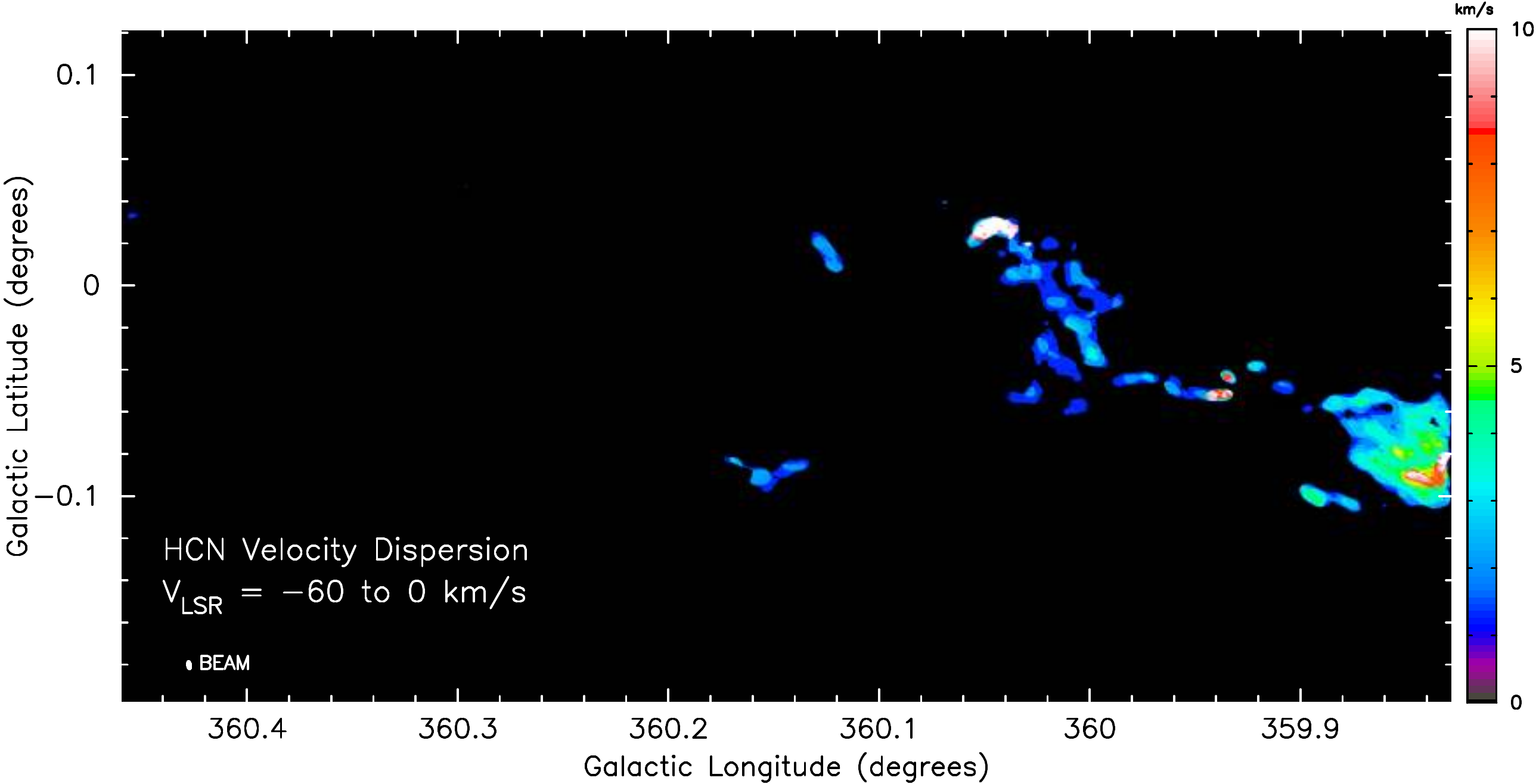} \\
\includegraphics[scale=0.56,angle=-90]{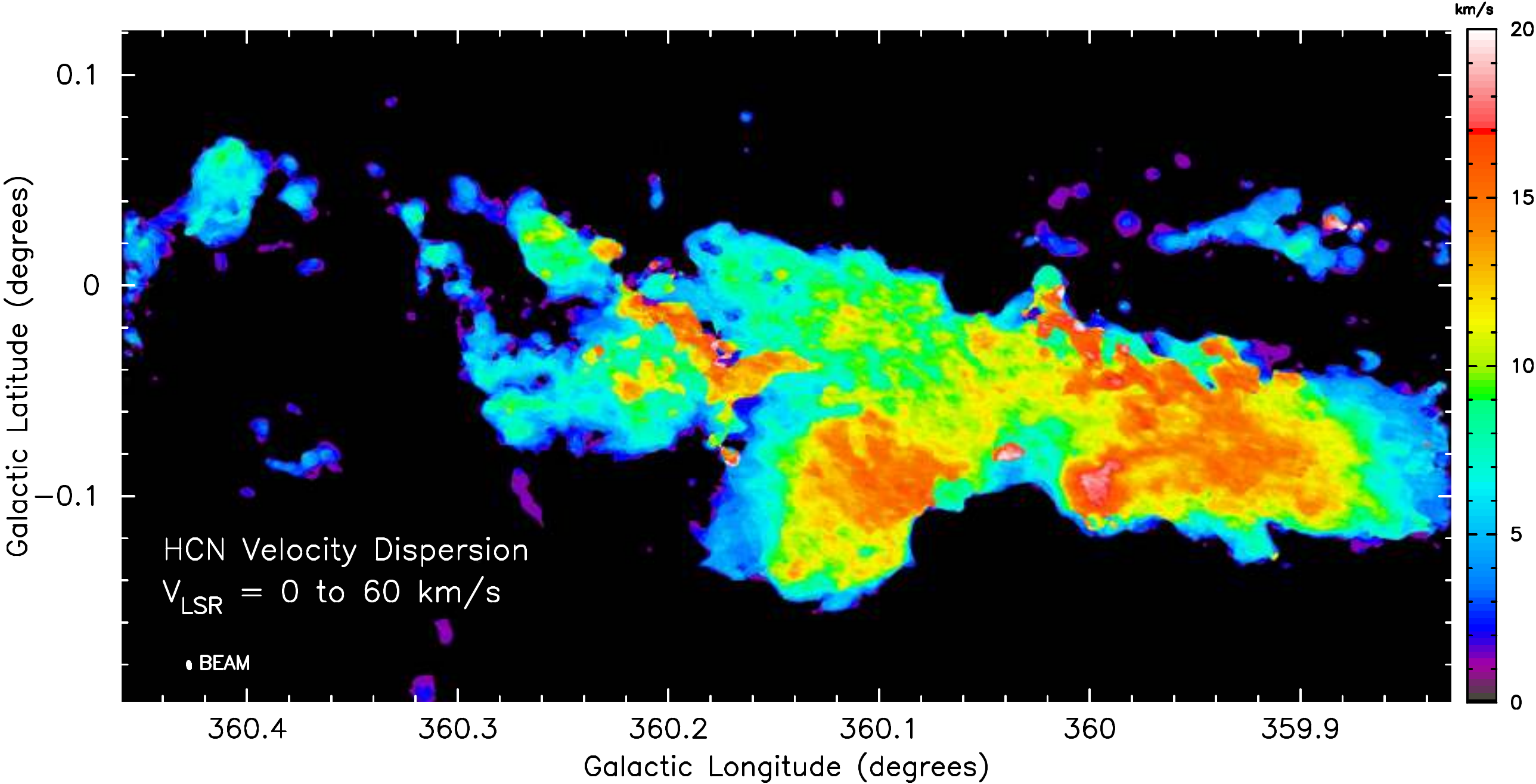} \\
\includegraphics[scale=0.56,angle=-90]{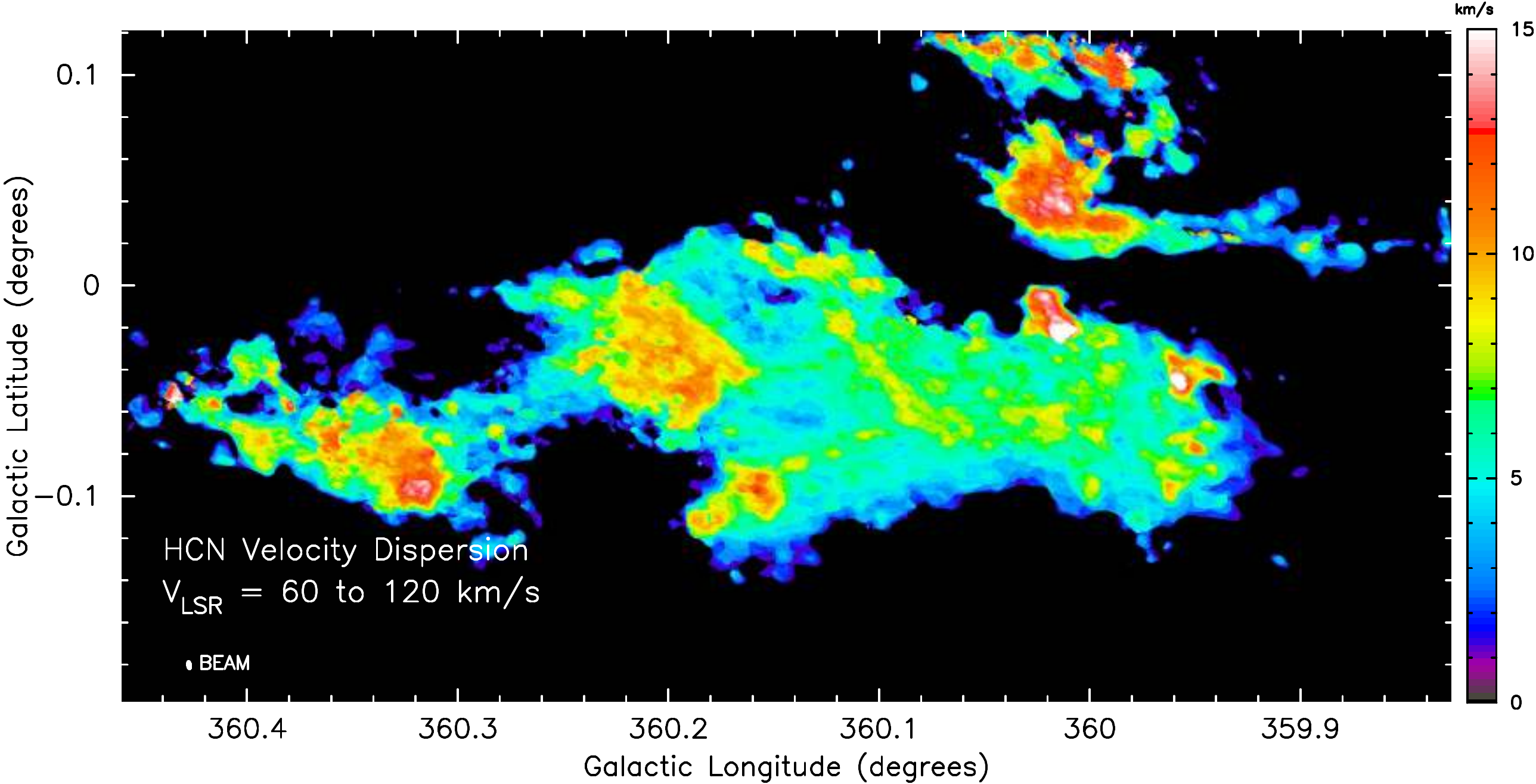} \\
\end{tabular}
\caption{\hcn\ one-dimensional velocity dispersion of the CARMA-15 plus Mopra data.
Panels show integrals over selected velocity intervals:
\textit{top)} \vlsr = -60 to 0 \kms; 
\textit{middle)} \vlsr = 0 to 60 \kms; 
\textit{bottom)} \vlsr = 60 to 120 \kms; 
and the color wedge indicates map velocity dispersion values.
}
\label{f-HCNdispersion}
\end{figure*}

\begin{figure*}
\begin{tabular}{c}
\includegraphics[scale=0.56,angle=-90]{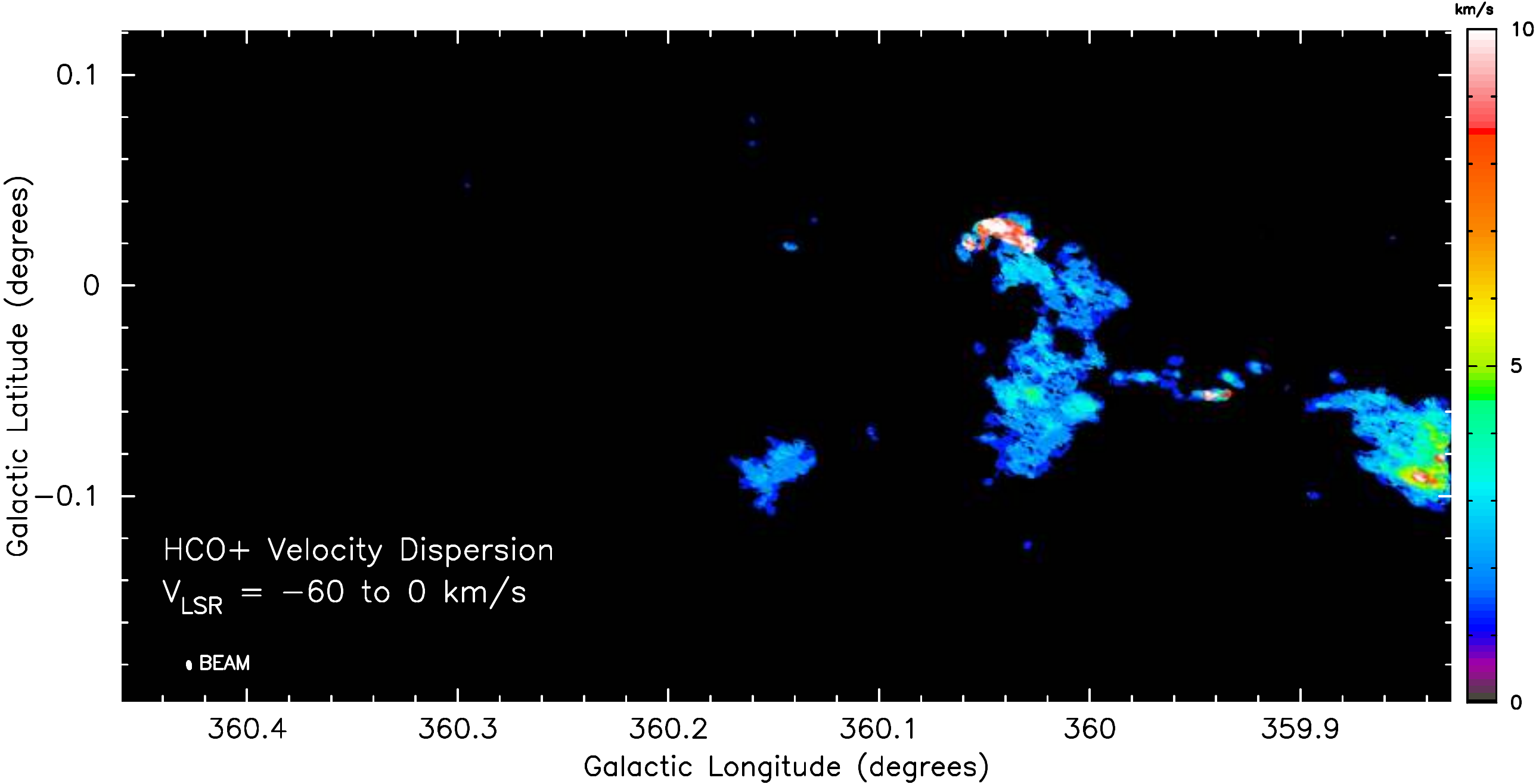} \\
\includegraphics[scale=0.56,angle=-90]{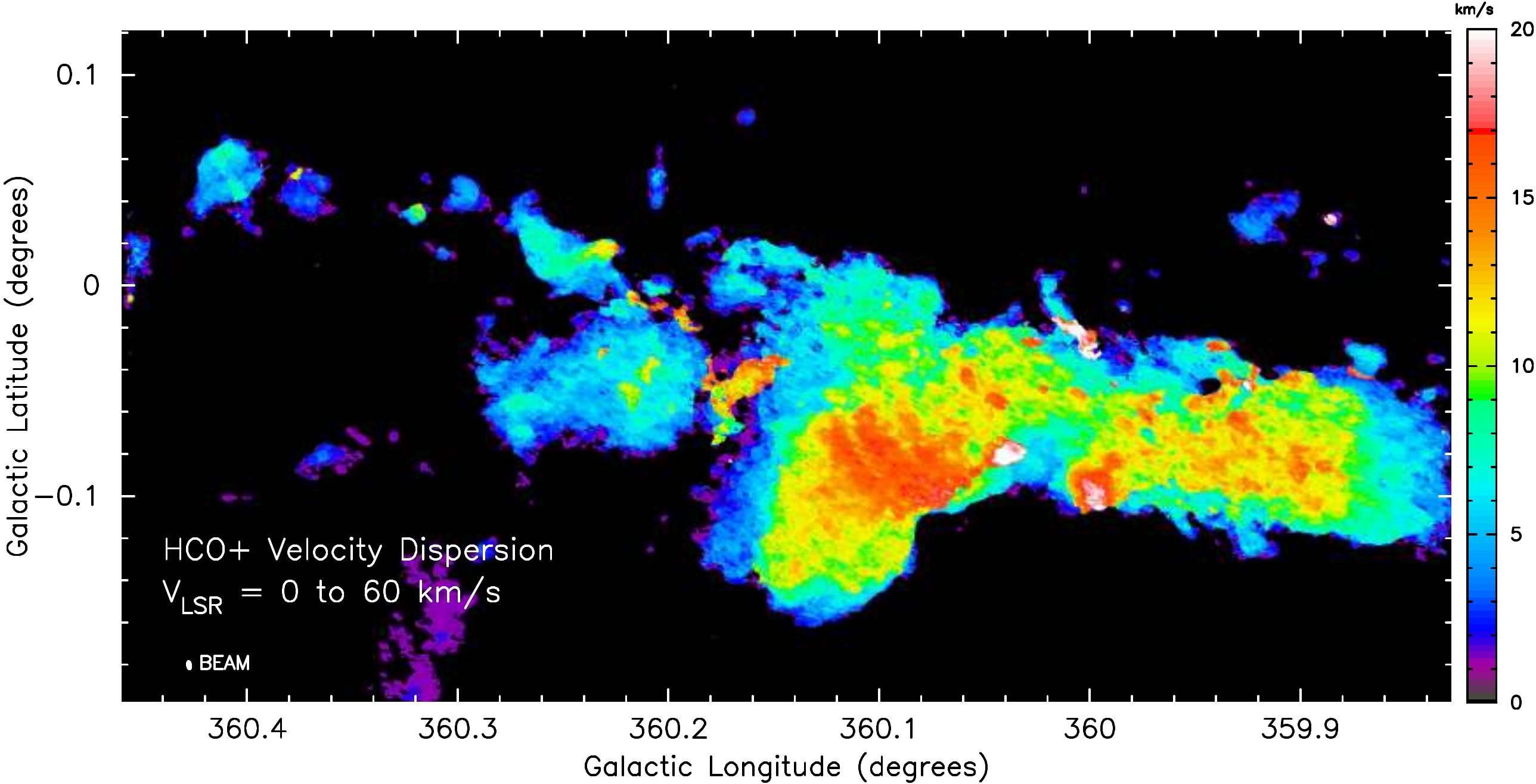}  \\
\includegraphics[scale=0.56,angle=-90]{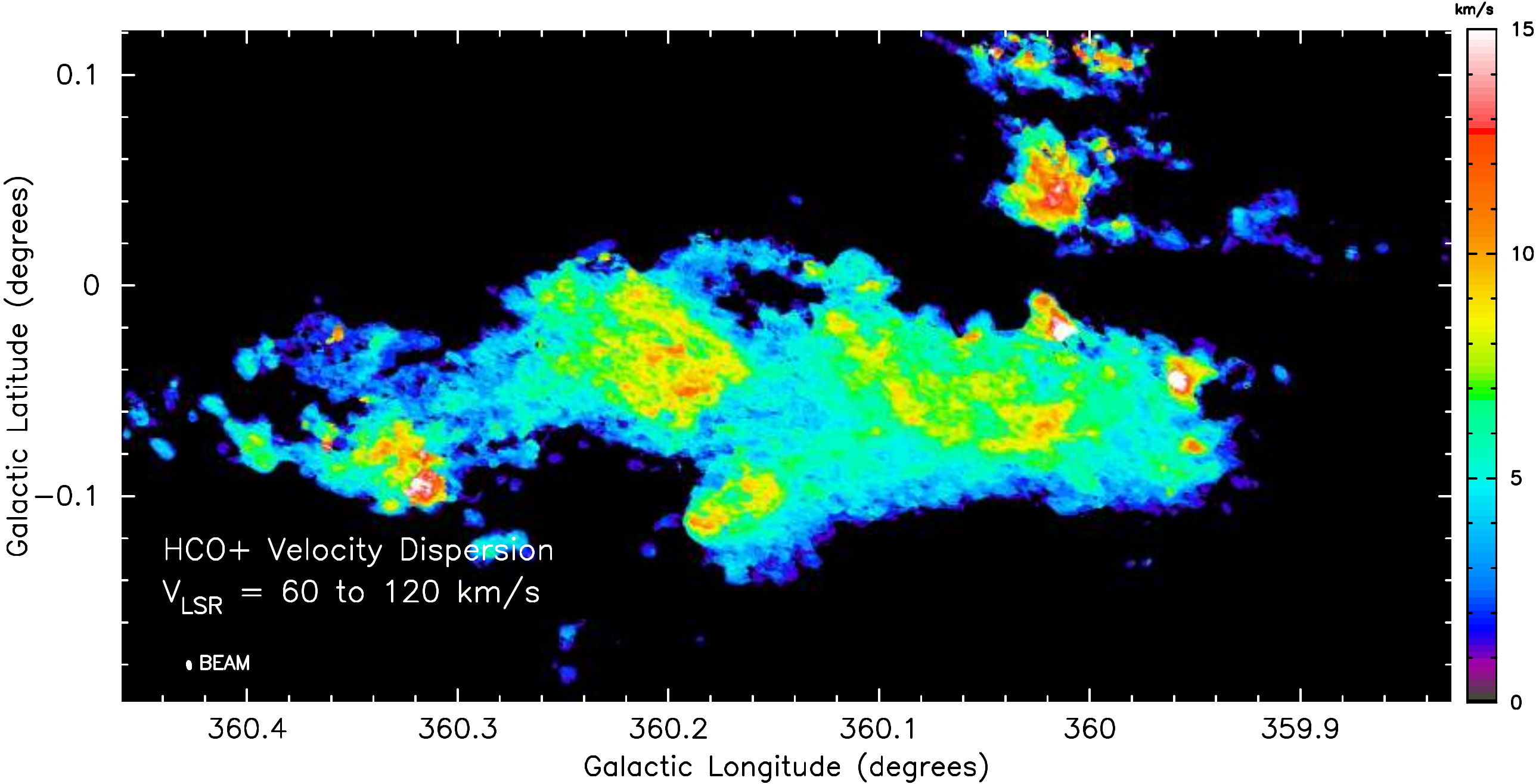} \\
\end{tabular}
\caption{\hcop\ one-dimensional velocity dispersion of the CARMA-15 plus Mopra data.
Panels show integrals over selected velocity intervals:
\textit{top)} \vlsr = -60 to 0 \kms; 
\textit{middle)} \vlsr = 0 to 60 \kms; 
\textit{bottom)} \vlsr = 60 to 120 \kms; 
and color wedge indicates map velocity dispersion values.
}
\label{f-HCOpdispersion}
\end{figure*}

\begin{figure*}
\begin{tabular}{c}
\includegraphics[scale=0.56,angle=-90]{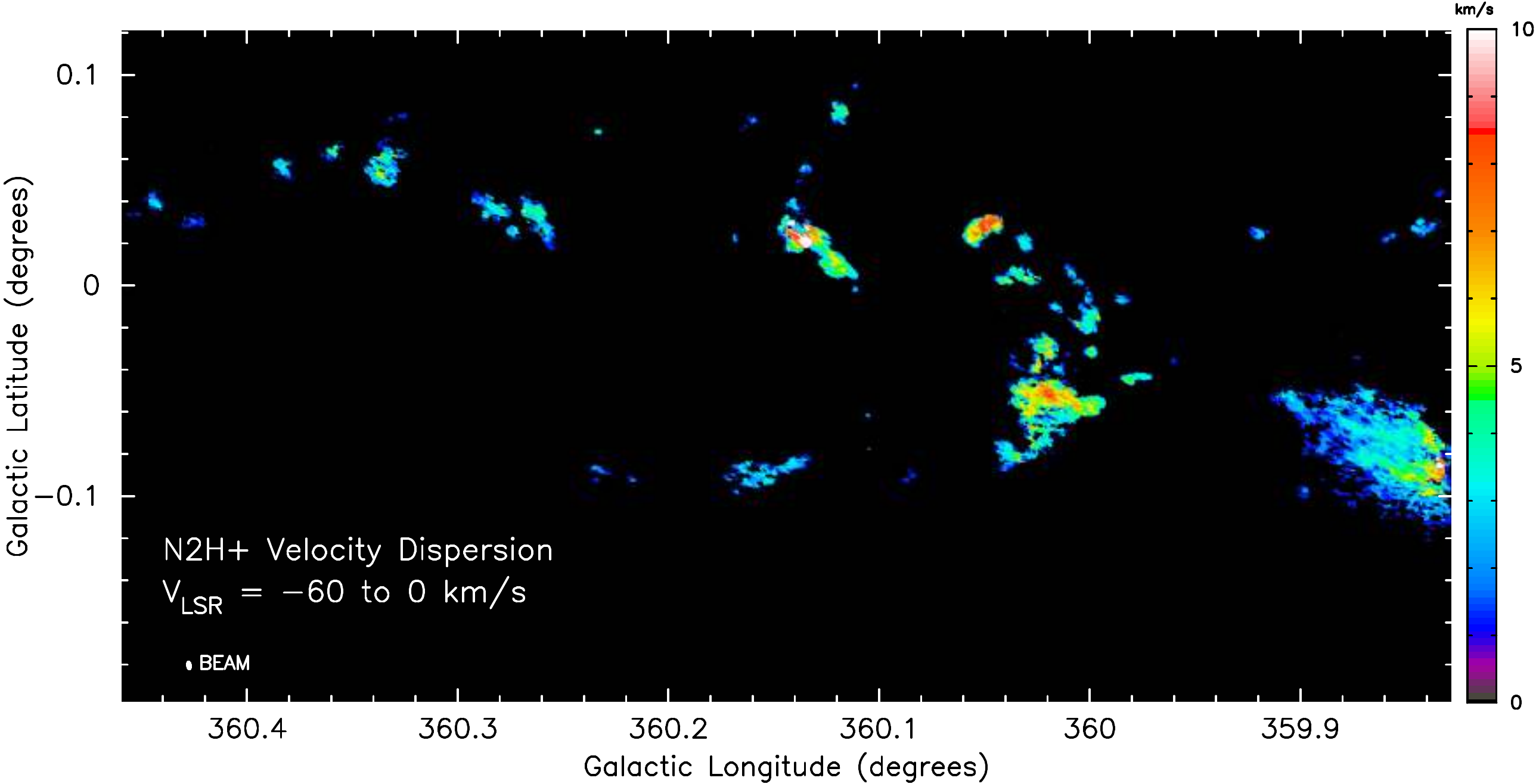} \\
\includegraphics[scale=0.56,angle=-90]{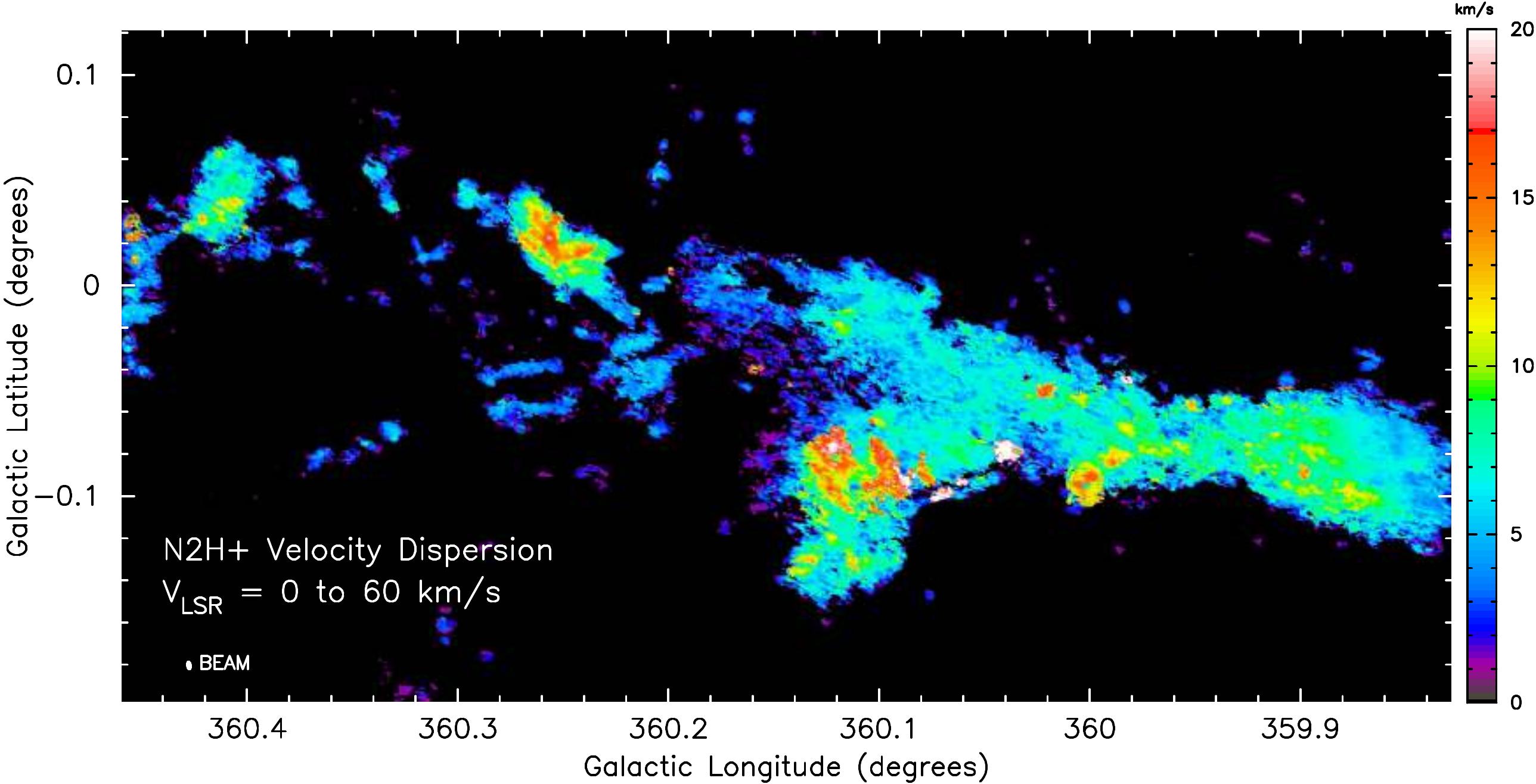} \\
\includegraphics[scale=0.56,angle=-90]{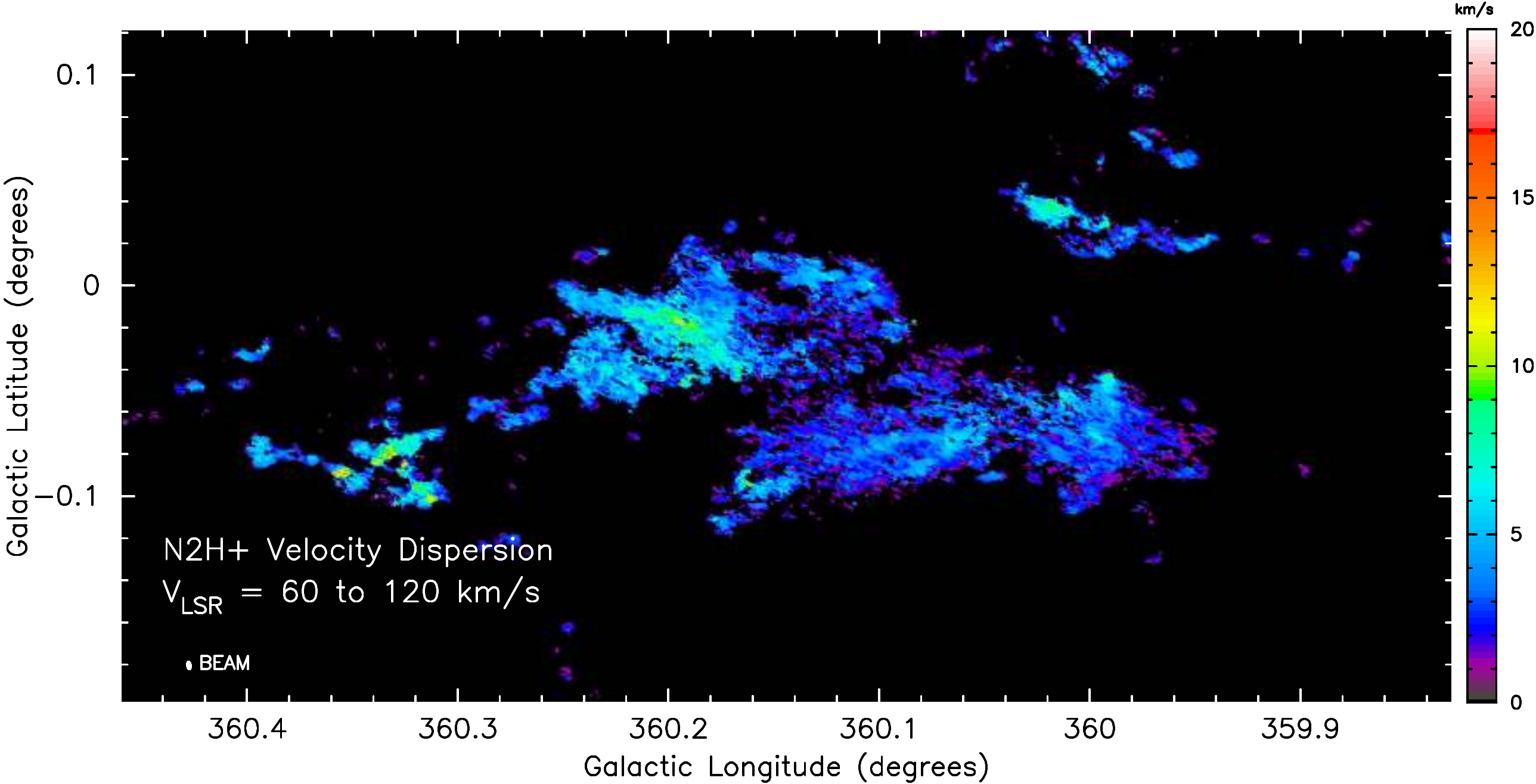} \\
\end{tabular}
\caption{\ntwohp\ one-dimensional velocity dispersion of the CARMA-15 plus Mopra data.
Panels show integrals over selected velocity intervals:
\textit{top)} \vlsr = -60 to 0 \kms; 
\textit{middle)} \vlsr = 0 to 60 \kms; 
\textit{bottom)} \vlsr = 60 to 120 \kms; 
and color wedge indicates map velocity dispersion values.
}
\label{f-N2Hpdispersion}
\end{figure*}

\begin{figure*}
\begin{tabular}{cc}
\includegraphics[trim={1.3cm 1.3cm 1.3cm 1.3cm},clip,scale=0.35,angle=0]{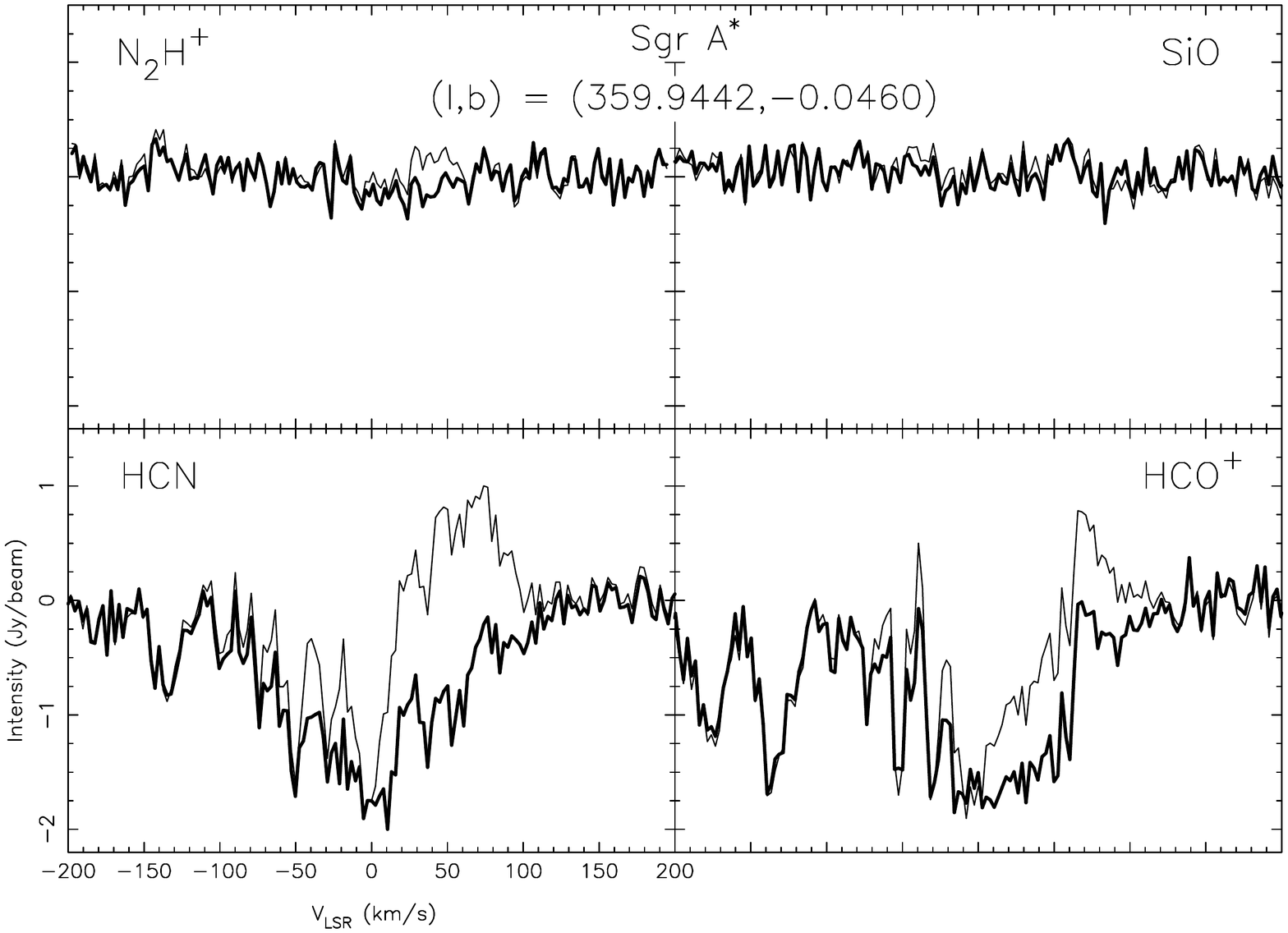} &
\includegraphics[trim={1.3cm 1.3cm 1.3cm 1.3cm},clip,scale=0.35,angle=0]{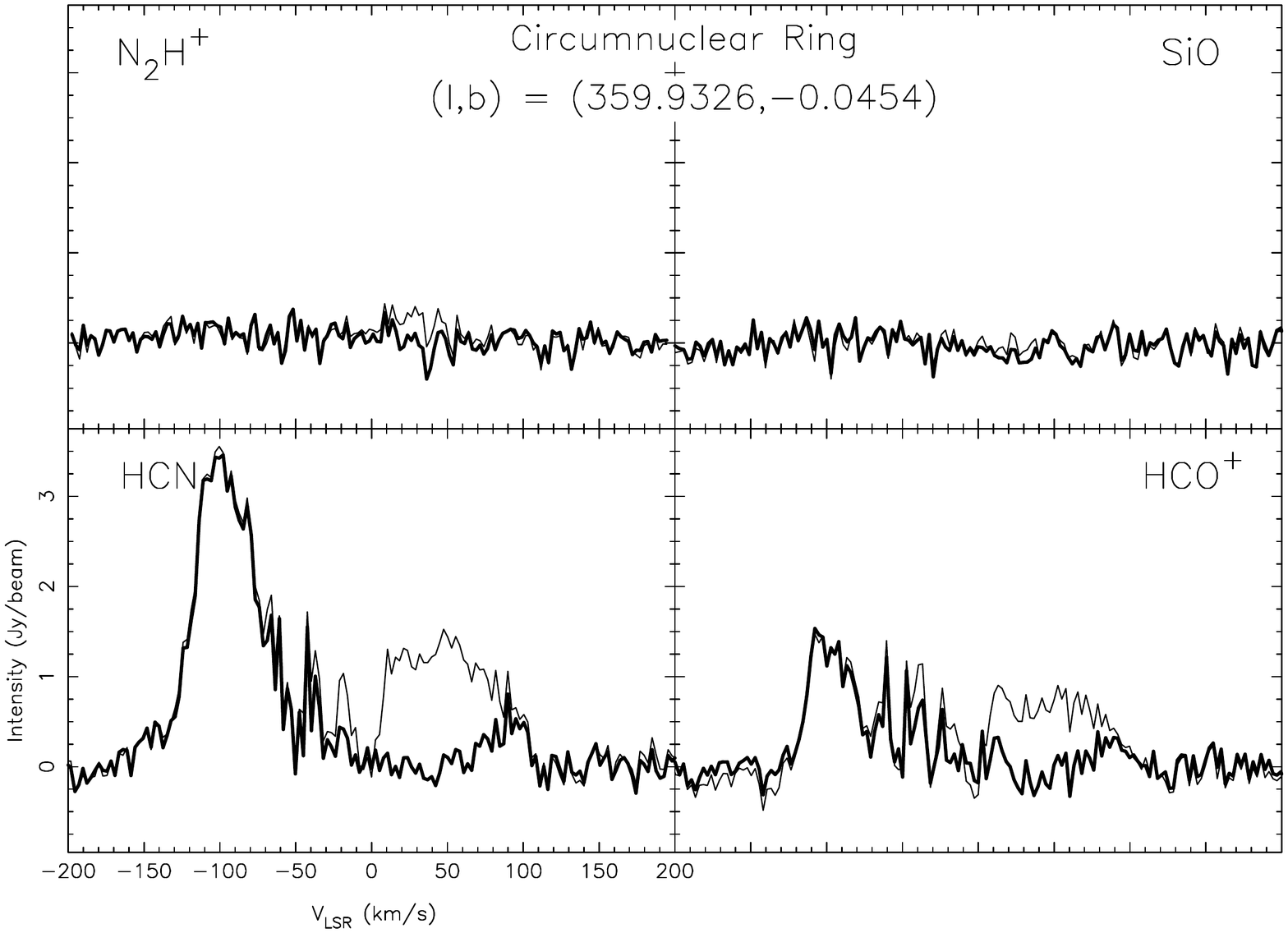} \\
\includegraphics[trim={1.3cm 1.3cm 1.3cm 1.3cm},clip,scale=0.35,angle=0]{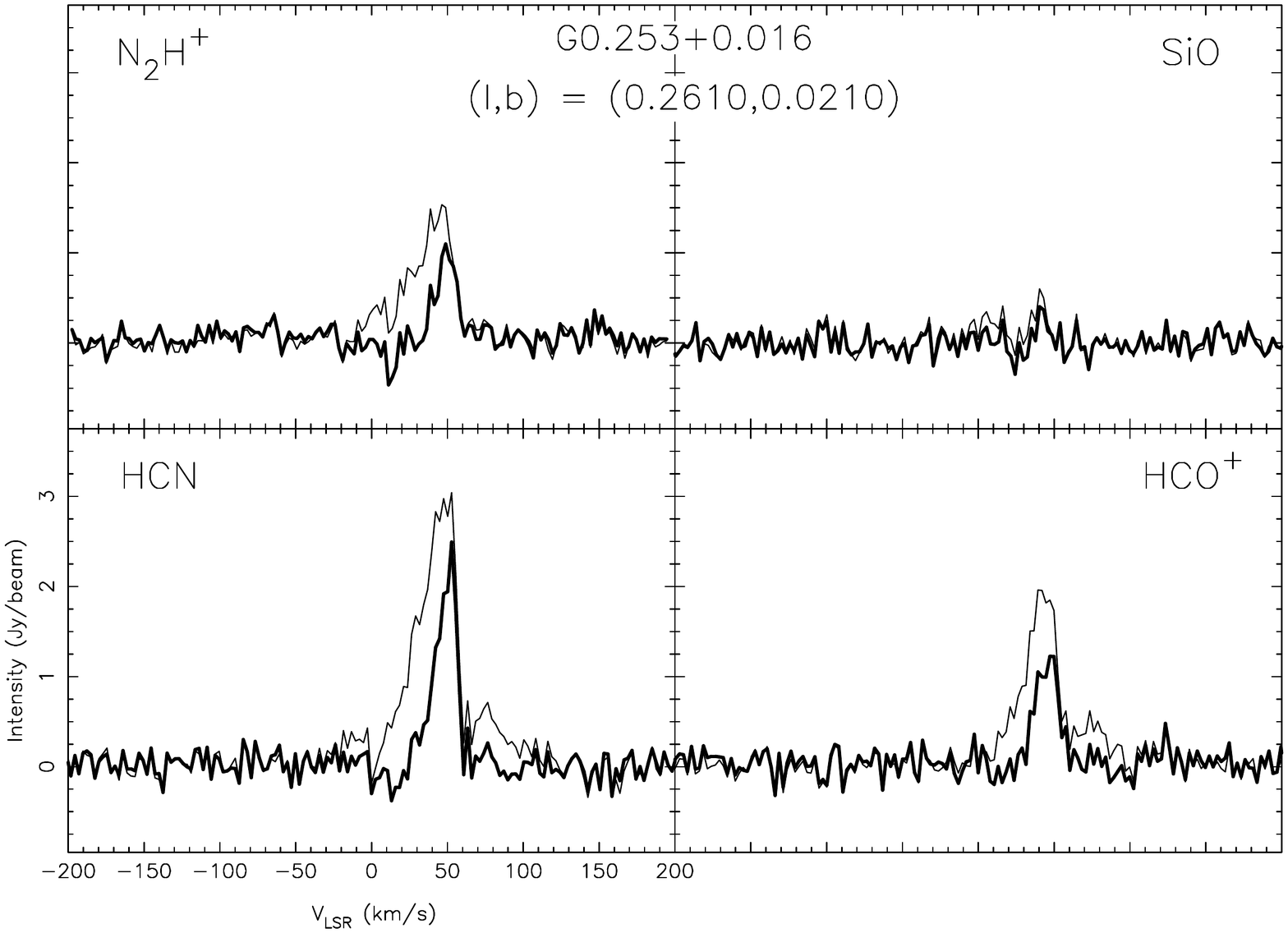} & 
\includegraphics[trim={1.3cm 1.3cm 1.3cm 1.3cm},clip,scale=0.35,angle=0]{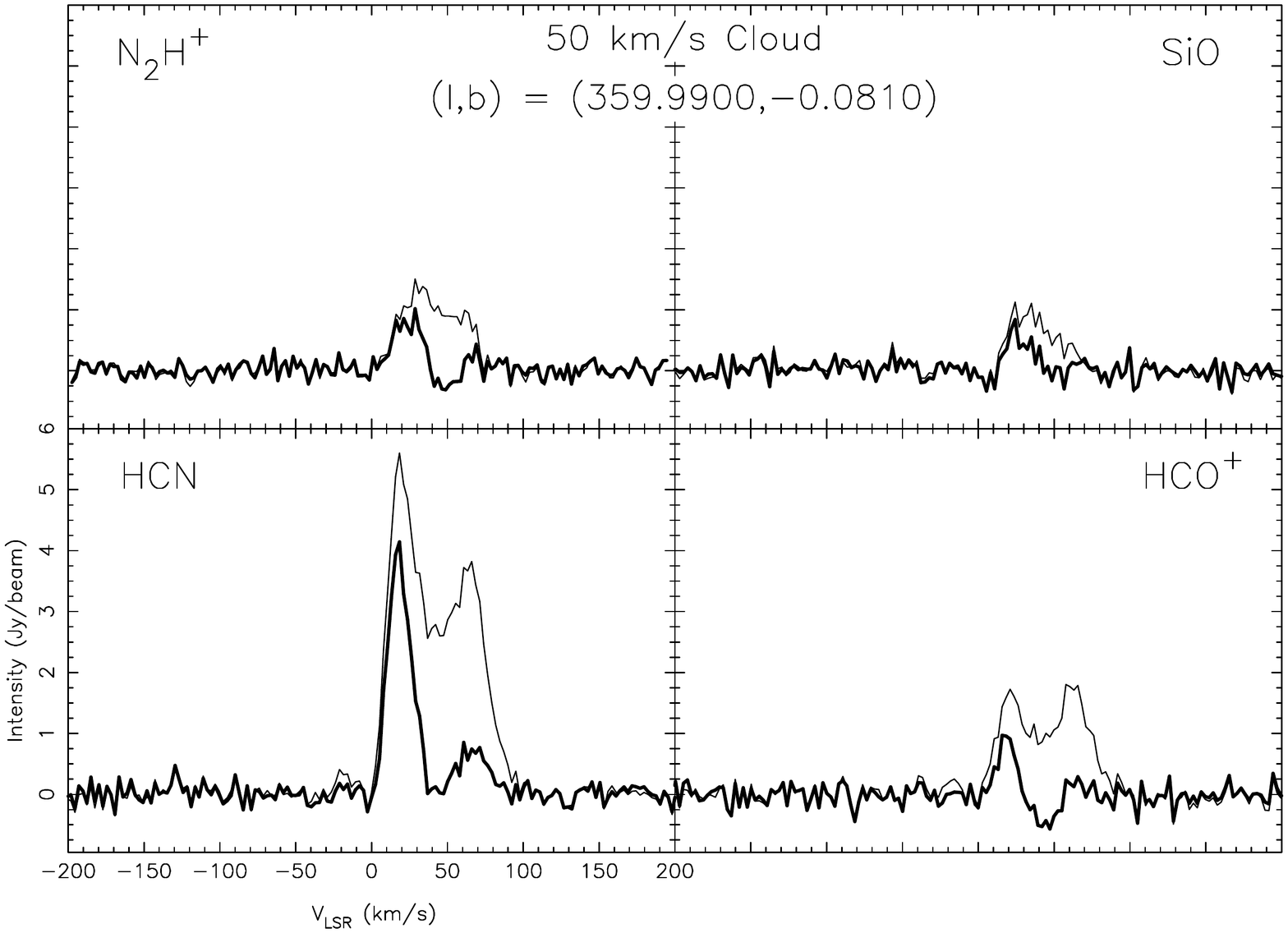} \\
\end{tabular}
\caption{Example spectra from four regions in the CMZ, clockwise from upper left: Sgr A*, the circumnuclear ring (centered on position I in fig. 8 of \protect{\citet{2001ApJ...551..254W}}), the 50 \kms\ cloud, and G0.253+0.016.  Spectra are averages of 11\arcsec $\times$ 11\arcsec\ boxes centered on the Galactic coordinates shown in each panel.Bold lines are CARMA-15 interferometric data, while thin lines are CARMA-15 plus Mopra data.
}
\label{f-examplespec}
\end{figure*}

Figs. \ref{f-SiOcentroid} to \ref{f-N2Hpcentroid} display the mean
velocity (first moment) in each of the spectral lines and Figs.
\ref{f-SiOdispersion} to \ref{f-N2Hpdispersion} display the one-dimensional
velocity dispersion (second moment).  We formed these moment maps using
the ``smooth-and-mask'' technique described by \cite{2003ApJS..145..259H}. 
This
technique can reveal low-level emission 
better than can be accomplished by simply calculating the moments using a fixed clip level,
and furthermore does not bias the noise statistics of the final image.
The purpose of this technique is to identify which pixels in a data cube should be included in the moment
calculations and which should not. This is done by spatially smoothing the data cube 
to bring out low S/N emission, then using that smoothed version of the data cube
to determine which pixels should be included in the moments (i.e., included in the mask).
For each of our spectral line data cubes, a mask was made by spatially smoothing the data with a Gaussian with FWHM =
20\arcsec\ and then including in the mask the pixels where the intensity
in the smoothed data was greater than 3 times the rms noise determined from
emission-free channels. Pixels that were
isolated on the velocity axis were eliminated from the mask to remove anomalous spikes outside the
velocity range of the main emission, which can skew the integrated
quantities.  The moments of the {\it unsmoothed} data were then computed using 
only pixels that fell within the mask.


In computing the moment maps, we have not attempted to deconvolve the three
hyperfine components of \hcnA\, which span $\sim$3.5~MHz or about 4.5
spectral channels, nor the seven hyperfine components of \ntwohpA, which
span $\sim$4.7~MHz or about 6 spectral channels.  The large line widths of
gas in the CMZ obscure much of the hyperfine structure and the similarity
between the HCN and \hcopA\ moments suggests that the contamination
to these quantities from the hyperfine lines is not significant.  We note that 
G0.253+0.016 stands out as having high velocity dispersion ($\sim 15 \kms$)
in \ntwohpA; this may be due to the multiple velocity components noted
by \citet{2013ApJ...765L..35K}.


Example spectra, both with and without the Mopra data added, are shown in
Fig. \ref{f-examplespec}. In compact regions, the CARMA-15 data capture
the entirety of the emission of the feature with the additional Mopra data
providing only the large scale foreground or background components.  For
instance, the spectra of the circumnuclear ring (Fig. \ref{f-examplespec},
upper right) are centered on component I in fig. 8 of \cite{2001ApJ...551..254W}.
This portion of the ring emits between $\vlsr=-150$ to $-50 \kms$ which
the CARMA data (bold line) fully recover.  However, at positive velocities
there is significant HCN and HCO$^+$ emission, likely from the 50 \kms\
cloud, that is mostly seen in the Mopra data (thin line).  This emission
is not detected by the interferometer because it spatially smooth on the
scale of $\sim 30\arcsec$.  In the spectra at the position of Sgr A*
(Fig. \ref{f-examplespec}, upper left), emission from the 50 \kms\
cloud is seen to ``fill in'' the HCN and \hcopA\ absorption features,
and is weakly visible in \ntwohpA.

In larger regions, such as G0.253+0.016 (Fig. \ref{f-examplespec},
lower left) and the 50 \kms\ cloud (Fig. \ref{f-examplespec}, lower
right) , the Mopra data increase the line width and peak intensity, or
add velocity components.  Towards G0.253+0.016, the spectrum is located on
a bright core near the center of cloud. Adding the Mopra data detects
more of the blueshifted, low emissivity, spatially smoother gas, and
enhances the \hcnA\ and \hcopA\ feature associated with a line-of-sight
cloud at 75~\kms\ \citep{2014ApJ...786..140R,2015ApJ...802..125R}.  The 50 \kms\
cloud is similarly a complex region with emission at several spatial
scales and velocities.  These spectra show the importance of recovering
all spatial scales when analyzing such regions.

\subsection{Correlations with 6.4 keV Iron Fluorescence}

Remarkably, the CMZ molecular clouds emit fluorescent \feka\ line emission
at 6.4 keV \citep{1996PASJ...48..249K}. One of our key motivations for mapping
the CMZ in different molecular species is to examine the relationship
between the distribution of the molecular line emission, the 6.4 keV line
emission, and nonthermal radio emission.  Broadly, there are two proposed
production mechanisms for the 6.4 keV \feka\ emission:
the irradiation of 
molecular cloud by X-rays \cite{2010ApJ...714..732P} or the impact of 100 MeV cosmic ray electrons 
with molecular gas  \citep{2002ApJ...568L.121Y}.
In the X-ray model,
a molecular cloud is subjected to strong external X-ray radiation which
is absorbed and reemited in flourescent iron lines.  
This model requires past X-ray flares from Sgr A* with $L_x \sim 10^{39}~{\rm erg~s^{-1}}$
\citep{1996PASJ...48..249K,2000ApJ...534..283M,2001ApJ...558..687M,2010ApJ...714..732P}.  
A principal piece of
evidence for this view is the observed short-term variability ($\sim 10$
years) of X-ray bright spots \citep{2010ApJ...714..732P,2014IAUS..303...94S}.

The cosmic ray picture requires metallicity  which is 2-3 times
solar. However, it is consistent with  enhanced cosmic ray density
and a high cosmic ray ionization rate, as traced in radio, infrared  and
$\gamma$-ray observations of the Galactic Center \citep{2002ApJ...568L.121Y}.
The electron population from diffuse synchrotron emitting relativistic particles 
 interacting with
molecular gas can heat and ionize molecular gas. In addition, \feka\ line
emission from neutral iron atoms can be produced by low-energy ($<100$~MeV)
cosmic ray bombardment of neutral gas. A three-way spatial correlation
between the distributions of molecular gas, nonthermal radio continuum as
the source of cosmic rays, and the 6.4 keV line emission is expected in the
context of the cosmic ray picture.  Our data reveal just such a correlation.

\begin{figure*}
\includegraphics[scale=0.7,angle=-90,trim={0cm 2cm 0cm 2cm}]{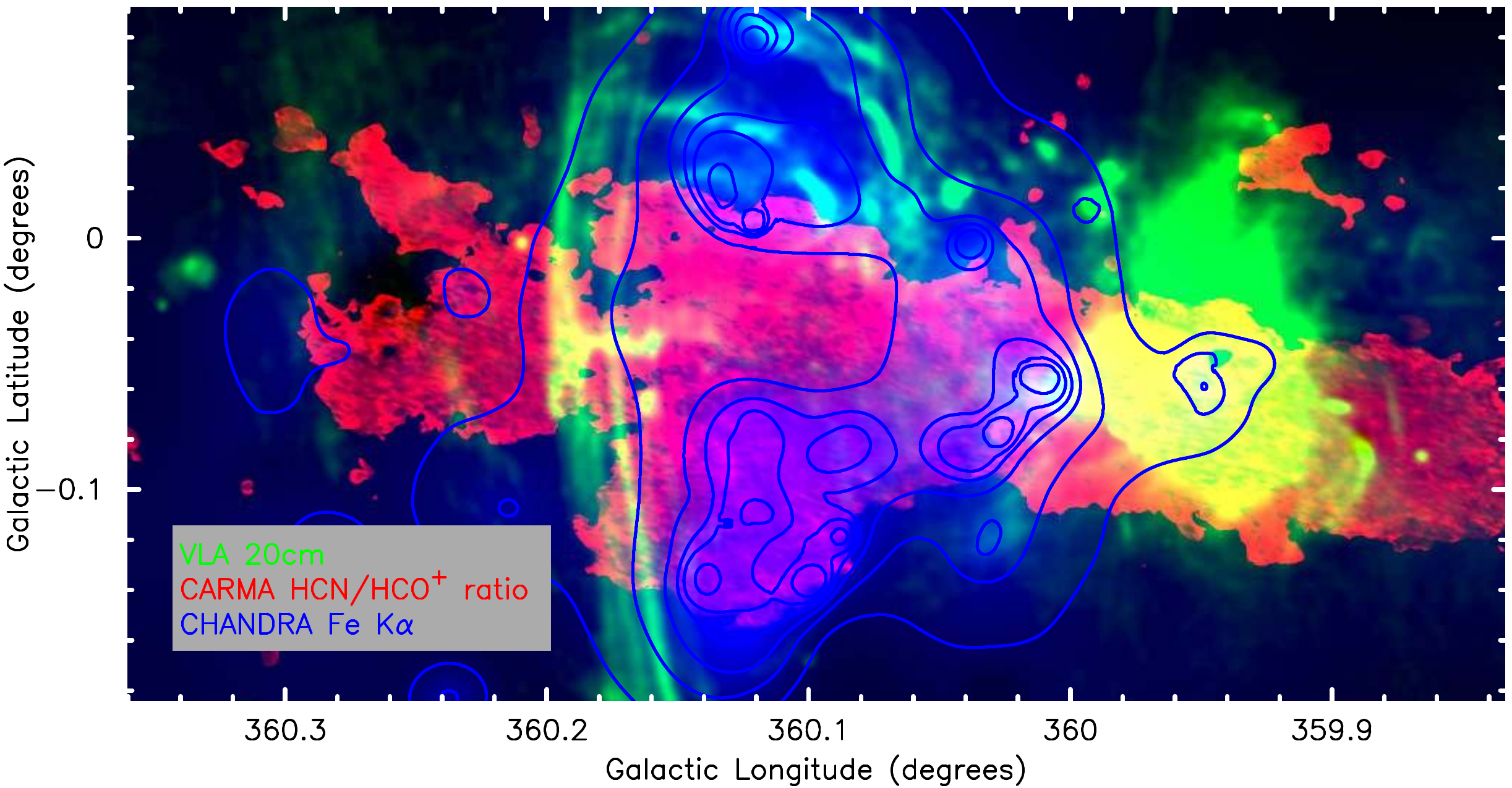}
\caption{Three-color composite showing ratio of integrated intensities
I(\hcnA)/I(\hcopA) integrated between \vlsr = 0 to 60 \kms\ 
[red] overlaid with 6.4KeV \feka\ intensity [blue], and
20~cm radio continuum [green].  
To form the ratio I(\hcnA)/I(\hcopA), we divided the maps shown in Figs. \ref{f-HCNmoments}
and \ref{f-HCOpmoments}, clipping where the I(\hcopA)$<0.1$ to avoid spuriously large
values.
The halftone range of I(\hcnA)/I(\hcopA) is 0.1 to 2.0 Jy beam$^{-1}$~\kms\ and the halftone range
of the 20~cm image is 0 to 0.1 Jy beam$^{-1}$.
Contours of  \feka\ equivalent width are 150, 225, 300, 375, 450, 600 eV.
A clear anti-correlation is seen between
I(\hcnA)/I(\hcopA)
and EW(\feka) in the region bounded by the nonthermal
radio emission. This anti-correlation is explored further in Fig. \ref{f-gluehcnhcop}. }
\label{f-rgbhcnhcop}
\end{figure*}

\begin{figure*}
\includegraphics[scale=0.71,angle=0,trim={0cm 4cm 2cm 4cm}]{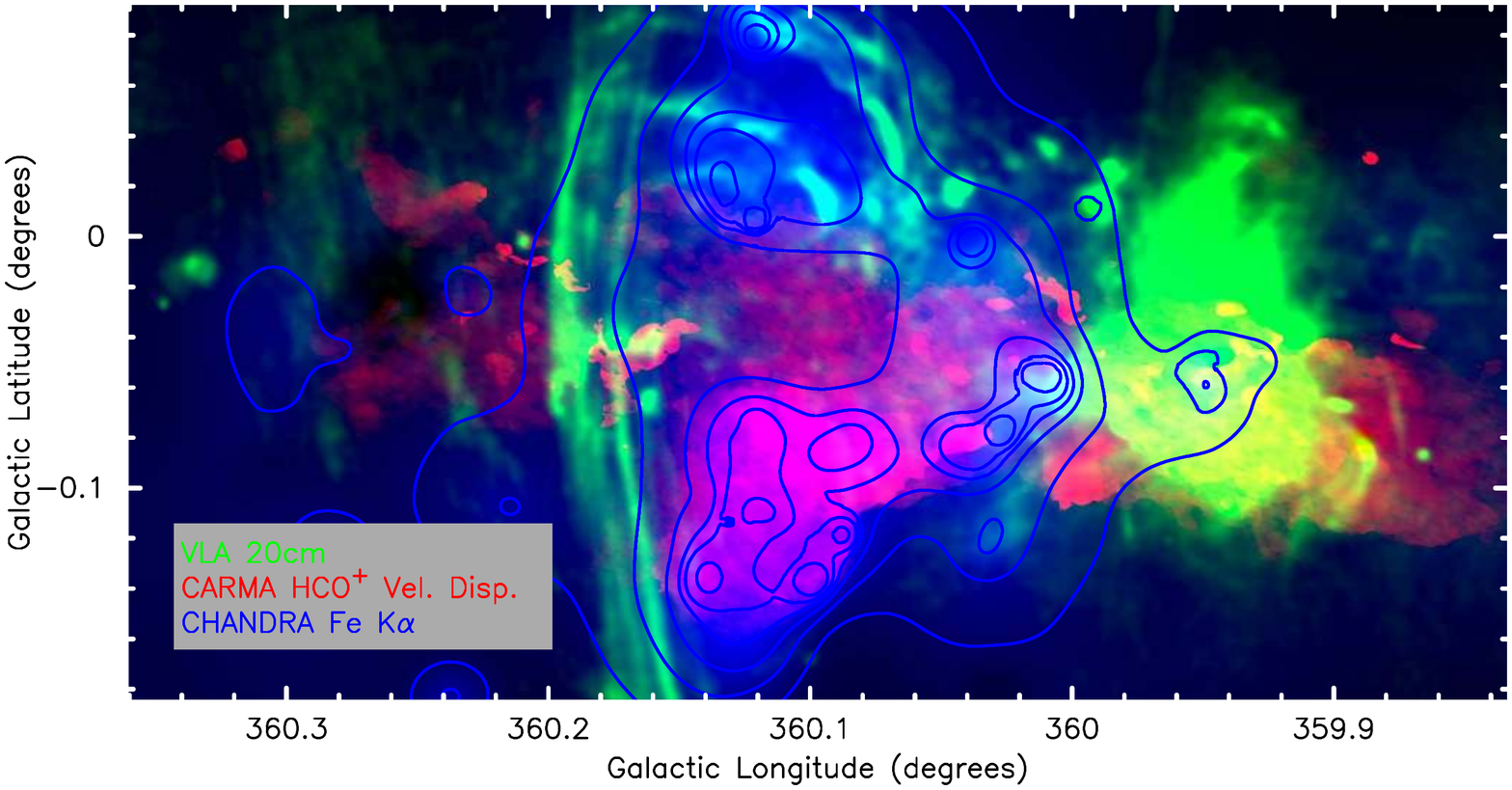}
\caption{Three-color composite showing 
velocity dispersion of \hcop\ [red] overlaid with 6.4KeV \feka\ intensity [blue], and
20~cm radio continuum [green].  
The halftone range of the velocity dispersion image is 2 to 15 \kms\ and the halftone range
of the 20~cm image is 0 to 0.1 Jy beam$^{-1}$.
Contours of  \feka\ equivalent width are 150, 225, 300, 375, 450, 600 eV.
In the region bounded by the nonthermal radio emission, the mean velocity dispersion and EW(\feka) appear positively correlated.  
}
\label{f-rgbhcopvdisp}
\end{figure*}

In Fig. \ref{f-rgbhcnhcop}, we show
the ratio of \hcn\ to \hcop\ integrated intensities  between \vlsr
= 0 to 60 \kms, overlaid with 6.4 keV line emission measured with Chandra
\citep{2007ApJ...656..847Y} and VLA 20~cm radio continuum emission 
\citep{2004ApJS..155..421Y}.  
In the region of strongest X-ray emission, bounded by the nonthermal Radio Arc (corresponding with the contour 
of equivalent width EW(\feka)$\sim 200~\rm{eV}$) and Sgr A,
there is a clear decrease in the I(HCN)/I(\hcopA) ratio compared with
the rest of the map.  In the same region
we observe an increase in \hcopA\ velocity dispersion, $\sigma_v(\hcopA)$, measured between \vlsr = 0 to 60 \kms\ 
(Fig. \ref{f-rgbhcopvdisp})
This region is known as one of the strongest Fe K$\alpha$ line emitting
region in the Galactic Center \citep{2007ApJ...656..847Y,2010ApJ...714..732P}.
Furthermore, the distribution of EW(\feka) shown here coincides not only
with enhanced \feka\ line emission but also with strong 74 MHz emission
tracing  nonthermal emission at low frequencies \citep{2013JPCA..117.9404Y}.

\begin{figure*}
\includegraphics[width=2\columnwidth,trim={1cm 1cm 1cm 4cm}]{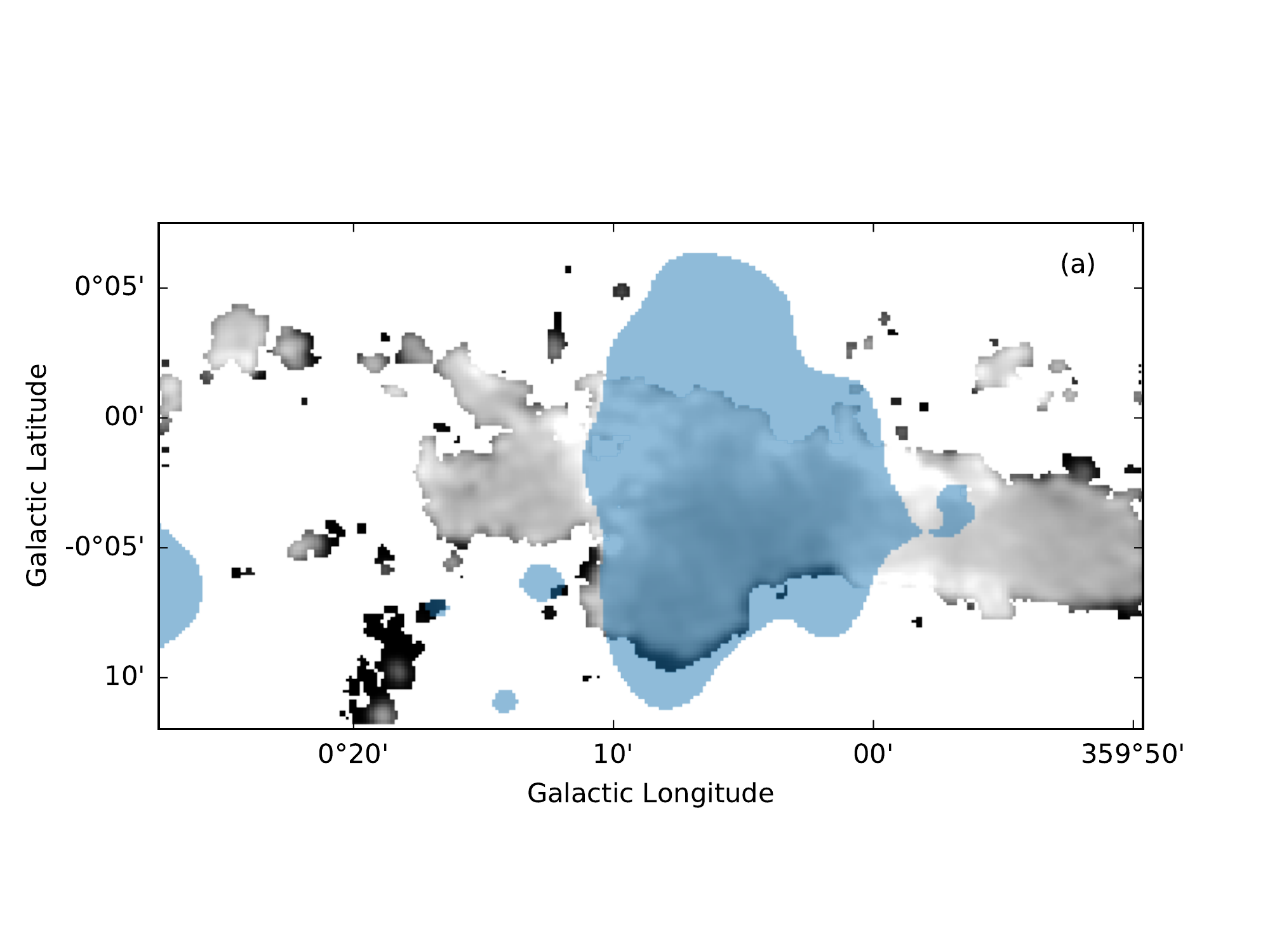} \\
\includegraphics[scale=.9,trim={1cm 7cm 1cm 7.25cm}]{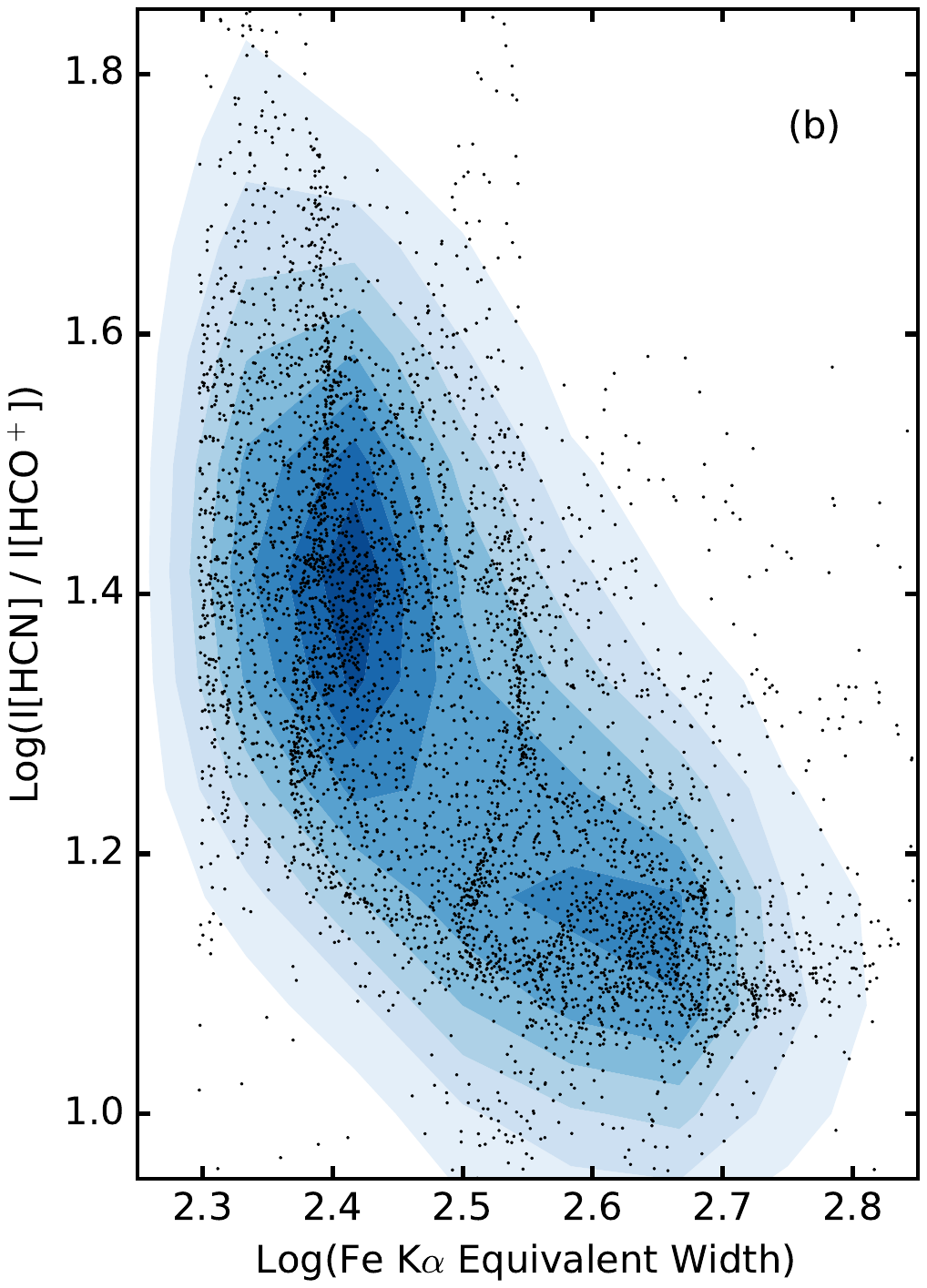}
\caption{Data exploration of the spatial anti-correlation of EW(\feka) and I(HCN)/I(\hcopA).
{\it a)} Data selection mask (blue) overlayed on a monochrome image of the
I(HCN)/I(\hcopA) ratio where darker regions indicate a lower ratio
(as in the red channel in Fig. \ref{f-rgbhcnhcop}).
The I(HCN)/I(\hcopA) map has been smoothed to 30\arcsec\ to match the resolution of the EW(\feka) map.  The blue mask highlights the spatial region where EW(\feka)$\ge$200.  {\it b)} Log-log scatter plot of EW(\feka)
vs. I(HCN)/I(\hcopA) (black points) from the regions highlighted in (a) where EW(\feka)$\ge$200.  Blue contours represent the Gaussian kernel density estimation of the data points which highlights the overall data trend: 
an anti-correlation between EW(\feka) and I(HCN)/I(\hcopA).
}
\label{f-gluehcnhcop}
\end{figure*}

In order to further demonstrate the effect, we have used the data 
exploration program Glue \citep{2014ascl.soft02002B, 2015ASPC..495..101B}
to make a linked comparison between the
EW(\feka) and  I(HCN)/I(\hcopA) 
maps.
To faciliate this comparison, we smoothed the integrated intensity ratio 
map to a 30\arcsec\ Gaussian beam.  This corresponds approximately to the resolution
of the EW(\feka) map which has variable resolution due to adaptive smoothing \citep{2007ApJ...656..847Y}.  
We then spatially rebinned the map I(HCN)/I(\hcopA) to 8\arcsec\ to match
the EW(\feka) map.   
Fig. \ref{f-gluehcnhcop} shows the comparison of
EW(\feka) and I(HCN)/I(\hcopA).  
We used Glue to select the spatial region where EW(\feka)$\ge$200 (blue);
the spatial domain plot (Fig. \ref{f-gluehcnhcop}a) displays the I(HCN)/I(\hcopA) as a grayscale monochrome image
masked (blue) by the regions where EW(\feka)$\ge$200, which corresponds roughly to the region bounded by the nonthermal radiation in Fig. \ref{f-rgbhcnhcop}.  
The scatter plot (Fig. \ref{f-gluehcnhcop}b) 
shows the pixel values of
EW(\feka) vs. I(HCN)/I(\hcopA) from the masked region (black points) and a Gaussian kernel density estimation of the data points shown in blue contours.
With this technique, the anti-correlation of
EW(\feka) and I(HCN)/I(\hcopA) within the areas of EW(\feka) $\gtrsim$ 200 and roughly bounded by the nonthermal sources is apparent.
The anti-correlation does not appear to be linear (note the axes are log scale).  Outside the region bounded by nonthermal sources (EW(\feka) $\lesssim$ 200), there is no correlation between these quantities.

One possible interpretation is that the decreased intensity
ratio traces a decreased abundance ratio.  Chemical
modelling by \citet{2013JPCA..117.9404Y} suggest that an increase in the cosmic
ray ionization rate results in a decrease to the [HCN/HCO$^+$]
abundance ratio.  In this scenario, the gas
velocity dispersion increases  due to heating of gas and damping
of MHD waves by cosmic ray electrons \citep{2013JPCA..117.9404Y}.  
We suggest an increased cosmic ray ionization rate due to the nearby
nonthermal sources may explain these observations.   Multi-transition
molecular line data are needed to determine the abundance ratio and test
chemical models of the interaction of cosmic rays and molecular gas.

\section{Summary}

There have been numerous large-scale, low-resolution  molecular line observations of 
the Galactic Center using single-dish telescopes over the last three decades.  
Here, we  showed  the first spectral line images of 
six different molecular species covering the  inner 90$\times 50$ pc of the Galactic Center region by 
combining interferometric (CARMA) and single-dish (Mopra) 
data.  
This work, which also includes 3 mm continuum survey, 
is expected to be useful for future multi-transition 
studies of
SiO, HCO$^+$, HCN, N$_2$H$^+$ and CS. The continuum data provided 
3 mm counterpart to a number of thermal, nonthermal and 
IRDCs found  toward the Galactic Center region. We qualititavely focussed on one aspect of the molecular line data
and argued that  the  molecular line ratios of \hcop\ and \hcn\ in
the region near the Radio Arc  $l \sim0.2^\circ$ and the molecular gas velocity dispersion  are anti-correlated 
with the presence of enhanced  Fe K$\alpha$ line at 6.4 keV  as well as  74 MHz emission. 
These measurements are consistent with the interaction of 
molecular gas with low-energy cosmic ray electrons  implying that the cosmic ray ionization 
rate must be higher in this region of the Galaxy. 
The high velocity dispersion of molecular gas could be explained by the higher fraction of 
cosmic ray  electrons 
coupling the magnetic field and the gas, thus increasing the damping time scale of the disturbed interacting region.

\section*{Acknowledgements}

Support for CARMA construction was derived from the states of Maryland,
California, and Illinois, the James S. McDonnell Foundation, the
Gordon and Betty Moore Foundation, the Kenneth T. and Eileen L. Norris
Foundation, the University of Chicago, the Associates of the California
Institute of Technology, and the National Science Foundation. CARMA
development and operations were supported by the National Science Foundation
under a cooperative agreement, and by the CARMA partner universities.
MWP acknowledges support from AST-2932160.  We thank Peter Teuben and Steve
Scott for their help in developing CARMA's on-the-fly mosaic capability.
This research has made use of the SIMBAD database, operated at CDS,
Strasbourg, France \citep{2000A&AS..143....9W}.  We thank Dr. Tracy Huard and the referee for helpful comments.

The continuum images and spectral line cubes described here are available in FITS from the Digital Repository at the University of Maryland, \url{http://hdl.handle.net/1903/20049}.   Data in MIRIAD visibility format are available upon request from MWP. 

\bibliographystyle{mnras}
\bibliography{ms}


\label{lastpage}
\end{document}